\documentclass[11pt]{cernrep}
\usepackage{graphicx,axodraw,epsfig}

  \renewenvironment{thebibliography}[1]{%
    \begin{oldthebibliography}{#1}%
      \setlength{\parskip}{0ex}%
      \setlength{\itemsep}{0ex}%

  }%
  {%
    \end{oldthebibliography}%
  }

\def\Ord{\lower .7ex\hbox{$\;\stackrel{\textstyle <}{\sim}\;$}}
\def\OOrd{\lower .7ex\hbox{$\;\stackrel{\textstyle >}{\sim}\;$}}

\begin{document}

%%\linenumbers

\begin{titlepage}
\def\thefootnote{\fnsymbol{footnote}}       % symbols for footnotes

\begin{center}
\mbox{ }

\end{center}
\begin{flushright}
%%% hep-ph/yymmnnn \\
%%%SHEP-06-36\\
%%%\today
%Submitted 31-8-2007
%Accepted  2-11-2007
Eur. Phys. J. C 53 (2008) 311\\
SHEP-06-36, LPT-ORSAY-07-103
\end{flushright}
\begin{center}
\vskip 2.0cm
{\LARGE\bf
{Studies of Spin Effects in Charged Higgs Boson Production\\
 with an Iterative Discriminant Analysis\\
 at the Tevatron and LHC\\}
}
\vskip 2cm
{\rm\Large 
           S. Hesselbach$^1$, 
           S. Moretti$^{1,2}$,
           J. Rathsman$^3$ and 
           A. Sopczak$^3$\footnote{Work carried out during leave from Lancaster University, Department of Physics,
           Lancaster LA1 4YW, UK.\\ E-mail: Andre.Sopczak@cern.ch}\\}
\bigskip
\bigskip
\bigskip
\large
$^1$School of Physics \& Astronomy, Southampton University, Highfield,\\
Southampton SO17 1BJ, UK\\
$^2$Laboratoire de Physique Th\'eorique, Paris XI, 91405 Orsay, France \\
$^3$High Energy Physics, Uppsala University, Box 535, S-75121 Uppsala, Sweden

\normalsize
\vskip 2.5cm
\centerline{\Large \bf Abstract}
\end{center}

\vskip 2.5cm
\hspace*{-2cm}
\begin{picture}(0.001,0.001)(0,0)
\put(0,-20){
\begin{minipage}{17cm}
\renewcommand{\baselinestretch} {1.2}
\normalsize
We report on detailed Monte Carlo comparisons of selection variables to
separate $ tbH^\pm$ signal events from the Standard Model $t\bar t$ background
using an Iterative Discriminant Analysis (IDA) method.
While kinematic differences exist between the two processes whenever 
$m_{H^\pm}\ne m_{W^\pm}$, the exploration of the spin difference between the
charged Higgs and the $W^\pm$ gauge boson becomes crucial in the particularly
challenging case of near degeneracy of the charged Higgs boson mass with the
$W^\pm$ mass.
The TAUOLA package is used to decay the tau leptons emerging from the charged Higgs 
and $W^\pm$ boson decays taking the spin difference properly into account. 
We demonstrate that, even if the individual selection variables have limited discriminant 
power, the IDA method achieves a significant separation between the expected signal and background.
%For several selection variables with limited discriminant power, the IDA
%method has proven to be very efficient.
%
For both Tevatron and LHC energies, the impact of the spin effects and $ H^\pm$ mass 
on the separation of signal and background has been studied quantitatively. The
effect of a hard transverse momentum cut to remove QCD background has been
studied and it is found that the spin effects remain important.
The separation is expressed in purity versus efficiency curves. The study is performed 
for charged Higgs boson masses between the $W^\pm$ mass and near the top mass.

\renewcommand{\baselinestretch} {1.}

\end{minipage}
}
\end{picture}

\end{titlepage}

\newpage
\normalsize
\setcounter{page}{1}
\pagestyle{plain}

\section{INTRODUCTION}
The importance of charged Higgs boson searches 
has in the recent years been emphasized~\cite{lep94,lc95,conveners,reviewLHC} 
for LEP, a future International Linear Collider (ILC), the Tevatron and  the
Large Hadron Collider (LHC),
as the detection of a charged Higgs boson would be a definite 
signal for the existence of New Physics beyond the Standard Model (SM). 
Charged Higgs bosons naturally arise in non-minimal Higgs scenarios, 
such as Two-Higgs Doublet Models (2HDMs). A Supersymmetric version of the
latter is the Minimal Supersymmetric Standard Model (MSSM). It is 
a Type II 2HDM with specific relations among neutral
and charged Higgs boson masses and couplings, dictated by
Supersymmetry (SUSY) \cite{guide}. 

The Tevatron collider at Fermilab is currently in its second stage
of operation, so-called
Run 2, with a center-of-mass (CM) energy of $\sqrt s=1.96$
TeV. This machine will be the first one to 
directly probe charged Higgs boson masses in the 
mass range up to $m_{H^\pm}\sim m_t$.
Starting from 2008, the LHC at CERN will be in a position
to confirm or rule out the existence of such a particle over a very
large portion of both the 2HDM and MSSM parameter space, 
$m_{H^\pm}\Ord 400$ GeV, depending on $\tan\beta$, 
the ratio of the vacuum expectation values of the two Higgs doublets 
(see the reviews~\cite{Roy:2004az,Roy:2004mm,Roy:2005yu} and a recent study~\cite{Mohn:2007fd}).

At present, a lower bound
on the charged Higgs boson mass exists from LEP \cite{leplimit},
$m_{H^\pm}\OOrd m_{W^\pm}$,
independently of the charged Higgs boson decay Branching Ratios (BRs).
This limit is valid within any Type II 2HDM whereas, in 
the low $\tan\beta$ region (below about 3), an indirect lower 
limit on $m_{H^\pm}$ can be derived  in the MSSM 
from the one on $m_{A}$ (the mass
of the pseudoscalar Higgs state of the model):
$m_{H^\pm}^2\approx m_{W^\pm}^2+m_{A}^2\OOrd (130~\mathrm{GeV})^2$.

If the charged Higgs boson mass $m_{H^\pm}$ satisfies 
$m_{H^\pm} < m_{t} - m_{b}$, where $ m_{t}$ is the top quark mass and 
$ m_{b}$ the bottom quark mass,
$H^\pm$ bosons could be produced in the decay of on-shell (i.e., $\Gamma_t\to0$)
top (anti-)quarks $ t \rightarrow bH^+$,
the latter being in turn produced in pairs via $gg$ fusion  and $q\bar q$ annihilation.
This approximation is the one customarily used in event generators
when $m_{H^\pm} \Ord m_{t}$. 
Throughout this study we adopt the same notation as in Ref.~\cite{Alwall:2003tc}: 
charged Higgs production is denoted by
$q\bar q$, $ gg \rightarrow t\bar t \rightarrow tbH^\pm$ if due to (anti-)top decays 
 and by $ q\bar q$, $ gg \rightarrow tbH^\pm$ if further production diagrams are included.
In fact, owing to the large top decay width ($ \Gamma_{t} \simeq 1.5$~GeV) and 
due to  the additional diagrams which do not proceed via direct $ t\bar t$ 
production~\cite{Borzumati:1999th,Miller:1999bm,Moretti:1999bw},
charged Higgs bosons could 
also be produced at and beyond the kinematic top decay threshold. 
The importance of these effects in the so-called `threshold' or 
`transition'  region ($m_{H^\pm}\approx m_t$) was emphasized in 
Les Houches proceedings~\cite{Cavalli:2002vs,Assamagan:2004mu}
as well as in Refs.~\cite{Alwall:2003tc,Guchait:2001pi,Moretti:2002ht,Assamagan:2004gv}, 
so that the calculations of Refs.~\cite{Borzumati:1999th,Miller:1999bm} 
(based on the appropriate $q\bar q,gg\to tb H^\pm$ description) 
are now implemented in 
HERWIG\,\cite{herwig,Corcella:2000bw,Corcella:2002jc,Moretti:2002eu}\,and 
PYTHIA\,\cite{pythia,Alwall:2004xw}. A comparison between
the two generators was carried out in Ref.~\cite{Alwall:2003tc}.
For any realistic simulation of $H^\pm$ production with  
$m_{H^\pm}\OOrd m_t$
the use of  these implementations is important.
In addition, 
in the mass region near the top quark mass, a matching of the calculations for the
$ q\bar q,~gg \rightarrow tbH^\pm$ and 
$ gb \rightarrow tH^\pm$ processes might be required
\cite{Alwall:2004xw}.

A charged Higgs boson with $m_{H^\pm}\Ord m_{t}$
decays predominantly into a $\tau$ lepton and
a neutrino. 
For large values of $\tan \beta$ ($\OOrd$ 5) the corresponding BR
is near 100\%. 
For $m_{H^\pm}\OOrd m_{t}$, $H^\pm\to \tau\nu_\tau$ is overtaken
by $H^\pm\to tb$, but the latter is much harder to disentangle from background
than the former.  
The associated top quark decays predominantly into a $W^\pm$ boson, or at times 
a second charged Higgs boson,
and a $b$ quark. 
The reaction
\begin{equation}
q\bar q, gg \to  tbH^\pm~~~(t\to bW)~~~(H^\pm \to \tau ^\pm \nu_{\tau})
\label{channel}
\end{equation}
is then a promising channel to search for a charged Higgs boson at both the Tevatron
(where the dominant production mode is $q\bar q$)
and the LHC (where $gg$ is the leading subprocess). If
the  $H^\pm\to\tau\nu_\tau$ decay channel is used to search for Higgs bosons, then
a key ingredient in the signal selection process should be the exploitation of
decay distributions that are sensitive to the spin nature of the particle
yielding the $\tau$ lepton ($H^\pm$ in the signal or $W^\pm$ in the
background), as advocated in 
Refs.~\cite{Roy:1991sf,Raychaudhuri:1995kv,Raychaudhuri:1995cc,Roy:1999xw}
(see also \cite{Assamagan:2002in,Assamagan:2002ne}).
The $\tau$ spin information affects both the energy and
the angular distribution of the $\tau$ decay products.

In the search for a charged Higgs boson signal
containing a $\tau$ lepton,
not only the magnitude of the production  
cross section is important, but also the efficiency of identifying the
$\tau$ lepton in the hadronic environment plays a crucial role.  
Since $\tau$ leptons have a very short life-time ($\sim10^{-6}$~s),
they decay within the detectors and can only be identified through
their decay products. In about 35\% of the cases they decay
leptonically and about 65\% of the times they do so hadronically. 
Both of these decay modes are usually addressed in charged Higgs boson
searches by employing dedicated $\tau$ lepton triggers. 
The identification of taus in hadronic $p\bar p$ collisions has recently
been studied, e.g. $Z\to\tau^+\tau^-$ events~\cite{Abazov:2004vd} and 
further details are given in~\cite{leshouches05}.

It is the purpose of this note to outline the possible improvements 
that can be achieved at the Tevatron and LHC in the search for charged 
Higgs bosons, with mass below the top mass and including the appropriate 
description of the spin effects in the $H^\pm\to\tau\nu_\tau$ decay.
%In order to quantify the spin effect an Iterative Discriminant Analysis (IDA)
%method has been applied. 
%%The IDA method is a powerful tool to separate
%%signal and background, in particular when several selection variables 
%%with limited discriminant power are present.
%In the following, the IDA method as a powerful tool to separate signal and 
%background, is applied.
In order to quantify the spin effect an Iterative Discriminant Analysis
(IDA) method has been applied, which is a powerful tool to separate signal 
and background, even in cases such as the one presently under study when 
several selection variables with limited discriminant power are present.

\section{TEVATRON ENERGY}

We start by studying charged Higgs production $q\bar q, gg \to  tbH^\pm$
with subsequent decays $t \to b W$, $H^\pm \to \tau \nu_\tau$
at the FNAL Tevatron with $\sqrt{s} = 1.96$~TeV.
In the following we analyze hadronic decays of
the $W^\pm$ boson and $\tau$ lepton
($W^\pm \to q\bar{q}'$, $\tau \to \mathrm{hadrons} + \nu_\tau$),
which results in the signature
$2 b + 2 j + \tau_\mathrm{jet} + p_t^{\rm miss}$
(2 $b$ jets, 2 light jets, 1 $\tau$ jet and missing transverse momentum).
The most important irreducible 
background process is $q\bar q, gg \to  t\bar{t}$
with the subsequent decays $t \to b W^+$ and $\bar{t} \to \bar{b} W^-$,
one $W^\pm$ boson decaying hadronically ($W^\pm \to q\bar{q}'$)
and one leptonically ($W^\mp \to \tau \nu_\tau$), which results in
the same final state particles as for the expected signal. 

\subsection{Simulation and Detector Response}
The signal process $q\bar q, gg \to  tbH^\pm$ is simulated with PYTHIA 
\cite{pythia}.
The subsequent decays $t \to b W^\pm$ (or its charge conjugate),
$W^\pm \to q\bar{q}'$ and
$H^\mp \to \tau \nu_\tau$ are also carried out within PYTHIA,
whereas the $\tau$ leptons are decayed externally with the program TAUOLA
\cite{Jadach:1990mz, Golonka:2003xt},
which includes the complete spin structure of the $\tau$ decay.
The background process $q\bar q, gg \to  t\bar{t}$ is also simulated
with PYTHIA with the built-in subroutines for $t\bar{t}$ production.
The decays of the top quarks and $W^\pm$ bosons are performed within PYTHIA
and that of the $\tau$ lepton within TAUOLA.

The momenta of the final $b$ and light quarks from the PYTHIA event
record are taken as the momenta of the corresponding jet, whereas for the
$\tau$ jet the sum of all non-leptonic final state particles as given by
TAUOLA is used.
The energy resolution of the detector and parton shower and hadronization effects
are emulated through a Gaussian
smearing $(\Delta(p_t)/p_t)^2 = (0.80/\sqrt{p_t})^2$
of the transverse momentum $p_t$
for all jets in the final state, including the $\tau$ jet~\cite{conveners}.
As typical for fast simulation studies, no effects of underlying events, are simulated.
Events are removed which contain jets with less than 20 GeV transverse 
momentum\footnote{In order to be largely independent of the specific 
detector performance, no requirement on the jet resolution is applied.}, 
corresponding to about $|\eta|>3$.
The transverse momentum of the leading charged pion in the $\tau$ jet
is assumed to be measured in the tracker independently of the
transverse momentum of the $\tau$ jet. The identification and momentum 
measurement of the pion is important to fully exploit the $\tau$ spin information.
In order to take into account the tracker performance
we apply  Gaussian smearing on $1/p_t^\pi$ with
$ \sigma(1/p_t^\pi)[\mathrm{TeV}^{-1}] =
 \sqrt{0.52^2 + 22^2/(p_t^\pi[\mathrm{GeV}])^2 \sin\theta_\pi}$,
where $\theta_\pi$ is the polar angle of the $\pi$.
The missing transverse momentum \mbox{$p_t^{\rm miss}$} is constructed from
the transverse momenta of all visible jets 
(including the visible $\tau$ decay products)
after taking the modelling of the detector into account.
The generic detector description is a good approximation for both Tevatron experiments,
CDF and D0.

\subsection{Expected Rates}
For completeness we present a brief discussion of the expected cross section 
of the charged Higgs boson signature under investigation.
The signal cross section has been calculated for
$\tan\beta = 30$
and $m_{H^\pm} = 80, 100, 130$ and $150$~GeV with PYTHIA, version 6.325,
using the implementation described in \cite{Alwall:2004xw},
in order to take the effects in the transition region into account. Furthermore,
it has been shown in \cite{Alwall:2003tc} that the signal cross section
for $tbH^\pm$ agrees with the one from the top-decay approximation
$t\bar t \to tbH^\pm$ for charged Higgs boson masses up to about 160~GeV
if the same factorization and renormalization scales are used.
Thus, we have used everywhere in this study the factorization scale $(m_t + m_{H^\pm})/4$ and the renormalization scale $m_{H^\pm}$ for both signal and background (i.e., those recommended in \cite{Alwall:2004xw} as most appropriate for the $tbH^\pm$ 
signal)\footnote{Clearly, for a proper
experimental study, factorization and renormalization scales for our background process 
$q\bar q$, $ gg \rightarrow t\bar t\to tbW^\pm$
ought to be chosen appropriately, i.e., unrelated to the charged Higgs boson mass.},
since the primary purpose of our study is to  single out variables that show a 
difference between our $W^\pm$ and $H^\pm$ data samples and that this
can unambiguously be ascribed to the different nature of the two kinds of bosons
(chiefly, their different mass and spin state).
In addition, the running $b$ quark mass entering in the Yukawa
coupling of the signal has been evaluated at $m_{H^\pm}$.
This procedure eventually results in a dependence of our background calculations on $\tan\beta$
and, especially, $m_{H^\pm}$ that is more marked than the one that would more naturally arise 
as only due to indirect effects through the top decay width.
Hence, the cross sections have been rescaled with a common factor
such that the total $t \bar t$ cross section is
$\sigma^{\rm prod}_{t\bar{t}} = 5.2$~pb \cite{ttbarxsec}.
To be more specific, we have first calculated the total cross section 
$\sigma^{\rm prod,PYTHIA}_{t\bar{t}}(m_{H^\pm})$ with the 
built-in routine for $t \bar t$ production
in PYTHIA for all $m_{H^\pm} = 80, 100, 130$ and $150$~GeV and then
calculated from this the respective rescaling factors
$c(m_{H^\pm}) = 
 5.2~\mathrm{pb}/\sigma^{\rm prod,PYTHIA}_{t\bar{t}}(m_{H^\pm})$
for each $m_{H^\pm}$.
Then we have calculated the background cross section 
for $m_{H^\pm}=80$~GeV into the final state with the signature
$2 b + 2 j + \tau_\mathrm{jet} + p_t^{\rm miss}$
by enforcing the respective decay channels in PYTHIA using the 
built-in routine for $t \bar t$ production and multiplied 
it with $c(80~\mathrm{GeV})$.
In the same manner we have calculated the signal cross sections with the
PYTHIA routines for $tbH^\pm$ production by enforcing the respective decay
channels in PYTHIA and multiplying with the rescaling factors
$c(m_{H^\pm})$ for $m_{H^\pm} = 80, 100, 130, 150$~GeV.
The resulting cross sections 
are given in Table~\ref{tab:Hpm:crosssec} before 
($\sigma^{\rm th}$) and after ($\sigma$) applying the basic cuts 
$p_t^{\rm jets} > 20$~GeV and the hard cut $p_t^{\rm miss} > 100$~GeV.
For the four signal masses, the $tbH^\pm$ and $ t\bar t \to tbH^\pm$
cross section calculations agree numerically.

\begin{table}[htbp]
\caption{\label{tab:Hpm:crosssec}
Tevatron cross sections of background $q\bar q, gg \to  t\bar{t}$
and signal $q\bar q, gg \to  tbH^\pm$
for $\tan\beta = 30$ and $m_{H^\pm} = 80, 100, 130$ and $150$~GeV
into the final state
$2 b + 2 j + \tau_\mathrm{jet} + p_t^{\rm miss}$
before ($\sigma^{\rm th}$) and after ($\sigma$)
the basic cuts ($p_t > 20$~GeV for all jets)
and the hard cut ($p_t^{\rm miss} > 100$ GeV).
}
\centering
\begin{tabular}{c|c|c|c|c|c}
 & $q\bar q, gg \to  t\bar{t}$ &
  \multicolumn{4}{c}{$q\bar q, gg \to  tbH^\pm$} \\
$m_{H^\pm}$ (GeV) & 80 & 80 & 100 & 130 & 150\\
  \hline
$\sigma^{\rm th}$ (fb) & 350 &  535    &  415    & 213 & 85  \\
$\sigma$ (fb) for $p_t^\mathrm{jets} > 20$ GeV 
         & 125   & 244     &   202   &  105 & 32 \\
$\sigma$ (fb) for $(p_t^\mathrm{jets},p_t^{\rm miss}) > (20,100)$ GeV 
         & 21   &   30   &  25    &  18 & 7 \\
\end{tabular}
\end{table}

\subsection{Event Preselection and Discussion of Discriminant Variables}
The expected cross sections of the 
$2 b + 2 j + \tau_\mathrm{jet} + p_t^{\rm miss}$ signature 
are of the same order of magnitude for the signal and background 
reactions, as shown in Table~\ref{tab:Hpm:crosssec}.
%
%For the analysis of different kinematic selection variables
%$5\cdot 10^5$ signal events have been simulated with PYTHIA
%for each charged Higgs mass at the Tevatron energy of 1.96~TeV 
%using the built-in $t\bar t$ routine in the 
%$t\bar t \rightarrow tbH^\pm$ approximation.
%For the $t\bar t$ background also $5\cdot 10^5$ events
%have been simulated using the built-in $t\bar t$ routine.
%Thus, the same number of signal and background events is assumed.
Thus, the same number of signal and background events is assumed for the 
analysis of different kinematic selection variables.
For the signal $5\cdot 10^5$ events have been simulated with PYTHIA for each 
charged Higgs mass at the Tevatron energy of 1.96~TeV using the built-in 
$t\bar t$ routine in the $t\bar t \rightarrow tbH^\pm$ approximation, 
while for the $t\bar t$ background also $5\cdot 10^5$ events have been 
simulated using the built-in $t\bar t$ routine.
Then the basic cuts $p_t^{\rm jets} > 20$~GeV are applied.
An additional hard cut on the missing transverse momentum
$p_t^{\rm miss} > 100$ GeV is used to suppress the QCD background, as 
for example demonstrated in Ref.~\cite{Assamagan:2002in}.
%
%The basic cuts imply a pseudo-rapidity region of about 
After the additional anti-QCD cut about 28000 to 42000 signal events,
depending on the simulated charged Higgs bosons mass, and about 30000 
$t\bar t$ background events remain.
Other background reactions, for example W+jet production, are expected to be 
negligible because they have either a much lower production cross section
or are strongly suppressed compared to $t\bar t$ background, 
as quantified for example in Ref.~\cite{Assamagan:2002in}.
In addition to the previous study 
(based on $5000\times \mathrm{BR}(\tau \to \mathrm{hadrons})$ events
each)~\cite{leshouches05},
the present one applies an IDA method~\cite{ida}
to explore efficiencies and purities. As already mentioned,
particular attention is devoted to the study of 
spin sensitive variables in the exploitation of polarization
effects for the separation of signal and background events.

\begin{figure}[htbp]
\epsfig{file=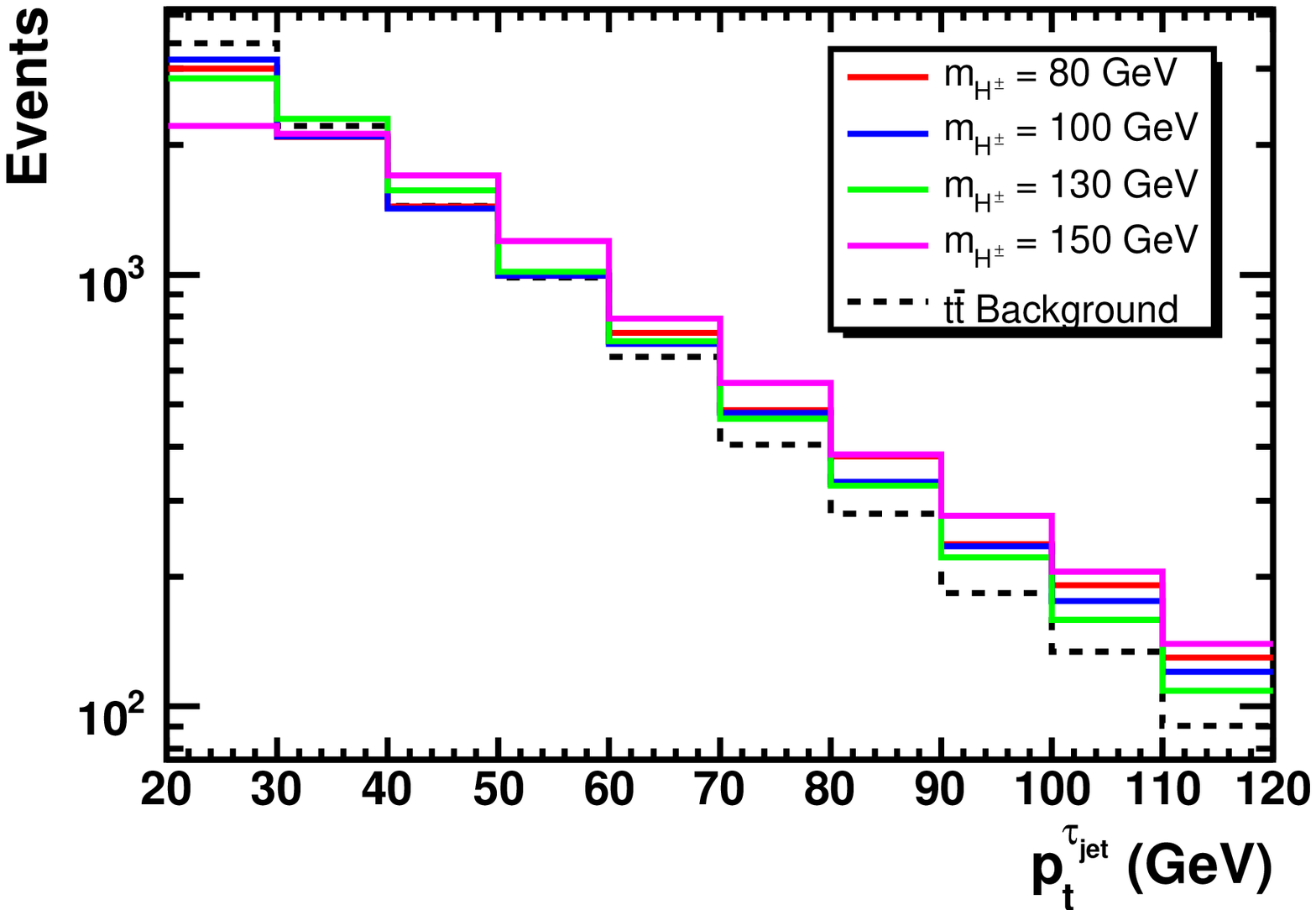, width=0.5\textwidth} \hfill
\epsfig{file=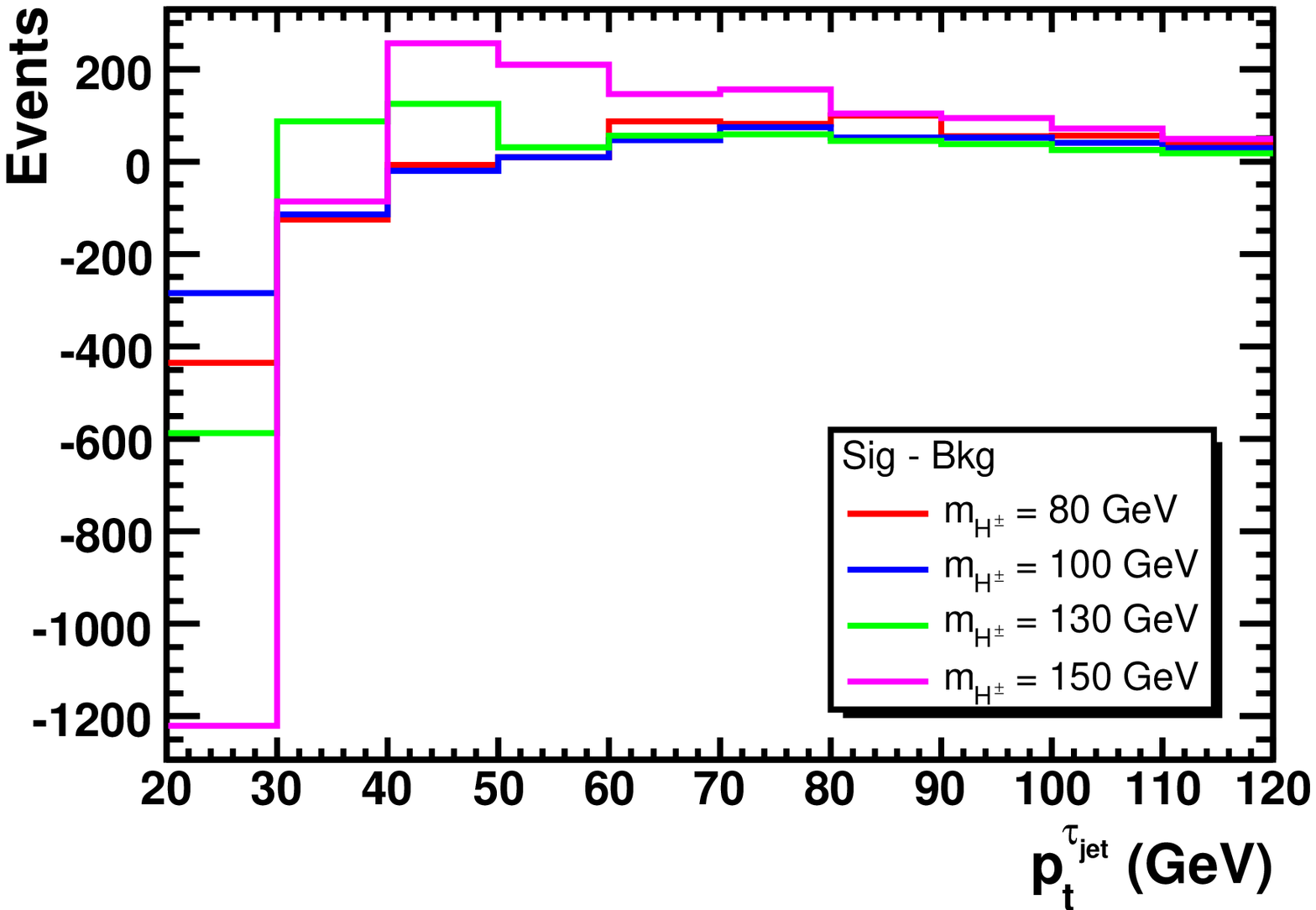, width=0.5\textwidth}
\epsfig{file=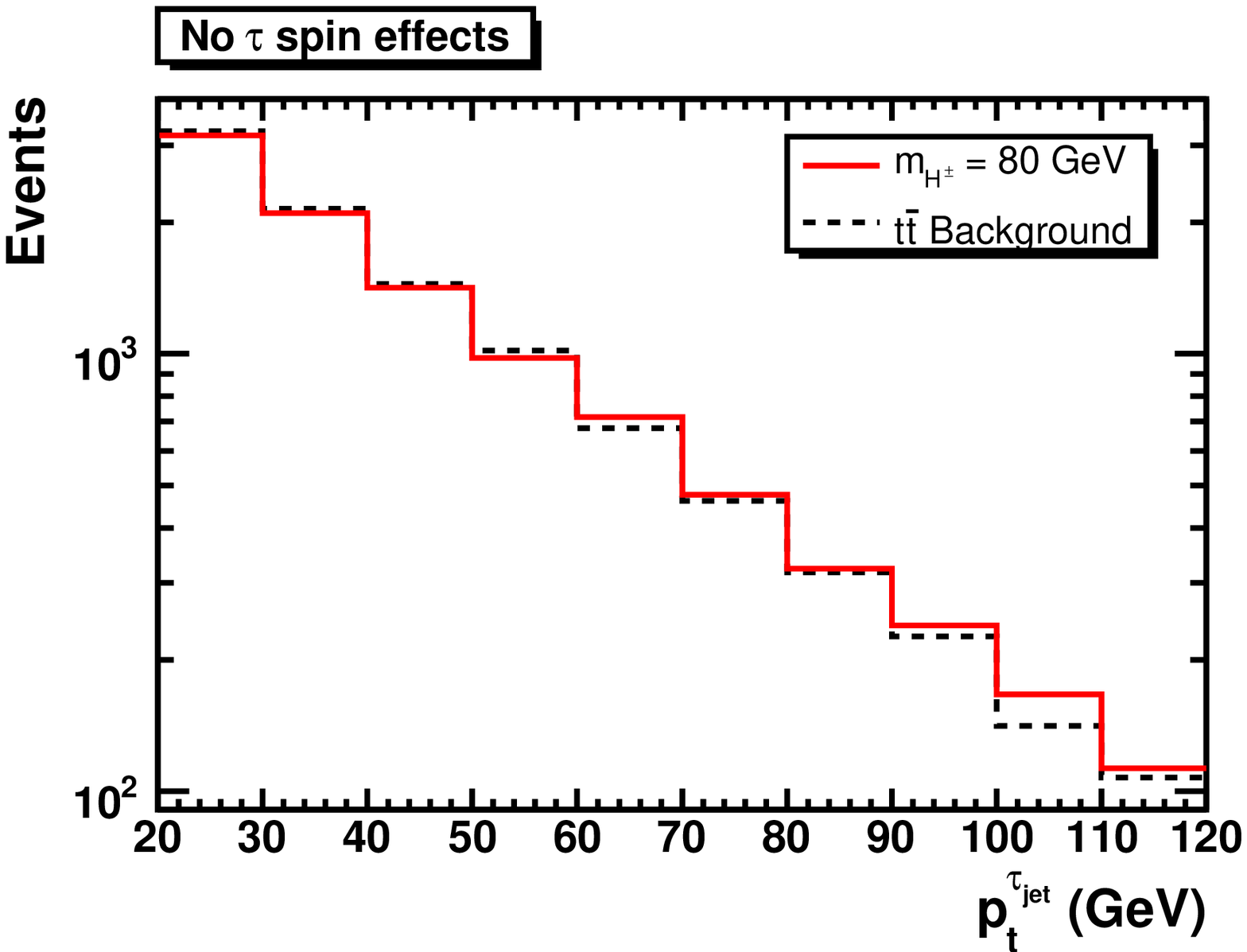, width=0.5\textwidth} \hfill
\epsfig{file=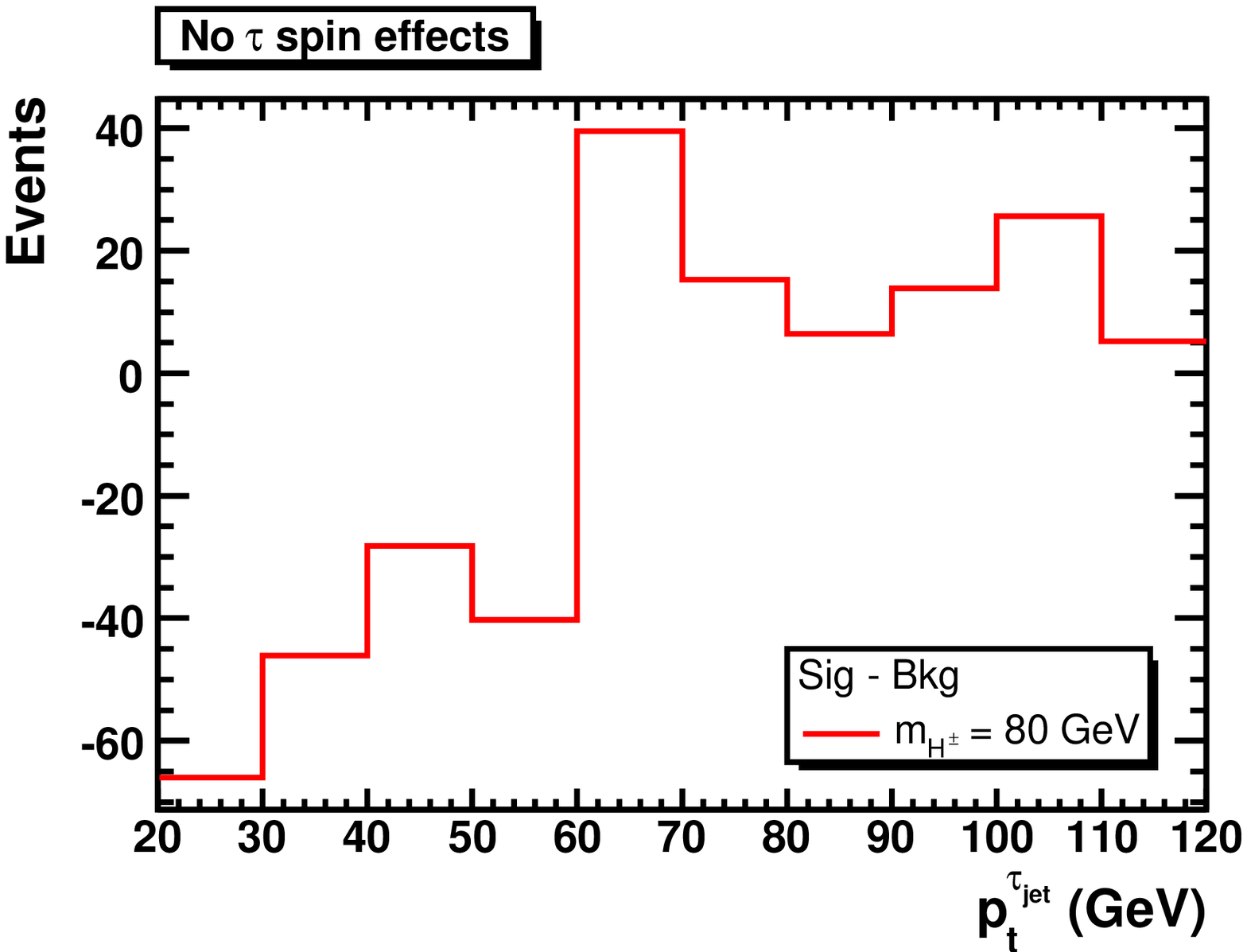, width=0.5\textwidth}
\caption{ 
$p_t$ distributions 
of the $\tau~{\mathrm{jet}}$ for the $tbH^\pm$ signal
and the $t\bar{t}$ background for $\sqrt{s}=1.96$~TeV (left)
and the respective differences between signal and background (right).
The lower plots show distributions without spin effects in the $\tau$
decays.
}
\label{fig:pttau}
\end{figure}

\begin{figure}[htbp]
\epsfig{file=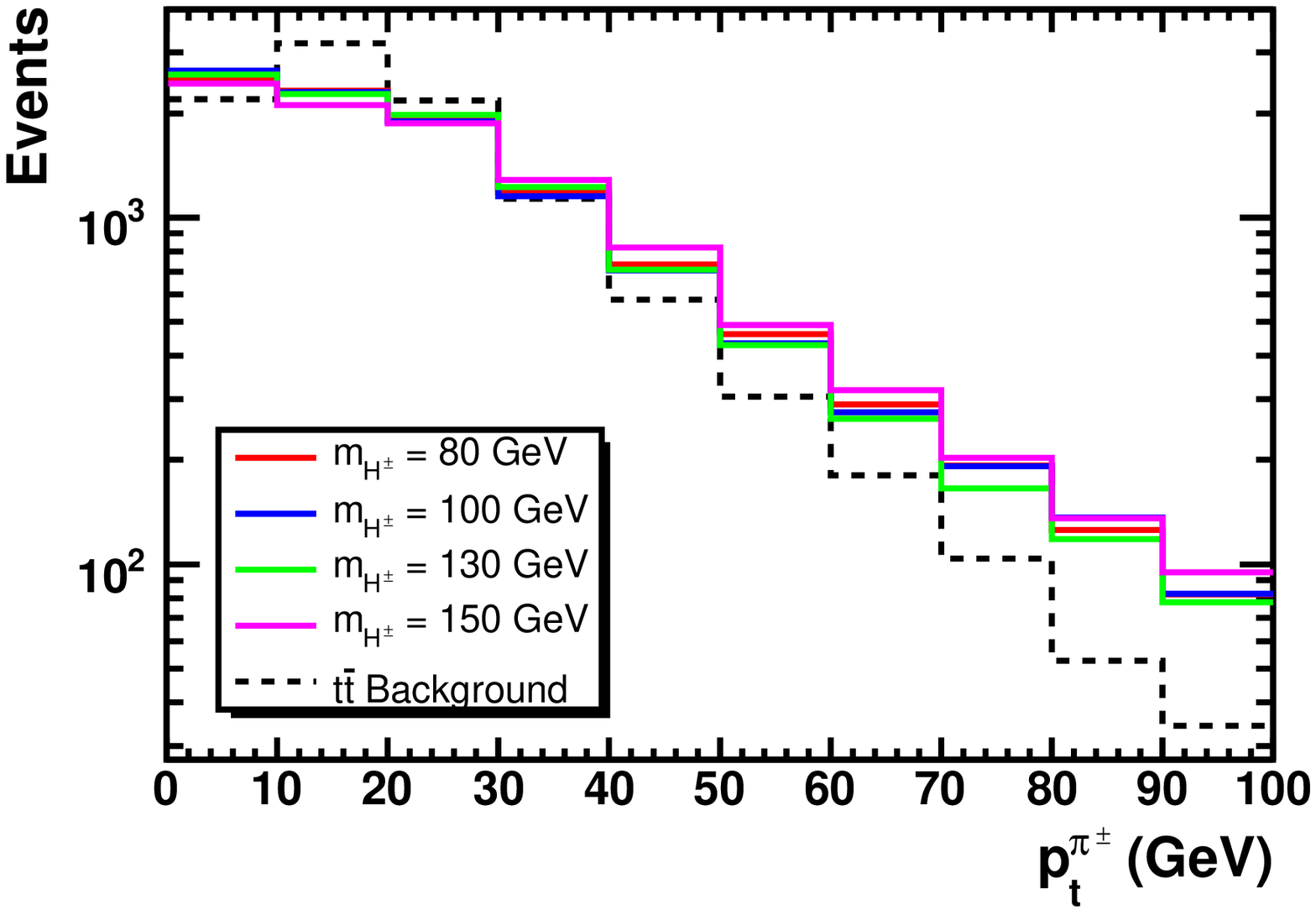, width=0.5\textwidth}  \hfill
\epsfig{file=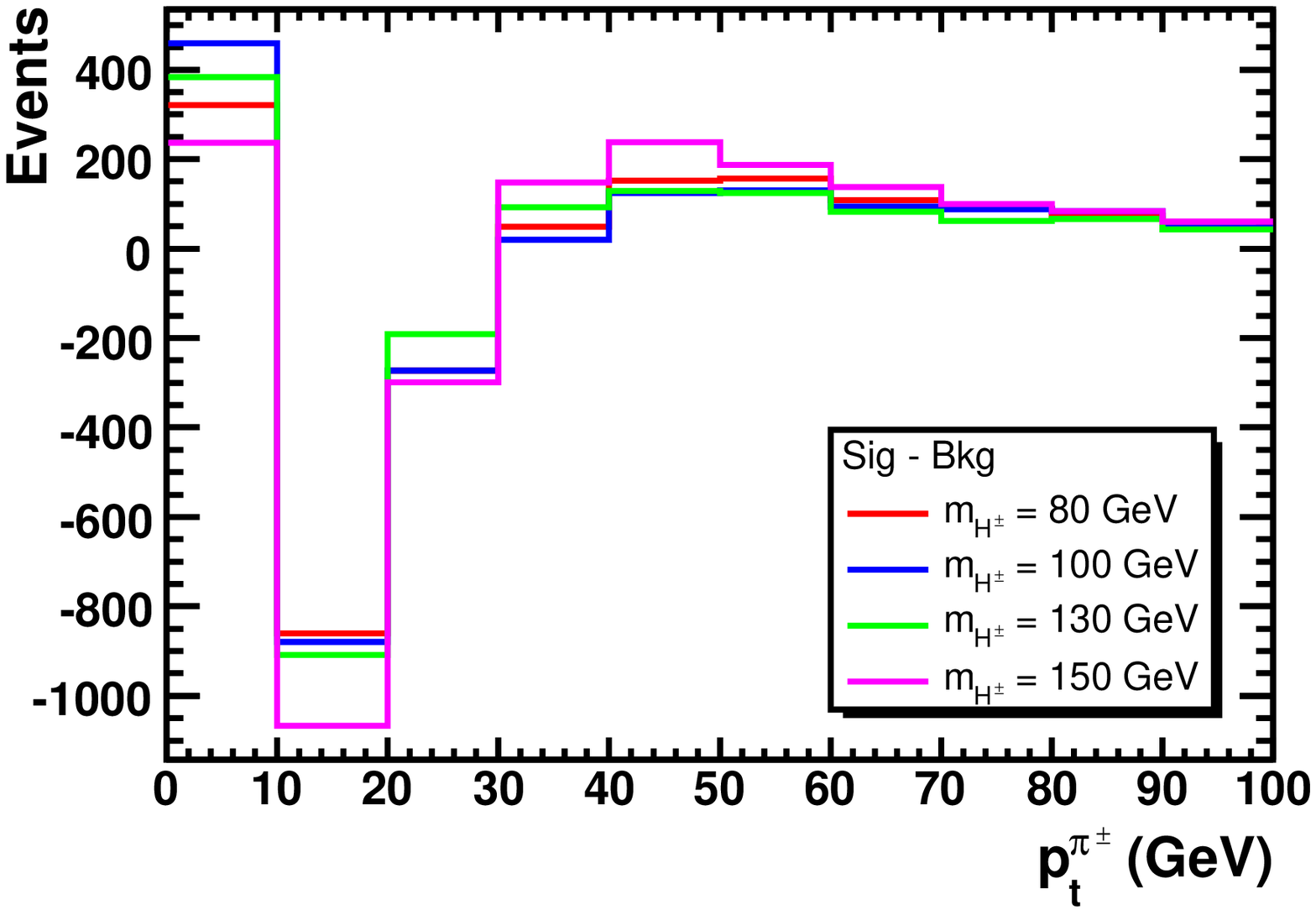, width=0.5\textwidth}
\epsfig{file=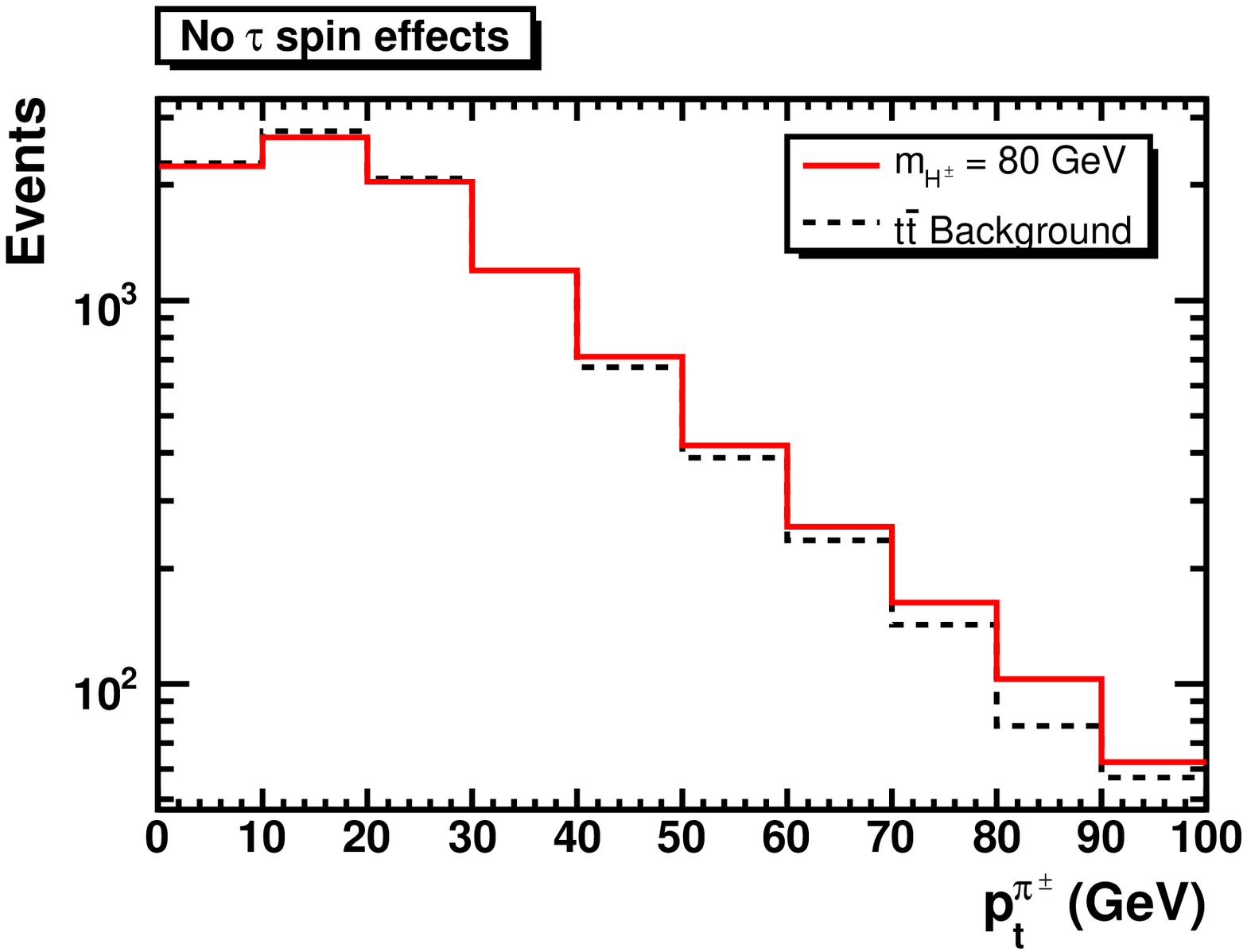, width=0.5\textwidth}  \hfill
\epsfig{file=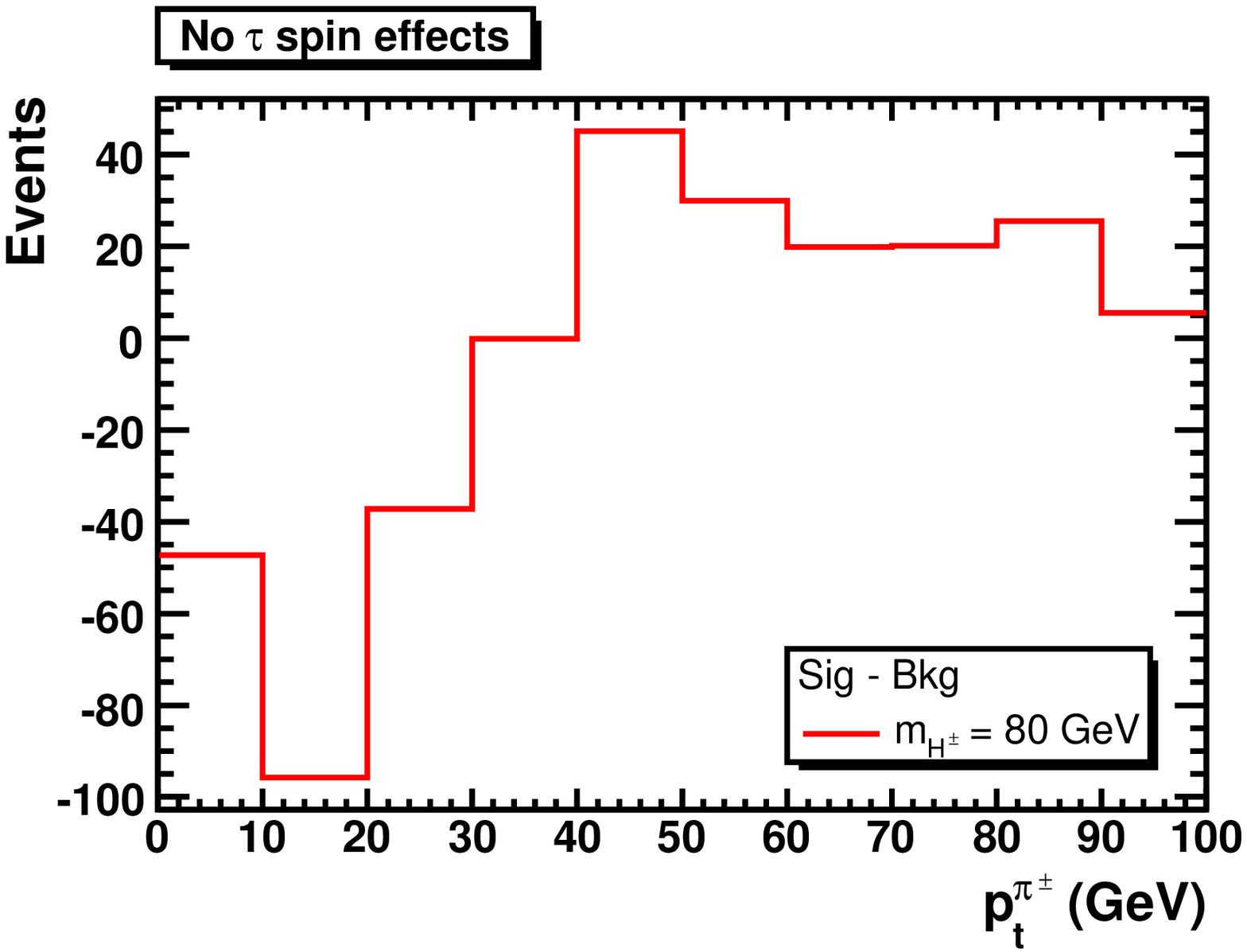, width=0.5\textwidth}
\caption{
$p_t$  distributions of the leading $\pi^\pm$ from the $\tau$ decay 
for the $tbH^\pm$ signal and the 
$t\bar{t}$ background for $\sqrt{s}=1.96$~TeV (left)
and the respective differences between signal and background (right).
The lower plots show distributions without spin effects in the $\tau$
decays.
}
\label{fig:ptpi}
\end{figure}

\begin{figure}[htbp]
\epsfig{file=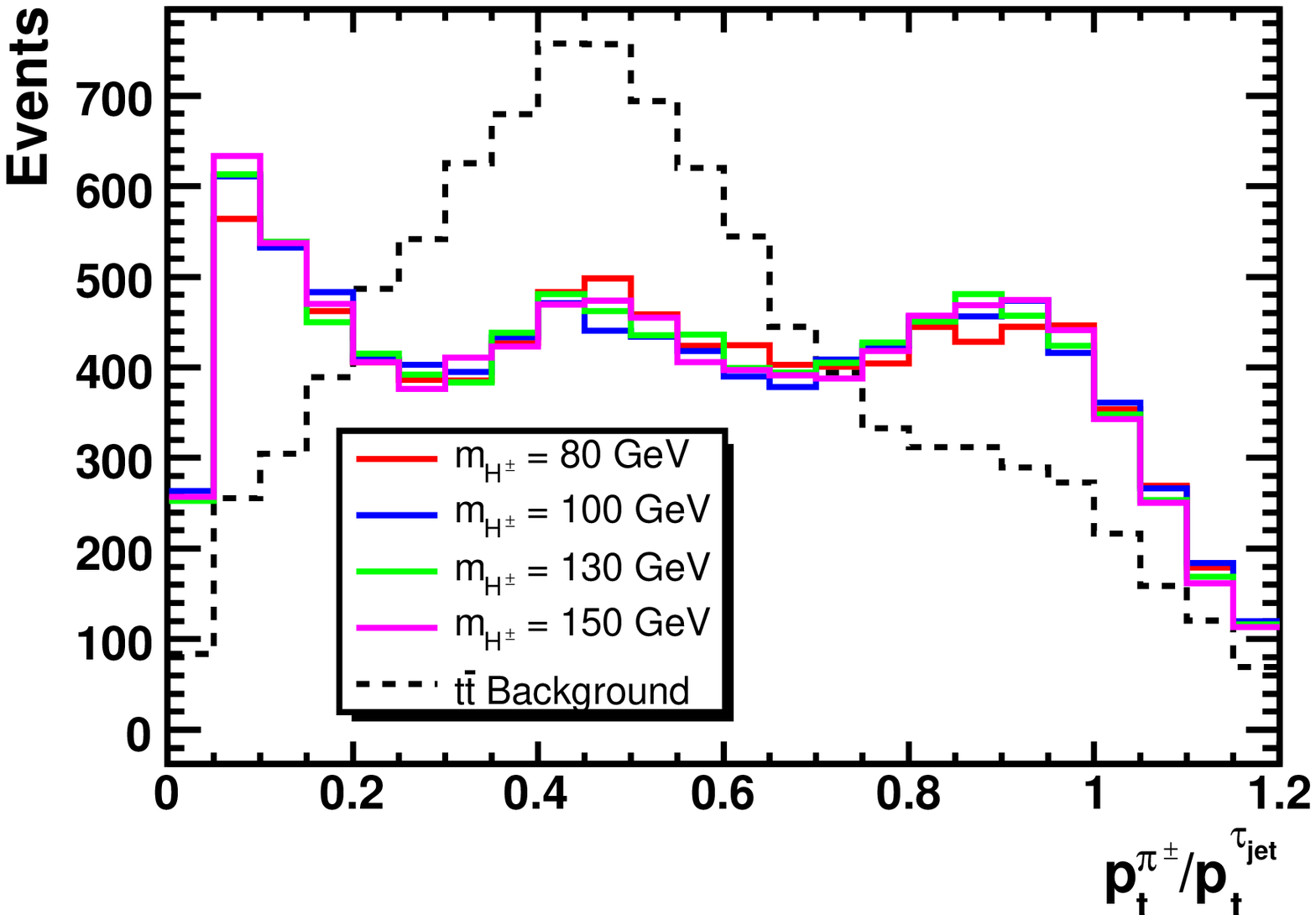, width=0.5\textwidth} \hfill
\epsfig{file=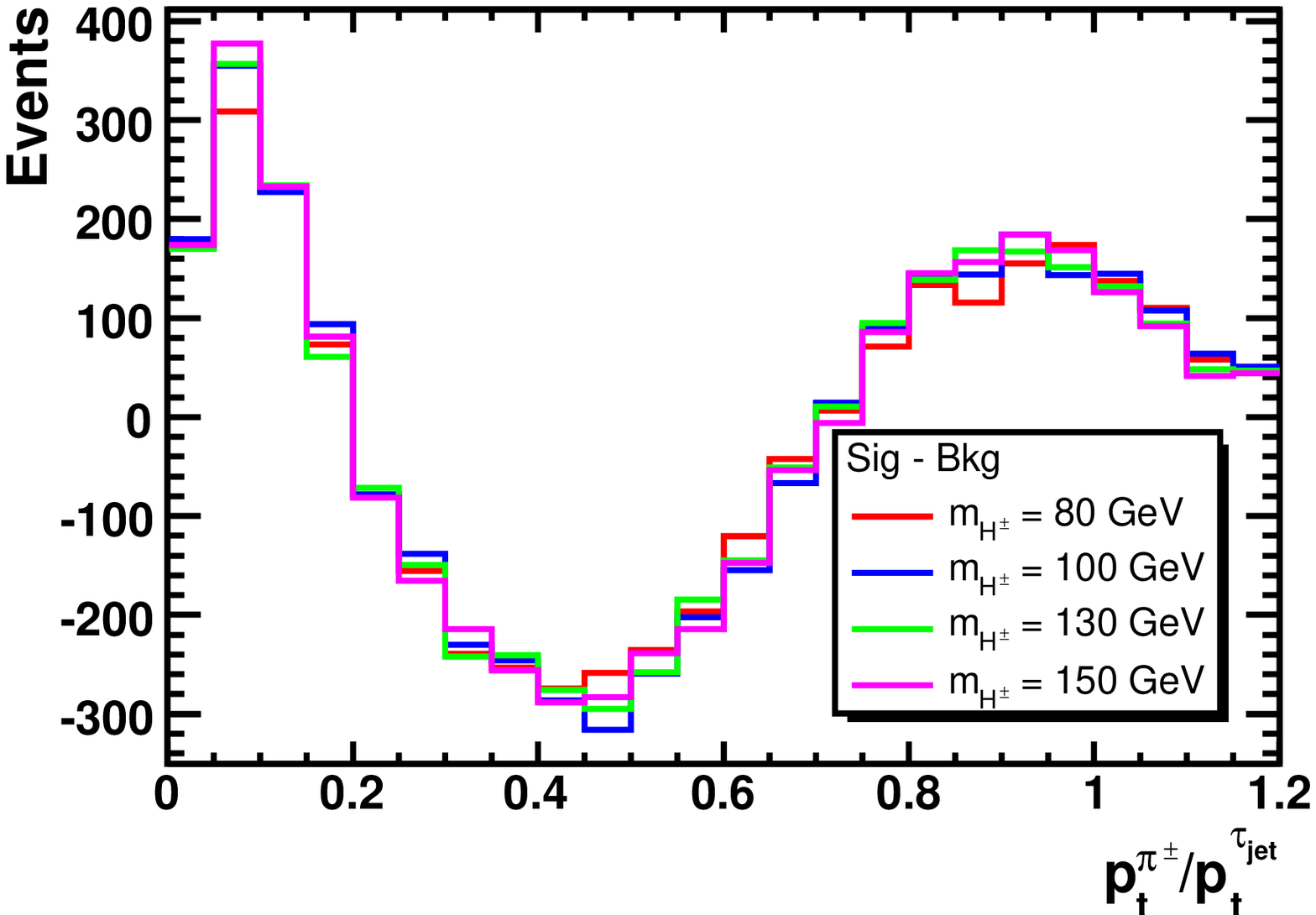, width=0.5\textwidth}
\epsfig{file=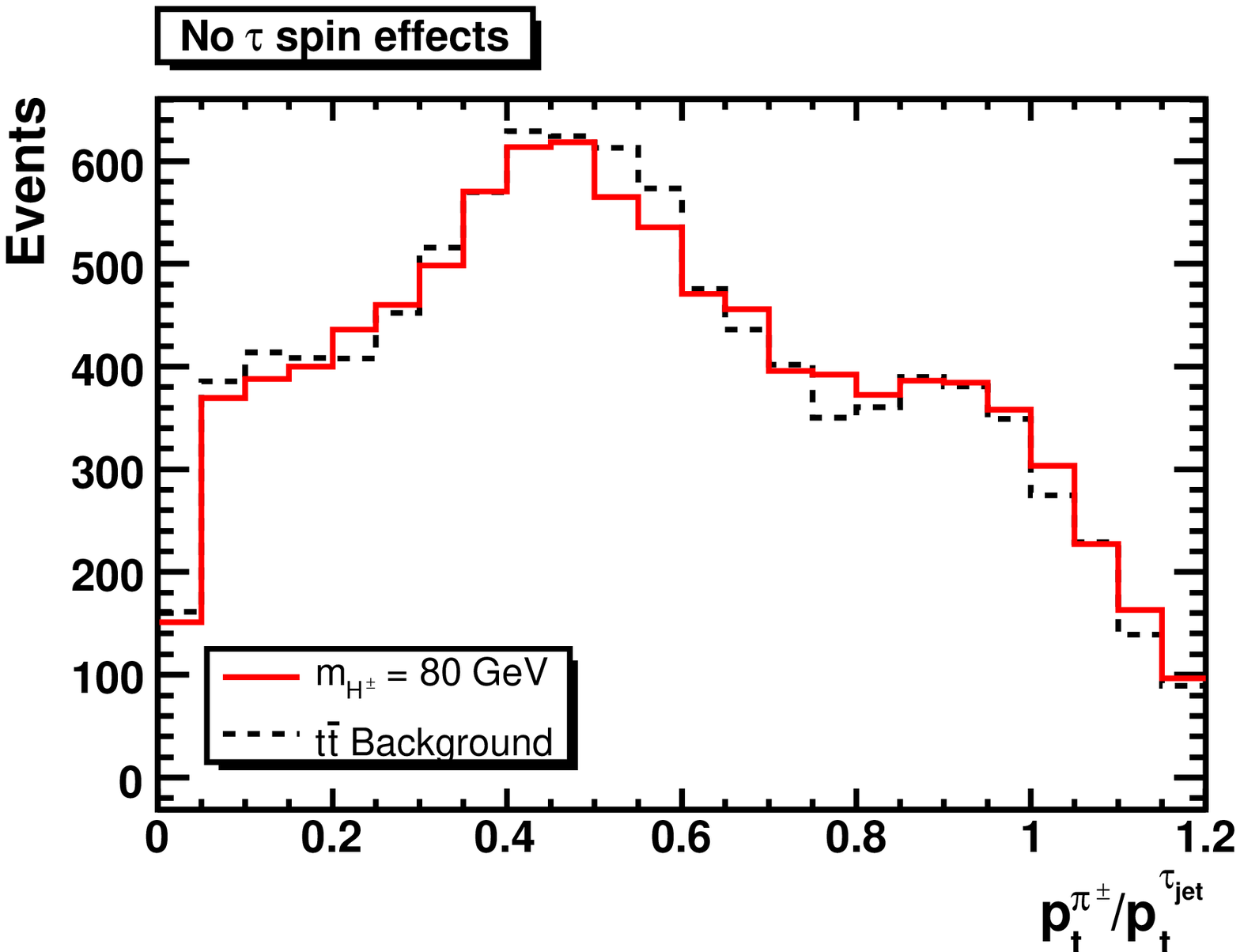, width=0.5\textwidth} \hfill
\epsfig{file=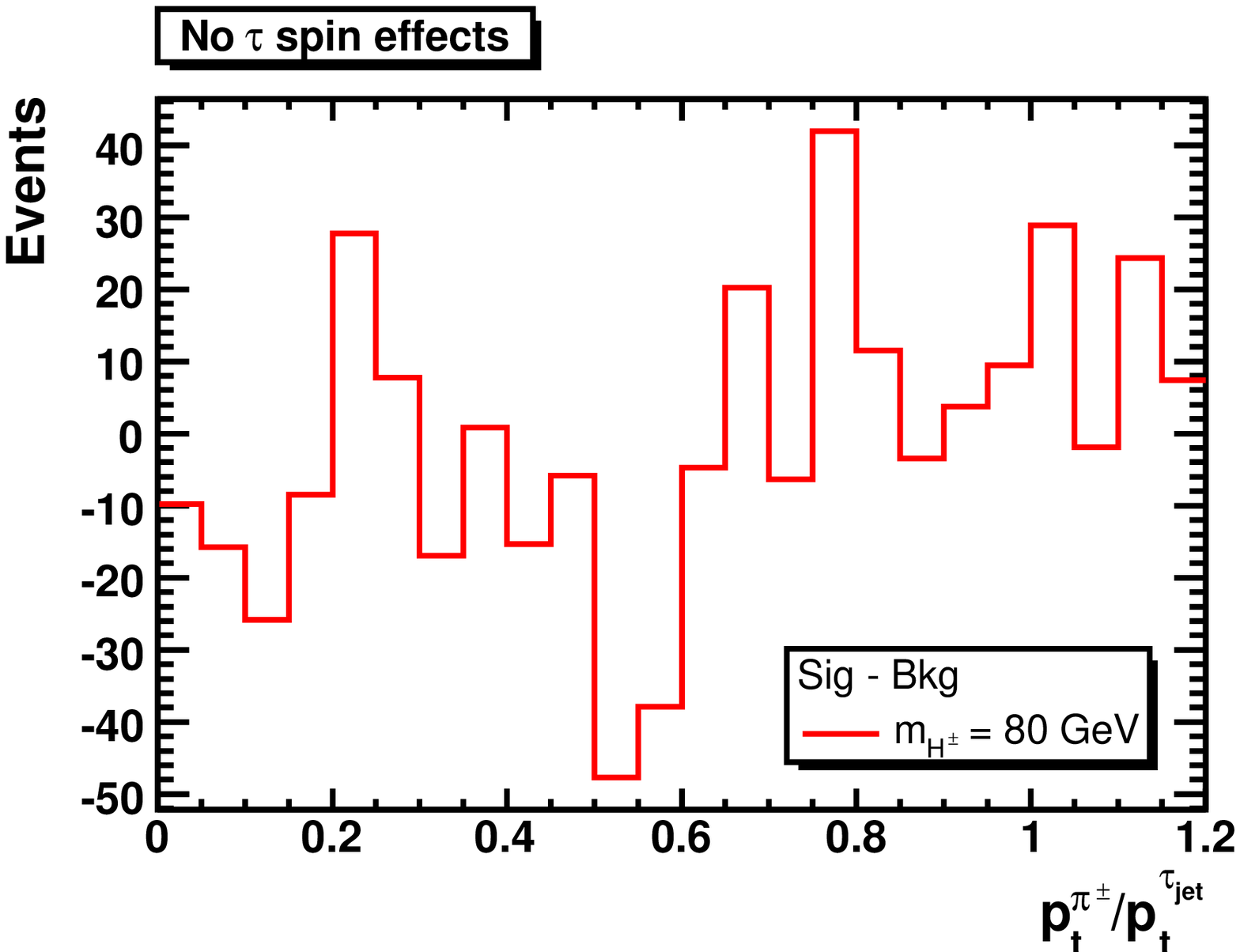, width=0.5\textwidth}
\caption{Distributions of the  
ratio $p_t^{\pi^\pm}/p_t^{\tau_\mathrm{jet}}$
for the $tbH^\pm$ signal
and the $t\bar{t}$ background for $\sqrt{s}=1.96$~TeV (left)
and the respective differences between signal and background (right).
The lower plots show distributions without spin effects in the $\tau$
decays.
}
\label{fig:r1}
\end{figure}

\begin{figure}[htbp]
\epsfig{file=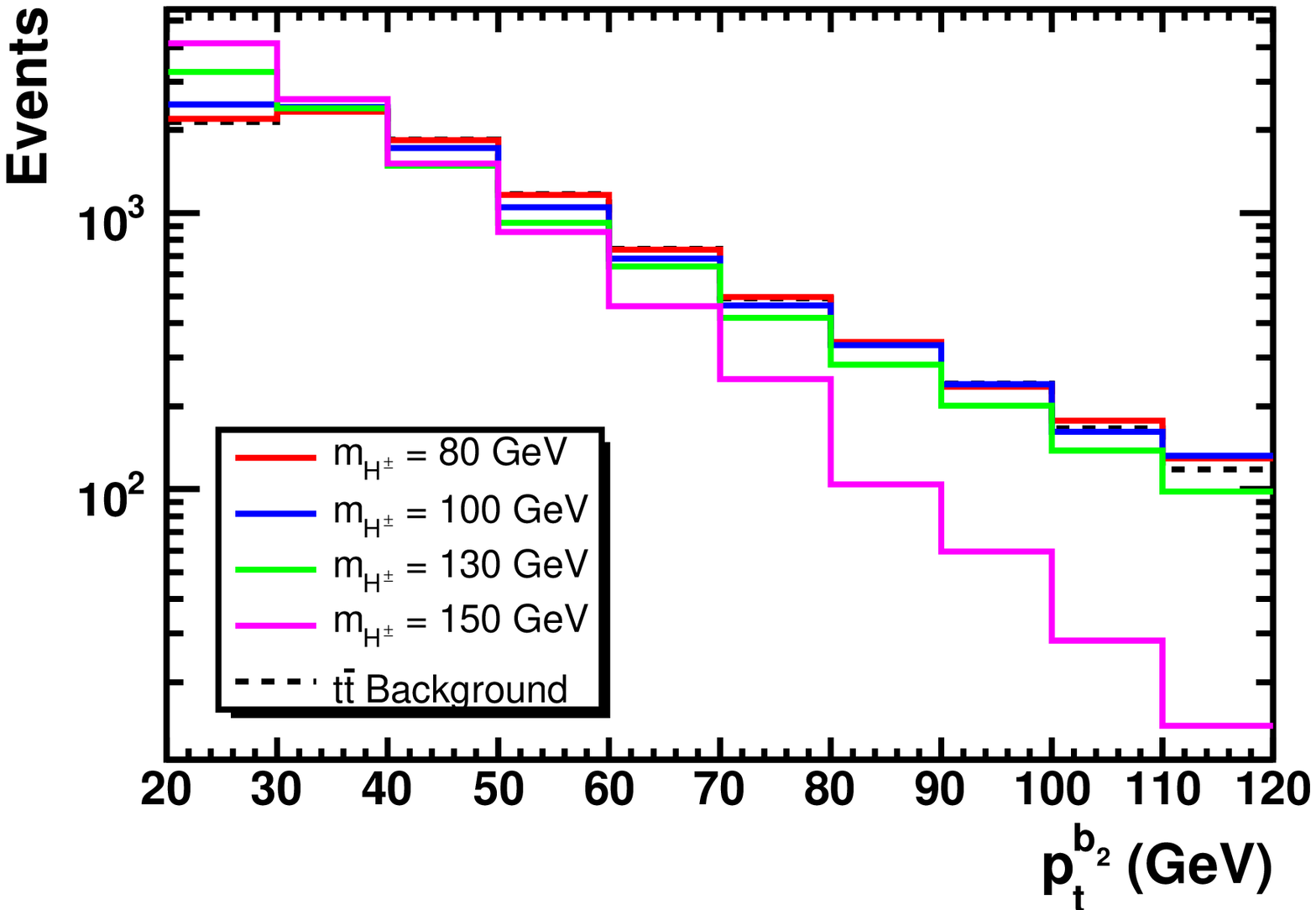, width=0.5\textwidth}  \hfill
\epsfig{file=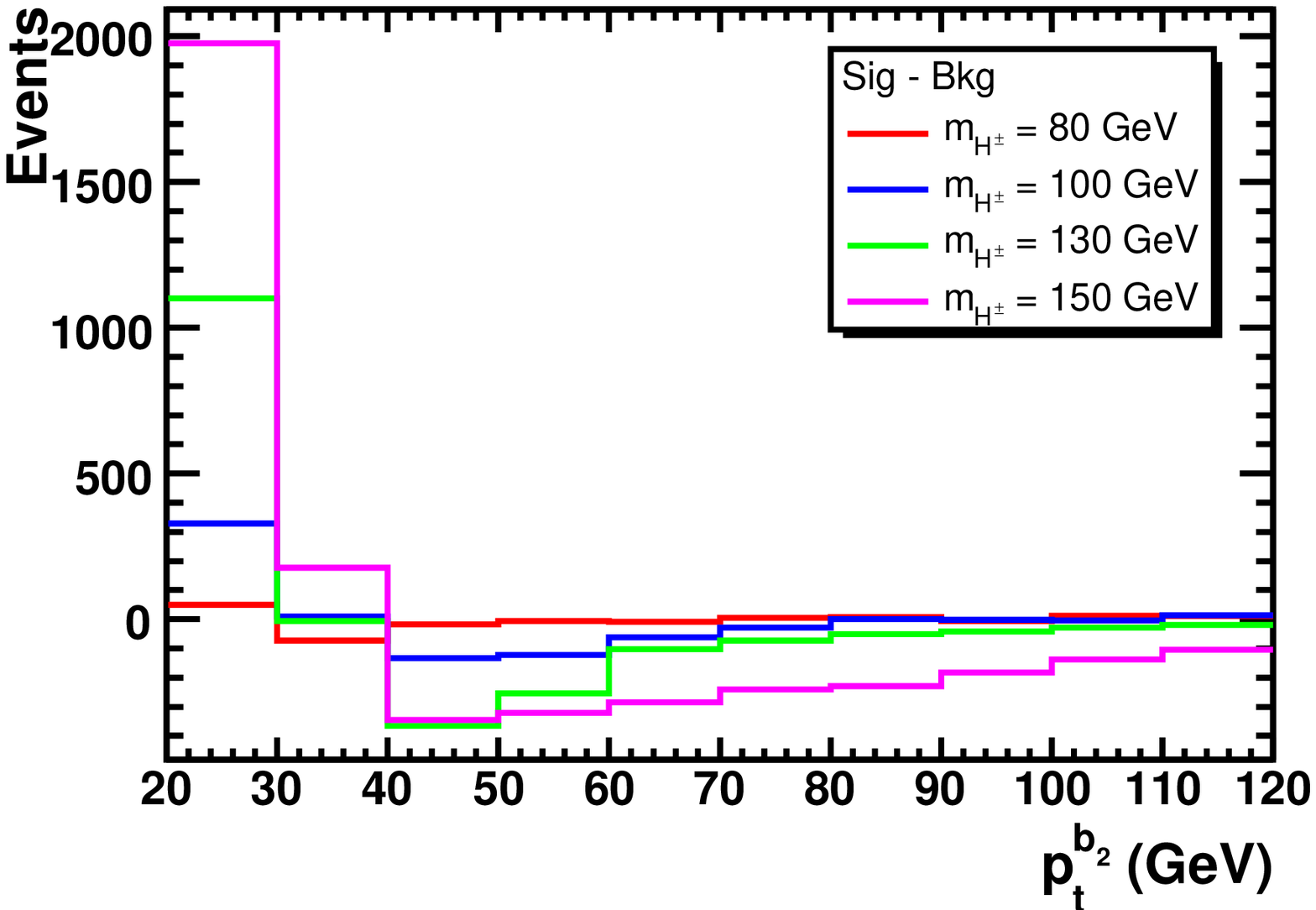, width=0.5\textwidth}
\caption{
$p_t$ distributions of the second (least energetic) $b$ quark jet
for the $tbH^\pm$ signal
and the $t\bar{t}$ background for $\sqrt{s}=1.96$~TeV (left)
and the respective differences between signal and background (right).
}
\label{fig:ptb2}
\end{figure}

\begin{figure}[htbp]
\epsfig{file=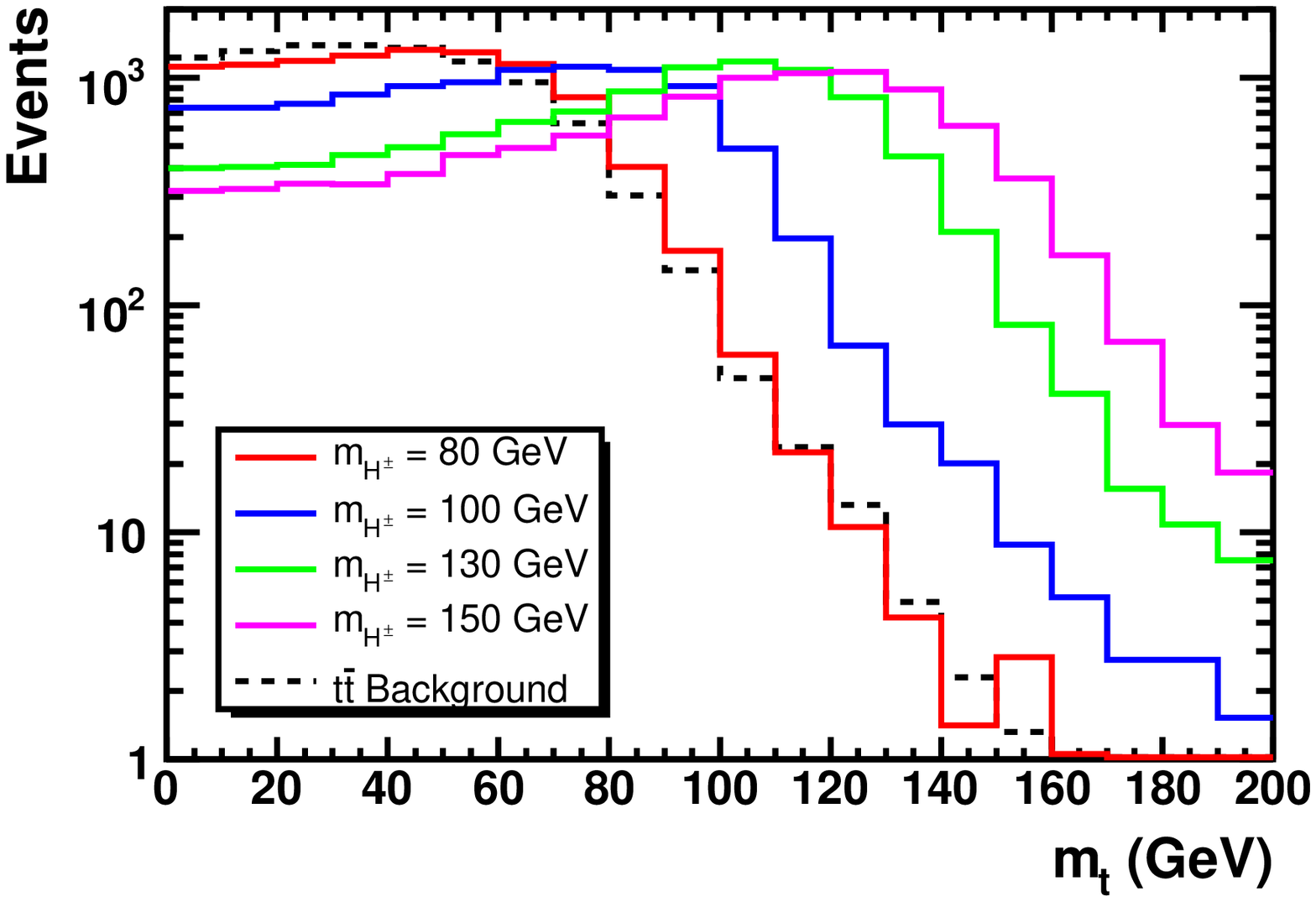, width=0.5\textwidth}  \hfill
\epsfig{file=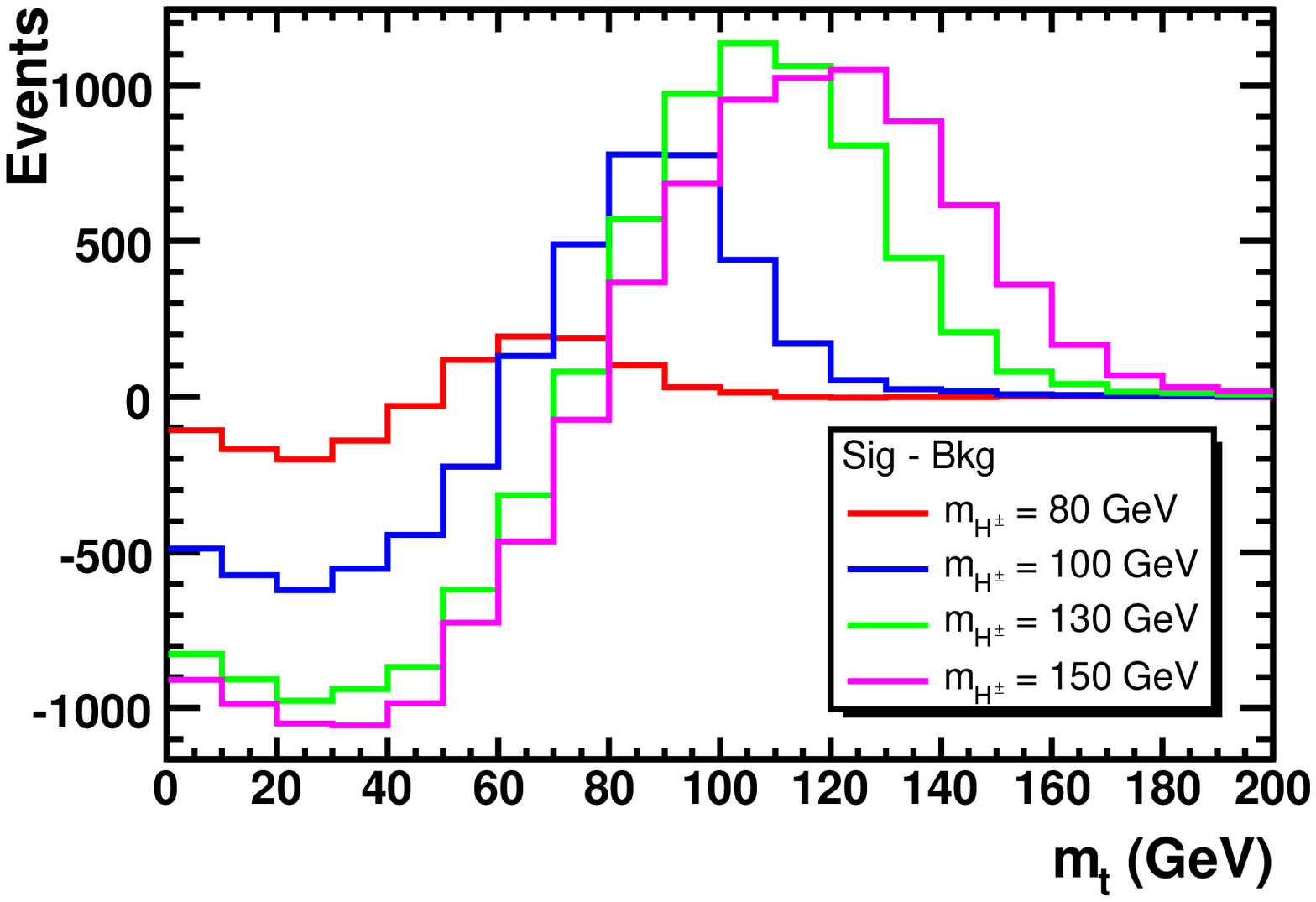, width=0.5\textwidth}
\caption{
Transverse mass
$m_t = \sqrt{2 p_t^{\tau_\mathrm{jet}} p_t^{\rm miss}
  [1-\cos(\Delta\phi)]}$
distributions of the
$\tau_{\rm jet} + p_t^{\rm miss}$ system
($\Delta\phi$ is the azimuthal angle
between $p_t^{\tau_\mathrm{jet}}$ and $p_t^{\rm miss}$)
for the $tbH^\pm$ signal
and the $t\bar{t}$ background for $\sqrt{s}=1.96$~TeV (left)
and the respective differences between signal and background (right).
}
\label{fig:mtransverse}
\end{figure}

\begin{figure}[htbp]
\epsfig{file=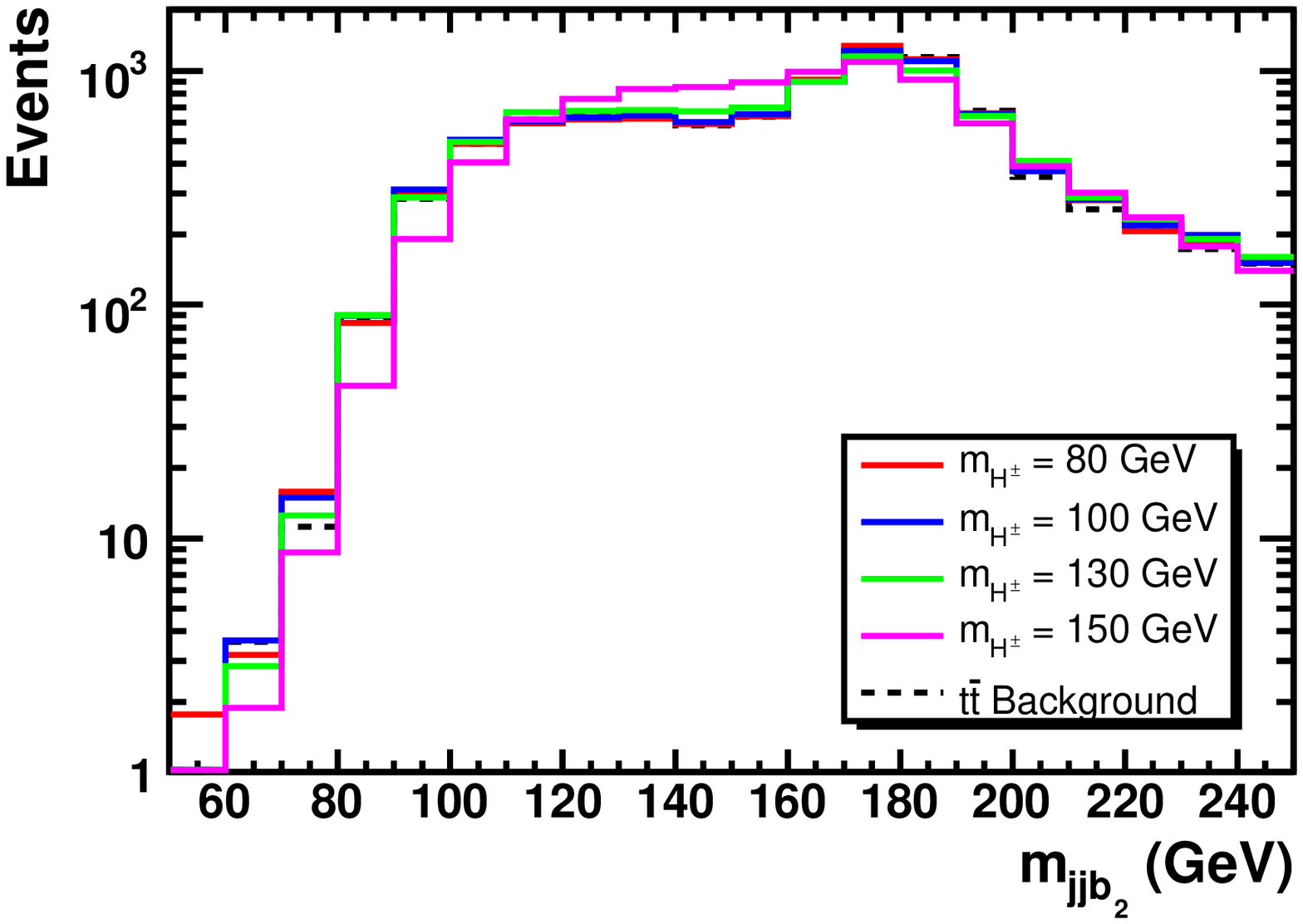, width=0.5\textwidth}  \hfill
\epsfig{file=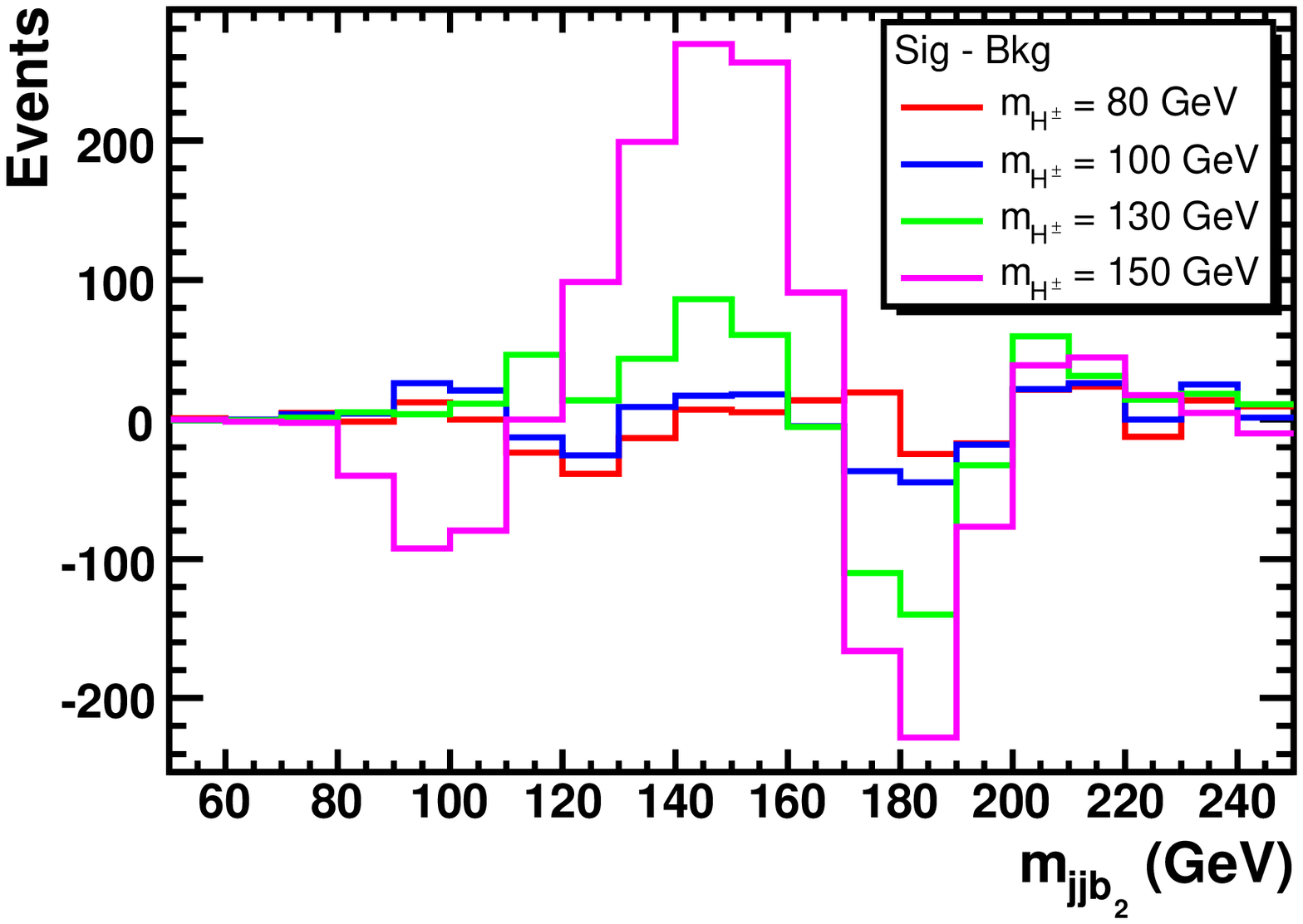, width=0.5\textwidth}
\caption{
Invariant mass distributions of the two light quark jets and the 
second (least energetic) $b$ quark jet
for the $tbH^\pm$ signal
and the $t\bar{t}$ background for $\sqrt{s}=1.96$~TeV (left)
and the respective differences between signal and background (right).
}
\label{fig:mjjb2}
\end{figure}

\begin{figure}[htbp]
\epsfig{file=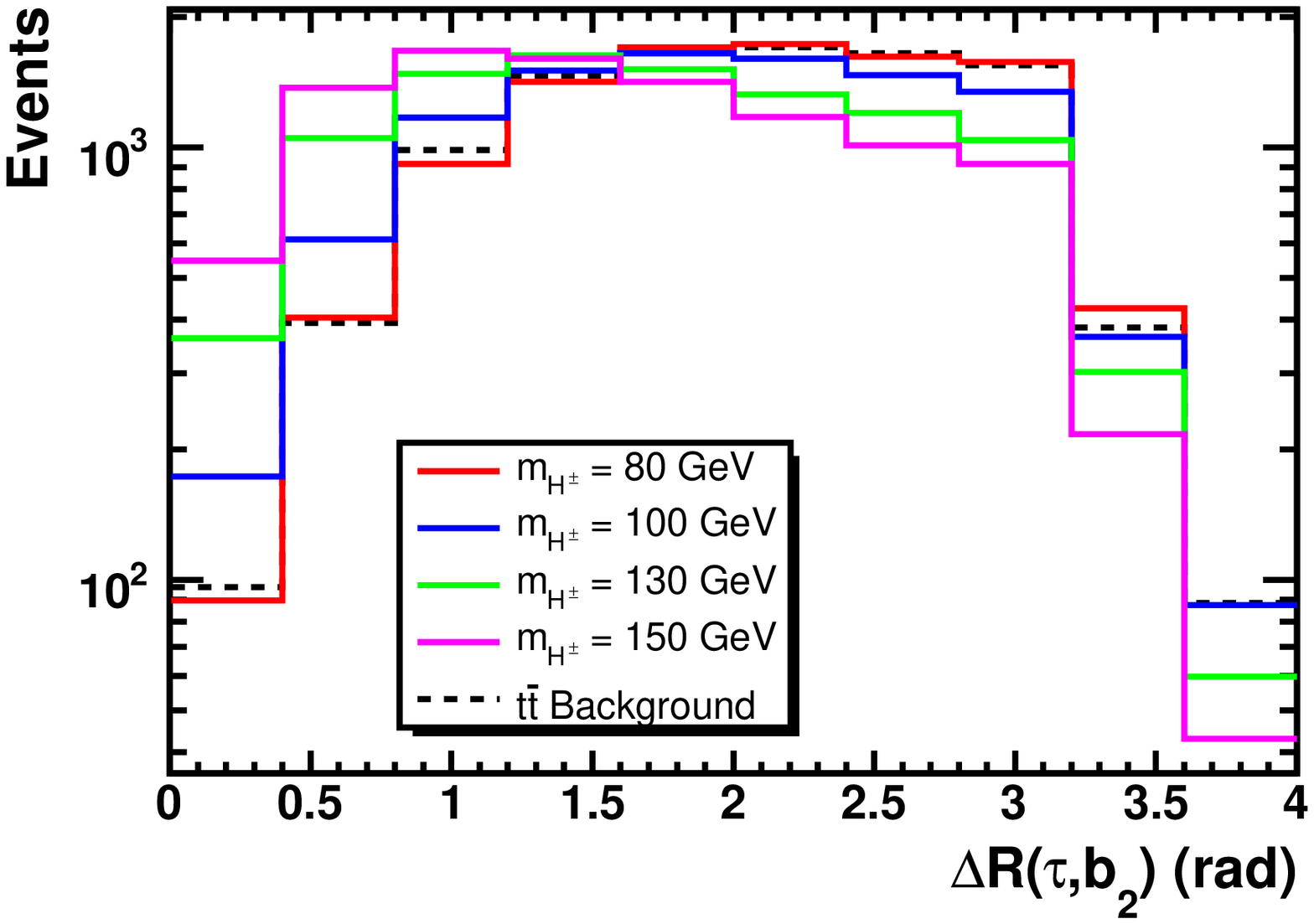, width=0.5\textwidth}  \hfill
\epsfig{file=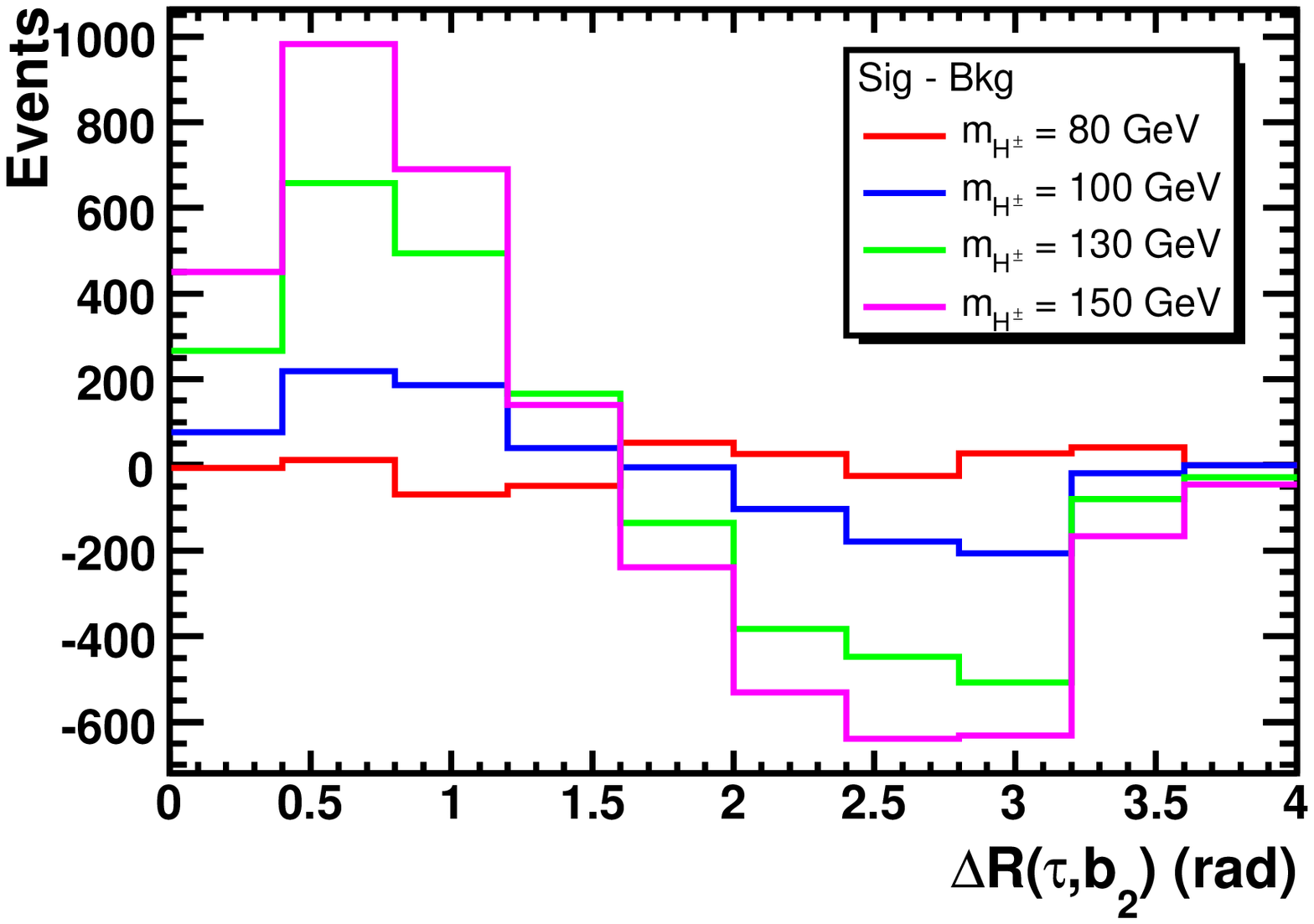, width=0.5\textwidth}
\caption{
Spatial distance
$\Delta R(\tau,b_2) = \sqrt{(\Delta\phi)^2 + (\Delta\eta)^2}$
distributions
(where $\Delta\phi$ is the azimuthal angle in rad between
the $\tau$ and $b$ jet)
for the $tbH^\pm$ signal
and the $t\bar{t}$ background for $\sqrt{s}=1.96$~TeV (left)
and the respective differences between signal and background (right).
}
\label{fig:distance-tau-b}
\end{figure}

\begin{figure}[htbp]
\epsfig{file=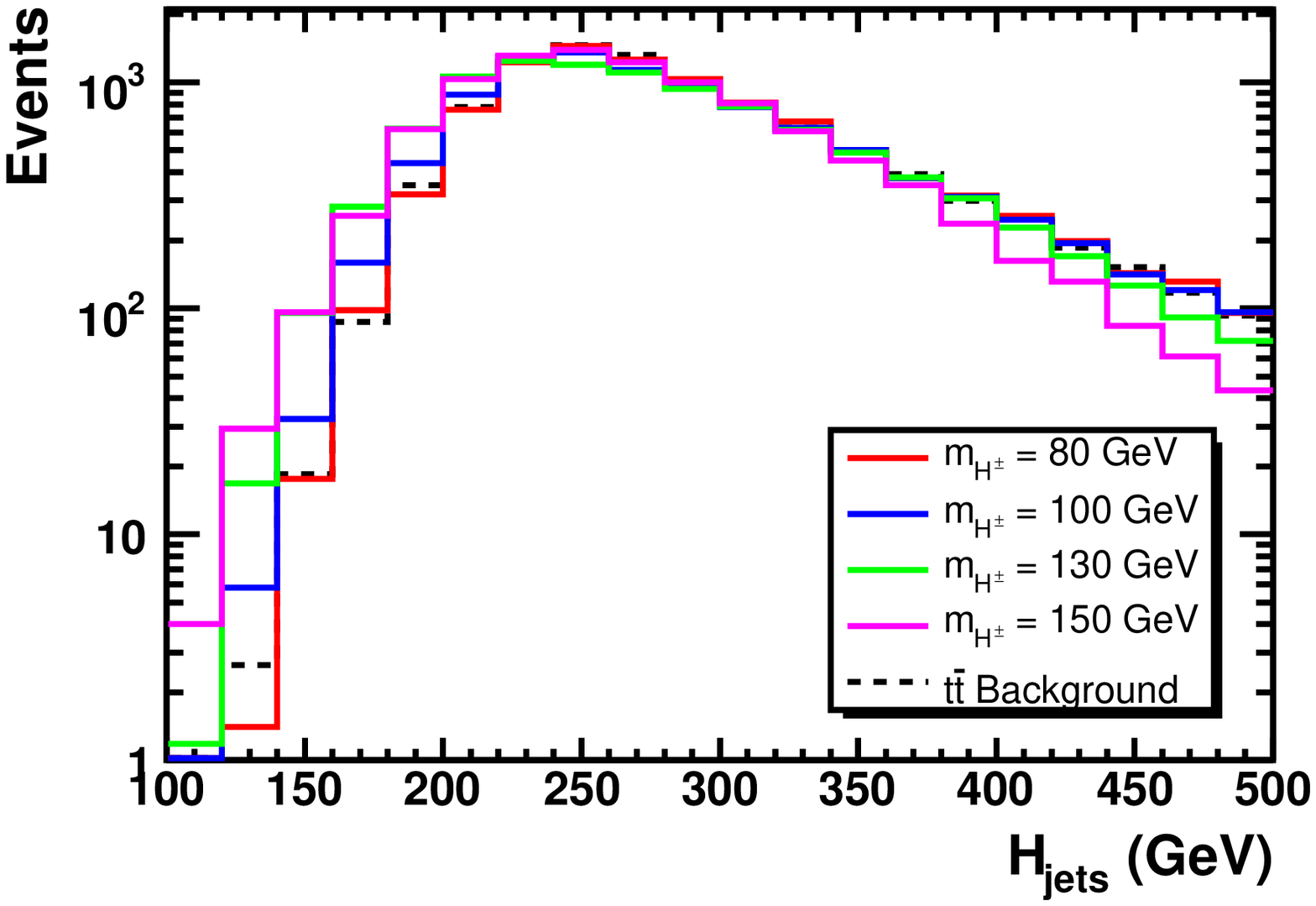, width=0.5\textwidth}  \hfill
\epsfig{file=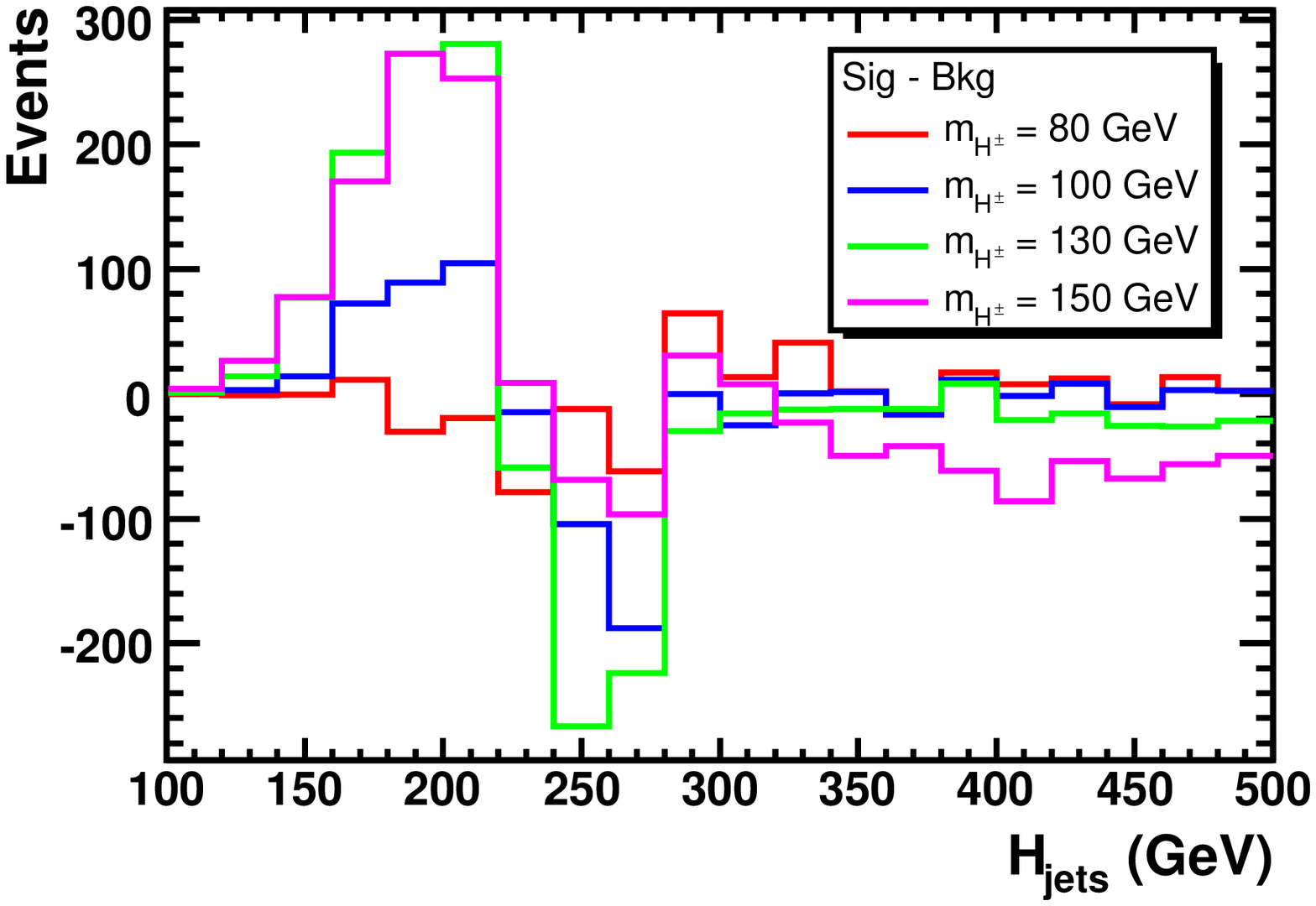, width=0.5\textwidth}
\caption{
Distributions of the
total transverse momentum of all quark jets,
$H_{\rm jets} = p_t^{j_1} + p_t^{j_2} + p_t^{b_1} + p_t^{b_2}$,
for the $tbH^\pm$ signal
and the $t\bar{t}$ background for $\sqrt{s}=1.96$~TeV (left)
and the respective differences between signal and background (right). 
}
\label{fig:hjet}
\end{figure}

Figures~\ref{fig:pttau}--\ref{fig:hjet} show examples of the signal and
background distributions of some of the kinematic 
variables used in the IDA method and the respective difference between
signal and background distributions, namely:
\begin{itemize}
\item the transverse momentum of the $\tau$ jet, $p_t^{\tau_{\rm jet}}$~(Fig.~\ref{fig:pttau}),
\item the transverse momentum of the leading $\pi^\pm$ in the $\tau$ jet, 
$p_t^{\pi^\pm}$ (Fig.~\ref{fig:ptpi})
\item the ratio $p_t^{\pi^\pm}/p_t^{\tau_{\rm jet}}$ (Fig.~\ref{fig:r1}),
\item the transverse momentum of the second (least energetic) $b$ quark jet, $p_t^{b_2}$ (Fig.~\ref{fig:ptb2}),
\item the transverse mass in the $\tau_{\rm jet} + p_t^{\rm miss}$ system,
      $m_t = \sqrt{2 p_t^{\tau_\mathrm{jet}} p_t^{\rm miss}
             [1-\cos(\Delta\phi)]}$, where $\Delta\phi$ is the azimuthal angle
      between $p_t^{\tau_\mathrm{jet}}$ and $p_t^{\rm miss}$ 
(Fig.~\ref{fig:mtransverse})\footnote{Strictly speaking this is not the transverse mass since there are two neutrinos in the decay chain of the charged Higgs boson we are considering, even so the characteristics of this mass are very similar to that of the true transverse mass.},
\item the invariant mass distribution of the two light quark jets and the
      second $b$ quark jet, $m_{jjb_2}$ (Fig.~\ref{fig:mjjb2}),
\item the spatial distance between the $\tau$ jet and the second $b$ quark jet, 
      $\Delta R(\tau,b_2) = \sqrt{(\Delta\phi)^2 + (\Delta\eta)^2}$,
      where $\Delta\phi$ is the azimuthal angle between the $\tau$ and $b$ jet (Fig.~\ref{fig:distance-tau-b})
      and 
\item the sum of the (scalar) transverse momenta of all the quark jets, 
      $H_{\rm jets} = p_t^{j_1} + p_t^{j_2} + p_t^{b_1} + p_t^{b_2}$ (Fig.~\ref{fig:hjet}).
\end{itemize}
The distributions of signal and background events are normalized 
to the same number of $10^4$ events, 
in order to make small differences better visible.

The signal and background distributions 
for the variables shown in Figs.~\ref{fig:ptb2}--\ref{fig:hjet} 
are as expected rather similar for $m_{H^\pm}=m_{W^\pm}$
and are hence mostly important to discriminate between signal and
background in the IDA for $m_{H^\pm} > m_{W^\pm}$.
Especially the transverse mass, Fig.~\ref{fig:mtransverse},
shows a large variation with the charged Higgs boson mass.
However, the different spin
of the charged Higgs boson and the $W^\pm$ boson has a large effect on
the $\tau$ jet variables $p_t^{\tau_{\rm jet}}$ and $p_t^{\pi^\pm}$
(Figs.~\ref{fig:pttau} and \ref{fig:ptpi}) resulting in
significantly different distributions of signal and background
even for $m_{H^\pm}=m_{W^\pm}$.
Moreover, the spin effects in the $p_t^{\tau_{\rm jet}}$ and $p_t^{\pi^\pm}$ 
distributions are correlated which can be seen in Fig.~\ref{fig:r1}
where the distributions of the ratio $p_t^{\pi^\pm}/p_t^{\tau_{\rm jet}}$
\cite{Roy:1991sf,Raychaudhuri:1995cc,Roy:1999xw}
show even larger differences.
This highlights the importance
of the additional variable $p_t^{\pi^\pm}$ (and hence
$p_t^{\pi^\pm}/p_t^{\tau_{\rm jet}}$),
compared to a previous study \cite{leshouches05}.
The large separation power of this variable
is indeed due to the different $\tau$ polarizations
in signal and background as can be inferred from the lower plots in
Figs.~\ref{fig:pttau}--\ref{fig:r1}.
There the signal and background distributions for $p_t^{\tau_{\rm jet}}$,
$p_t^{\pi^\pm}$ and $p_t^{\pi^\pm}/p_t^{\tau_{\rm jet}}$ are shown
for reference samples where the $\tau$ decay has been performed without the
inclusion of spin effects with the built-in routines of PYTHIA and
hence the differences between signal and background nearly vanish.

\subsection{Iterative Discriminant Analysis (IDA)}

The IDA method is a modified Fisher Discriminant Analysis~\cite{ida}
and is characterized by 
the use of a quadratic, instead of a linear, discriminant function and 
also involves iterations in order to 
enhance the separation between signal and background.

In order to analyze our events with the IDA method, signal
and background  have been split in two samples of equal size.
With the first set of samples the IDA training has been performed and then
the second set of samples has been analyzed.
We have used the following 20 variables in the IDA study:
the transverse momenta 
$p_t^{\tau_{\rm jet}}$, $p_t^{\pi^\pm}$,
$p_t^{\rm miss}$,
$p_t^{b_1}$, $p_t^{b_2}$, $p_t^{j_1}$, $p_t^{j_2}$, $p_t^{j j}$;
the transverse mass $m_t$;
the invariant masses
$m_{jj}$, $m_{jjb_1}$, $m_{jjb_2}$, $m_{bb}$ and
$\hat{s}=m_{jjbb\tau}$;
the spatial distances $\Delta R(\tau,b_1)$, $\Delta R(\tau,b_2)$,
$\Delta R(\tau,j_1)$, $\Delta R(\tau,j_2)$;
the total transverse momenta of all quark jets $H_{\rm jets}$ 
and of all jets $H_{\rm all} = H_{\rm jets} + p_t^{\tau_{\rm jet}}$.
In the analysis of real data, b-quark tagging probabilities and 
the reconstruction of $t$ and $W$ masses could be used to improve
the jet pairing, and replace the allocation of least and most energetic
$b$-jet by a probabilistic analysis.

\begin{figure}[htbp]
\epsfig{file=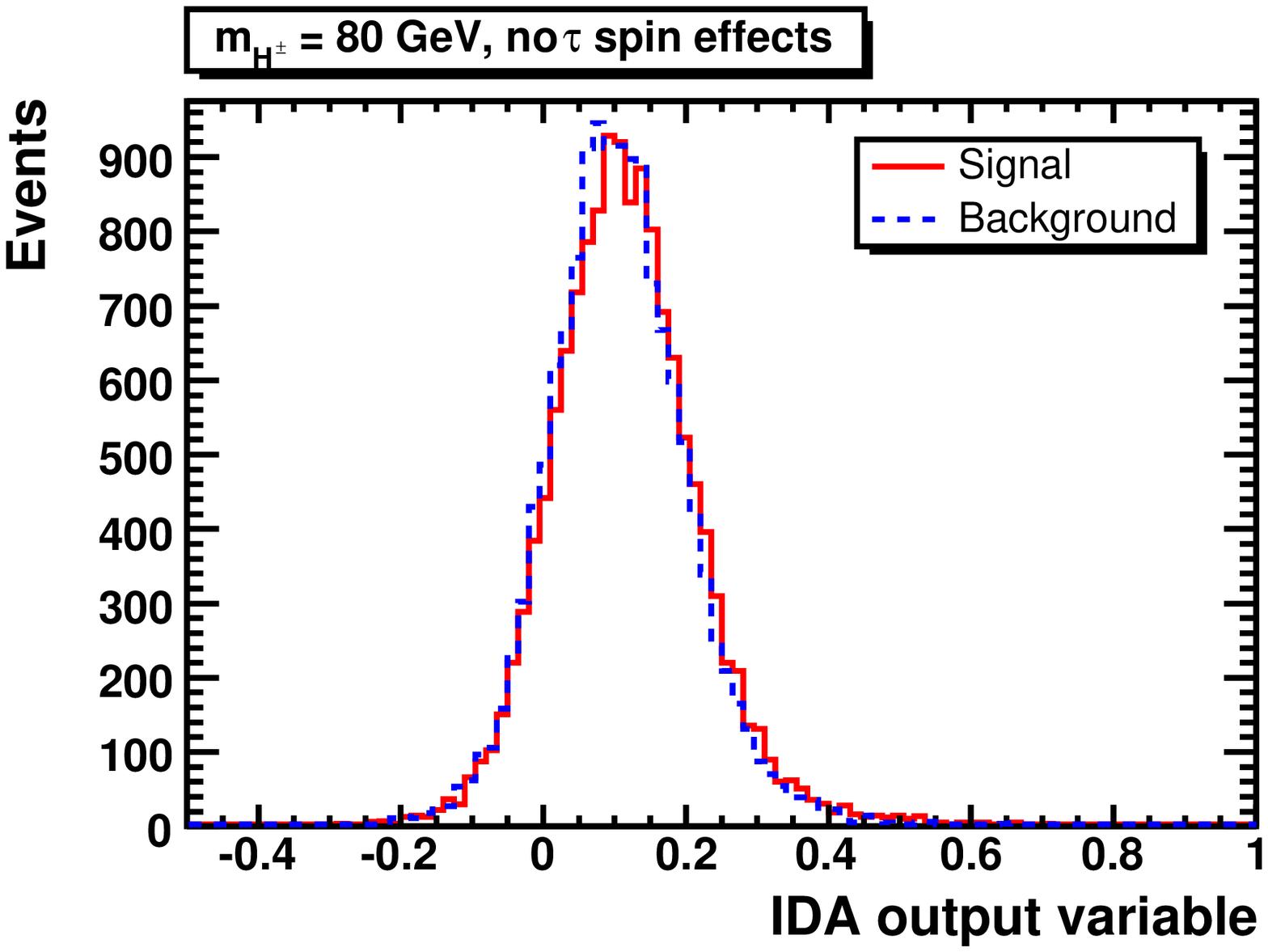, width=0.5\textwidth}  \hfill
\epsfig{file=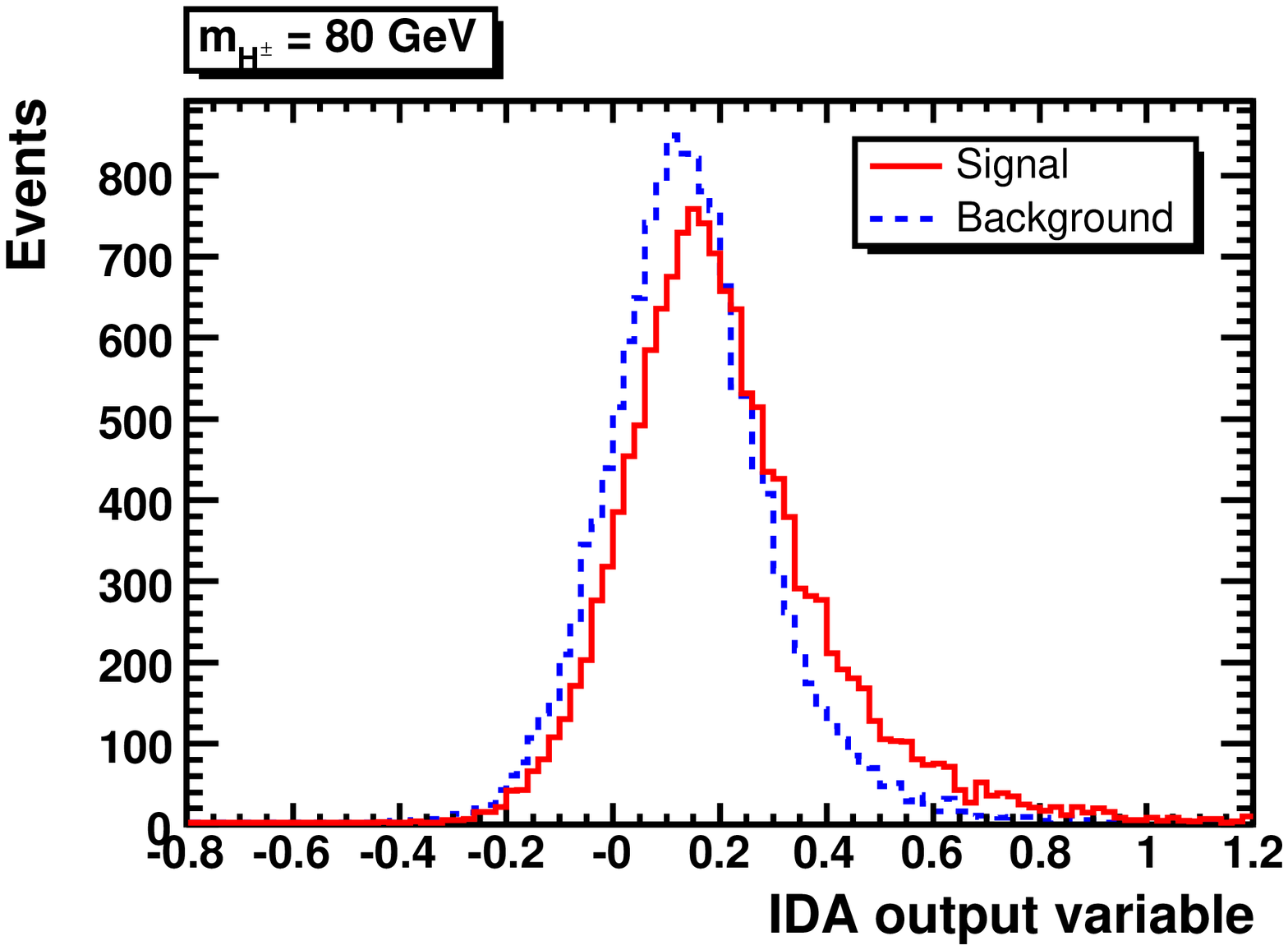, width=0.5\textwidth}
\epsfig{file=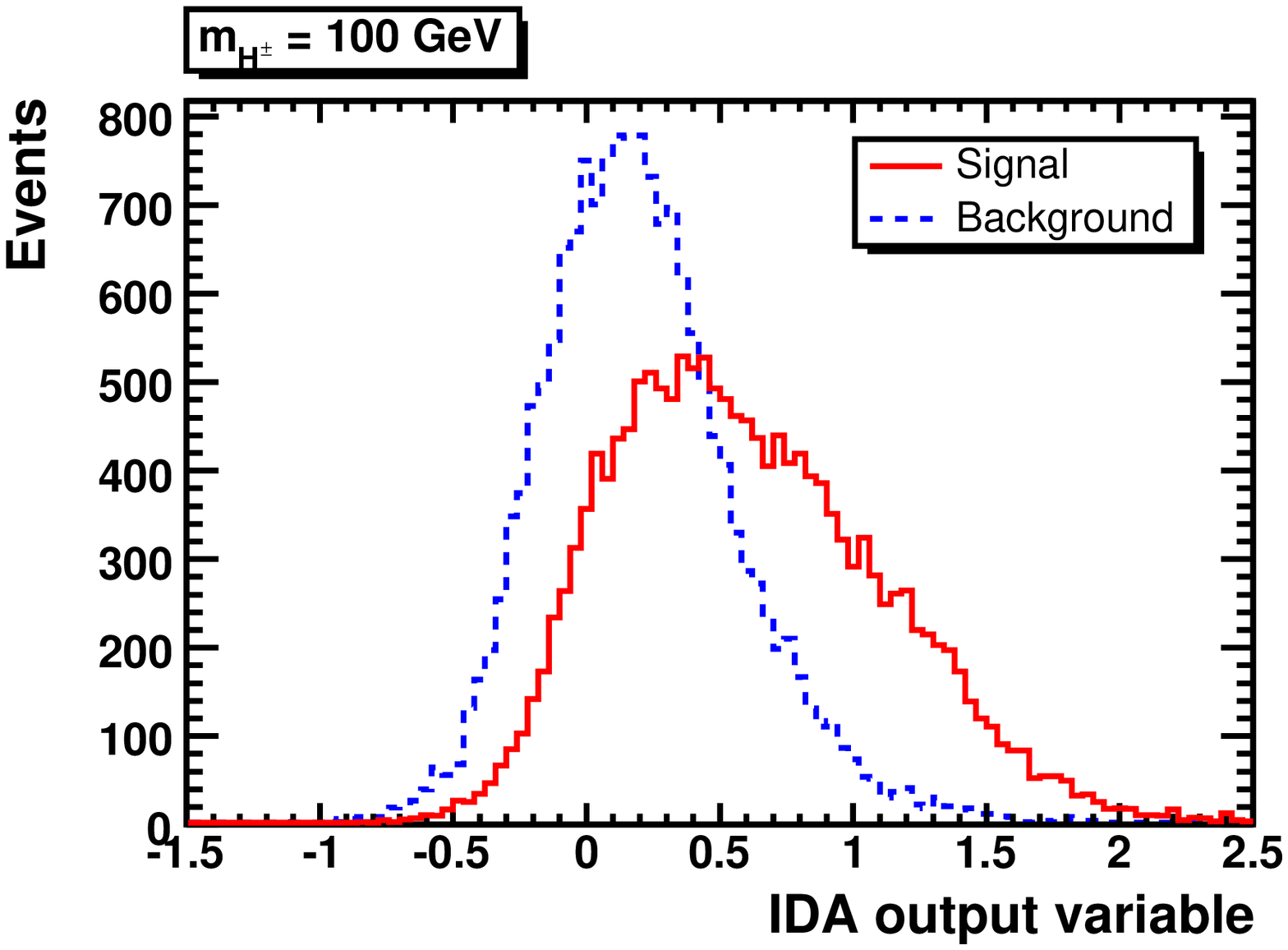, width=0.5\textwidth}  \hfill
\epsfig{file=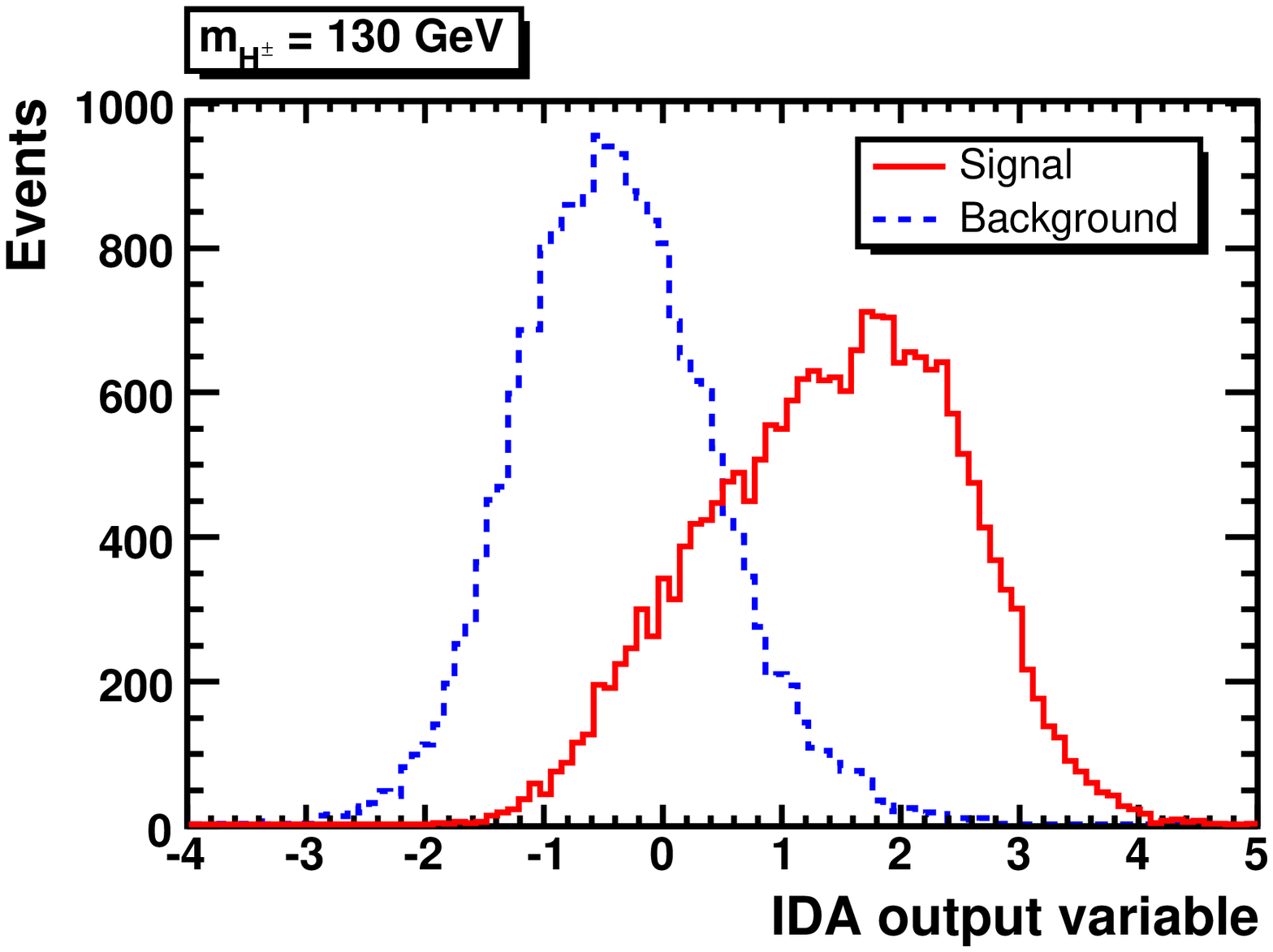, width=0.5\textwidth}
\epsfig{file=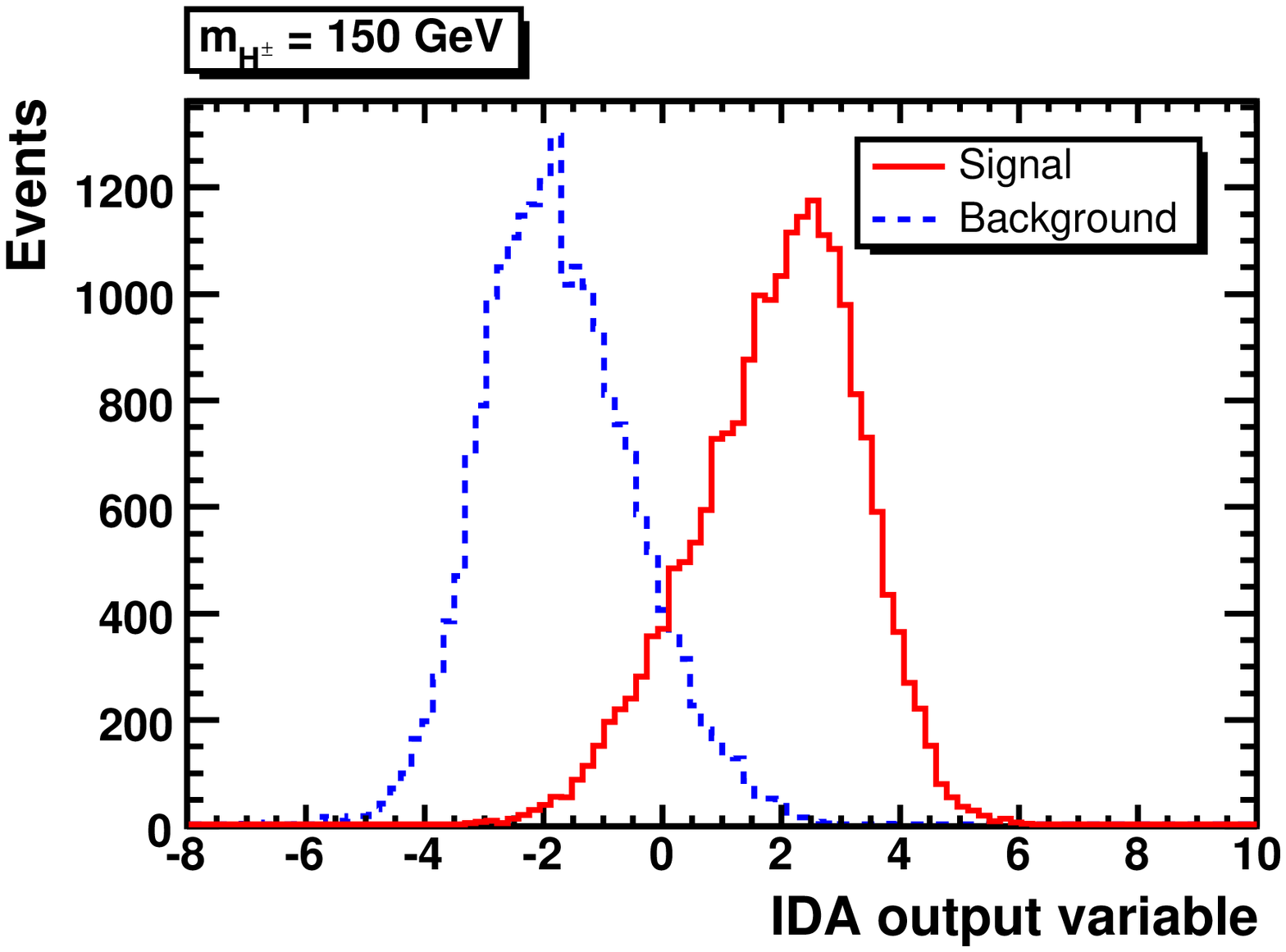, width=0.5\textwidth}  \hfill
\caption{
Distributions of the IDA output variable in the first IDA
step for the $tbH^\pm$ signal (solid, red)
and the $t\bar{t}$ background (dashed, blue) for $\sqrt{s}=1.96$~TeV.
}
\label{fig:ida1}
\end{figure}

\begin{figure}[htbp]
\epsfig{file=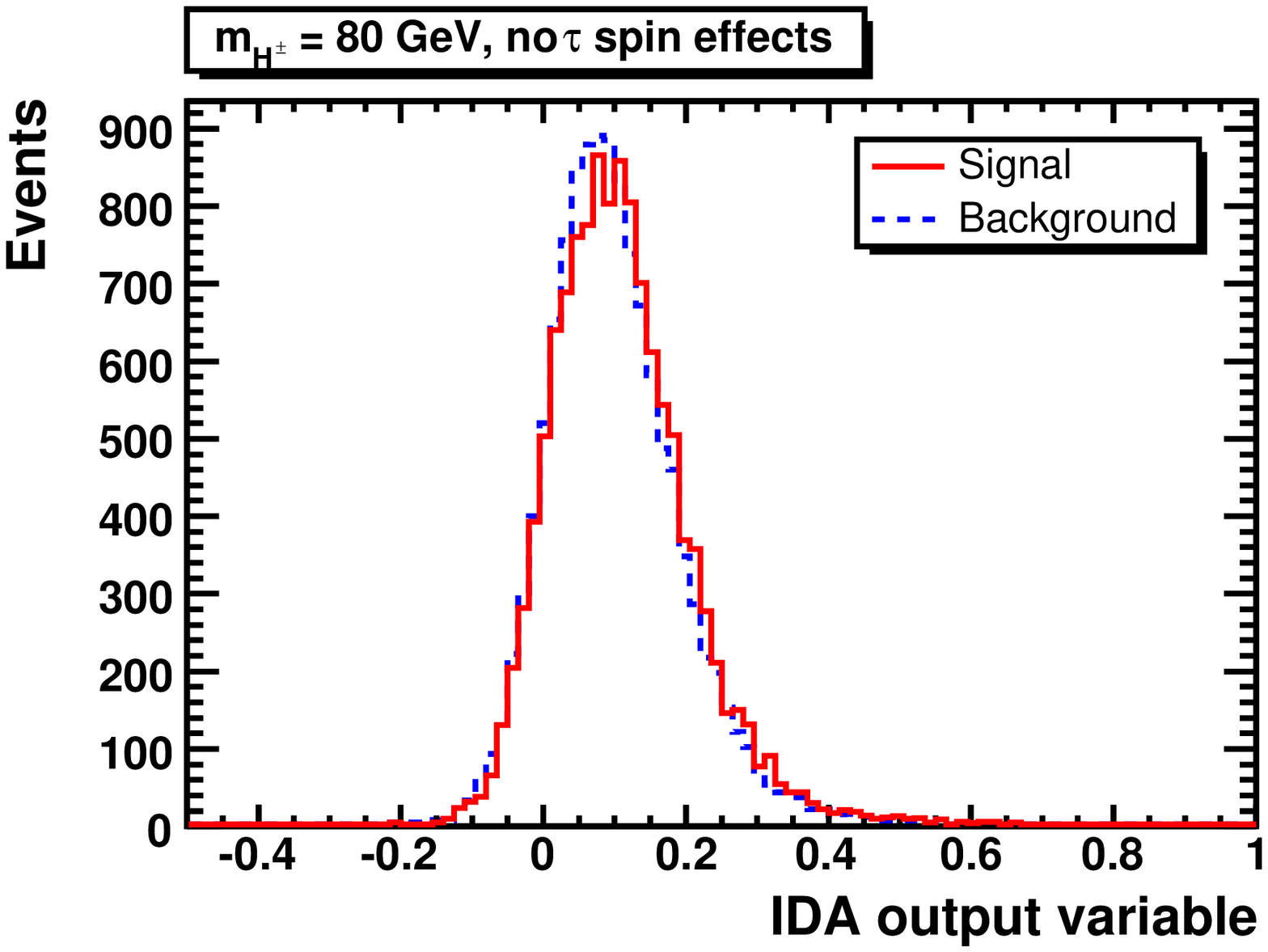, width=0.5\textwidth}  \hfill
\epsfig{file=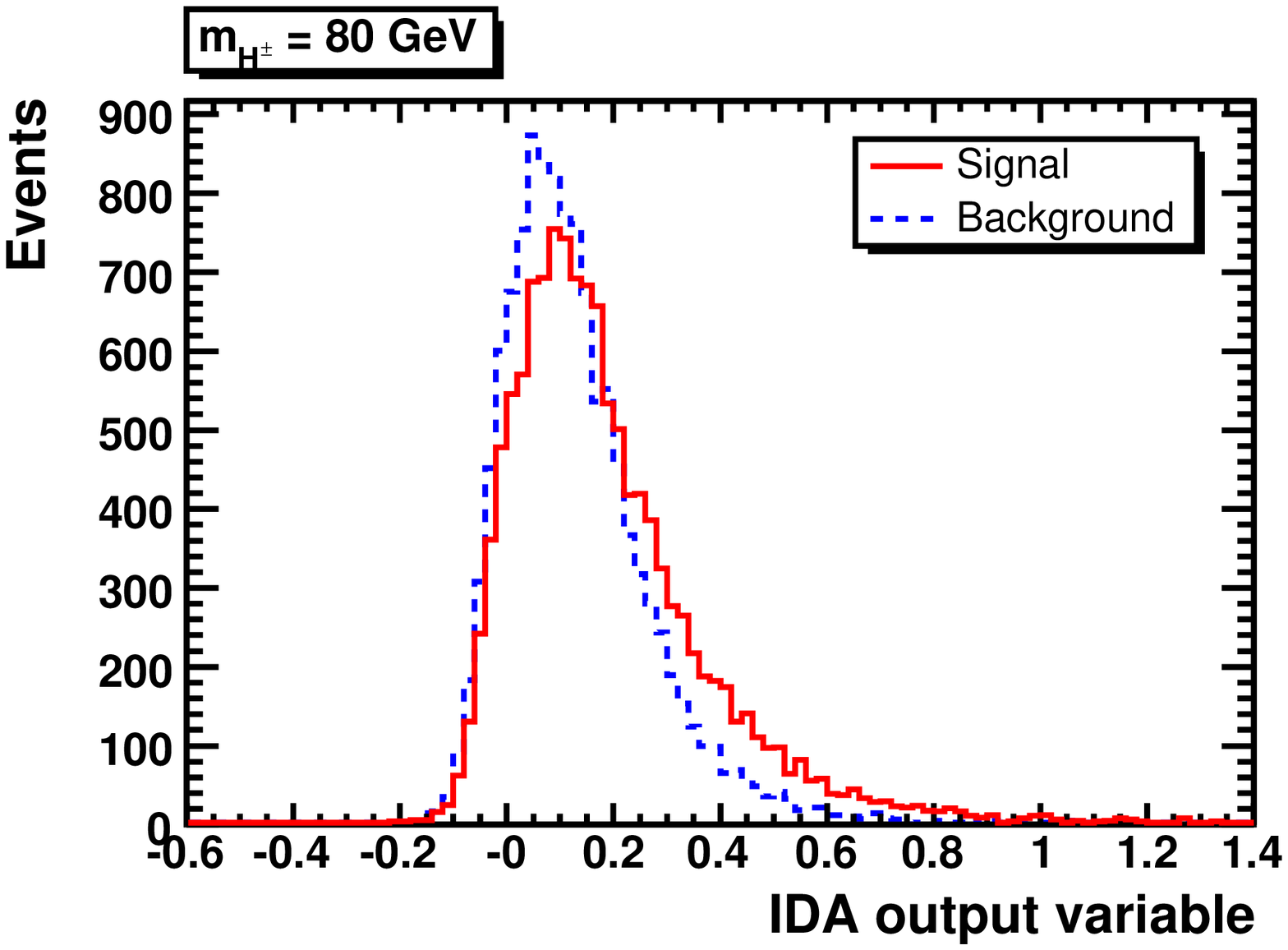, width=0.5\textwidth}
\epsfig{file=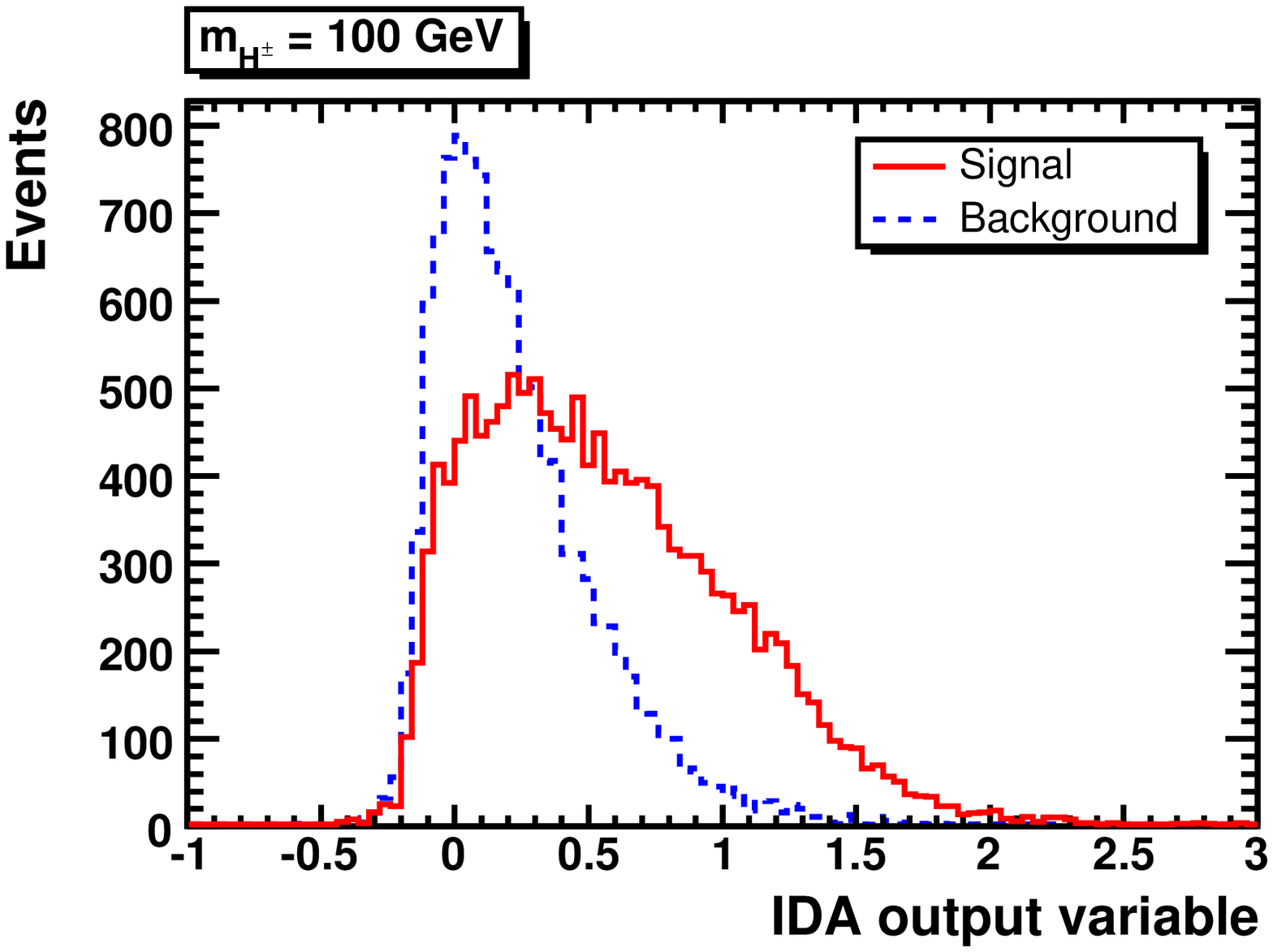, width=0.5\textwidth}  \hfill
\epsfig{file=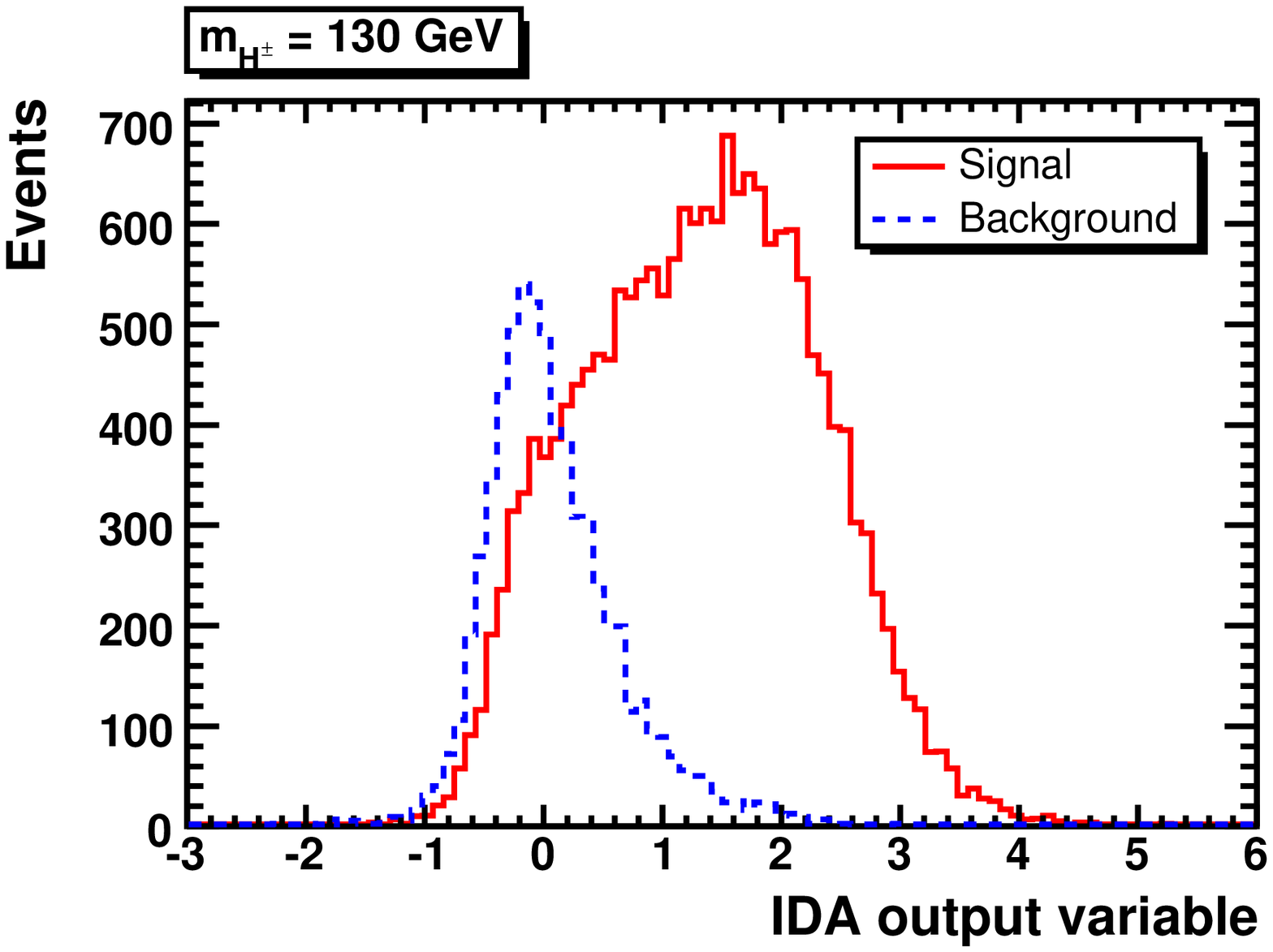, width=0.5\textwidth}
\epsfig{file=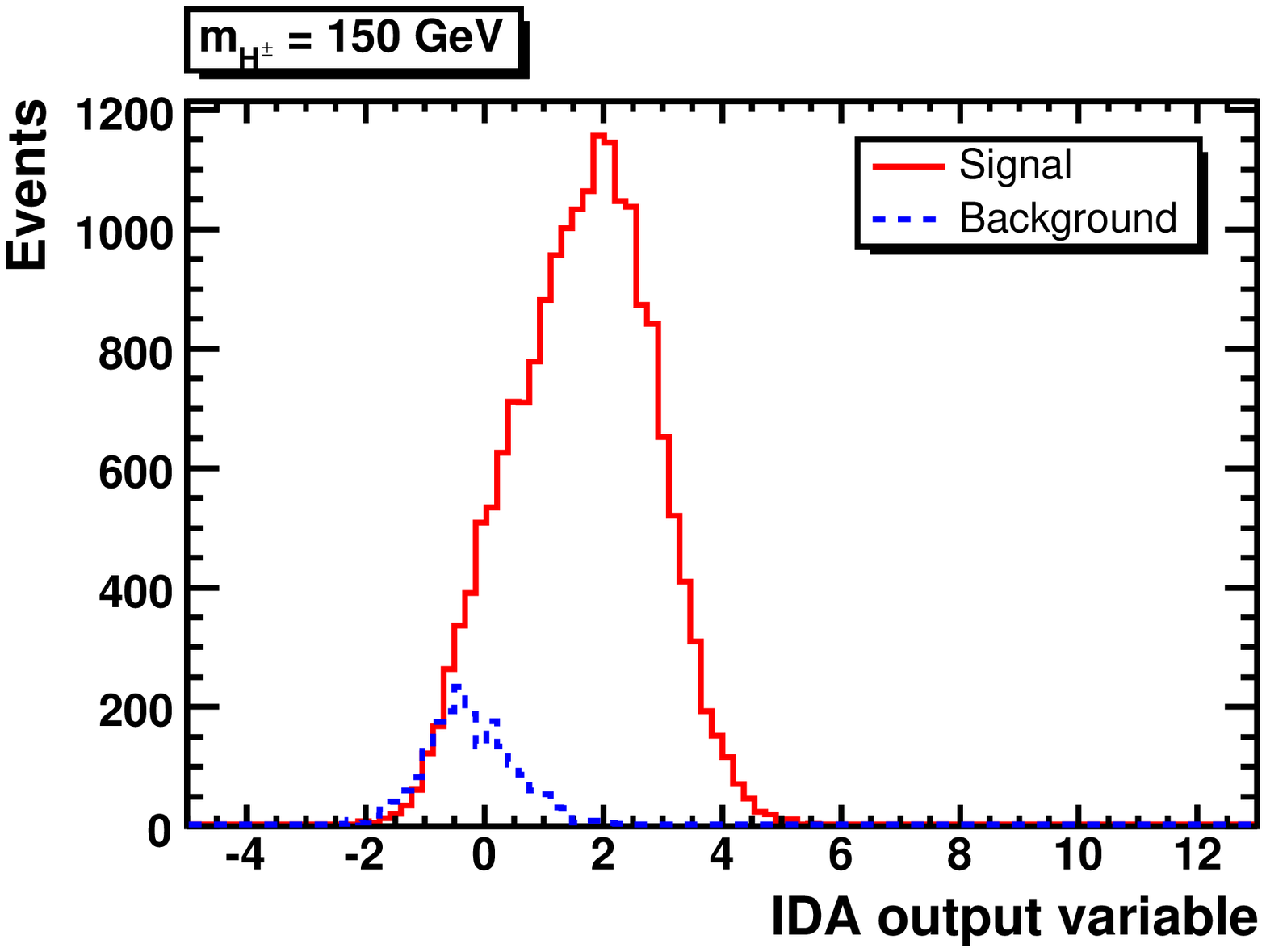, width=0.5\textwidth}  \hfill
\epsfig{file=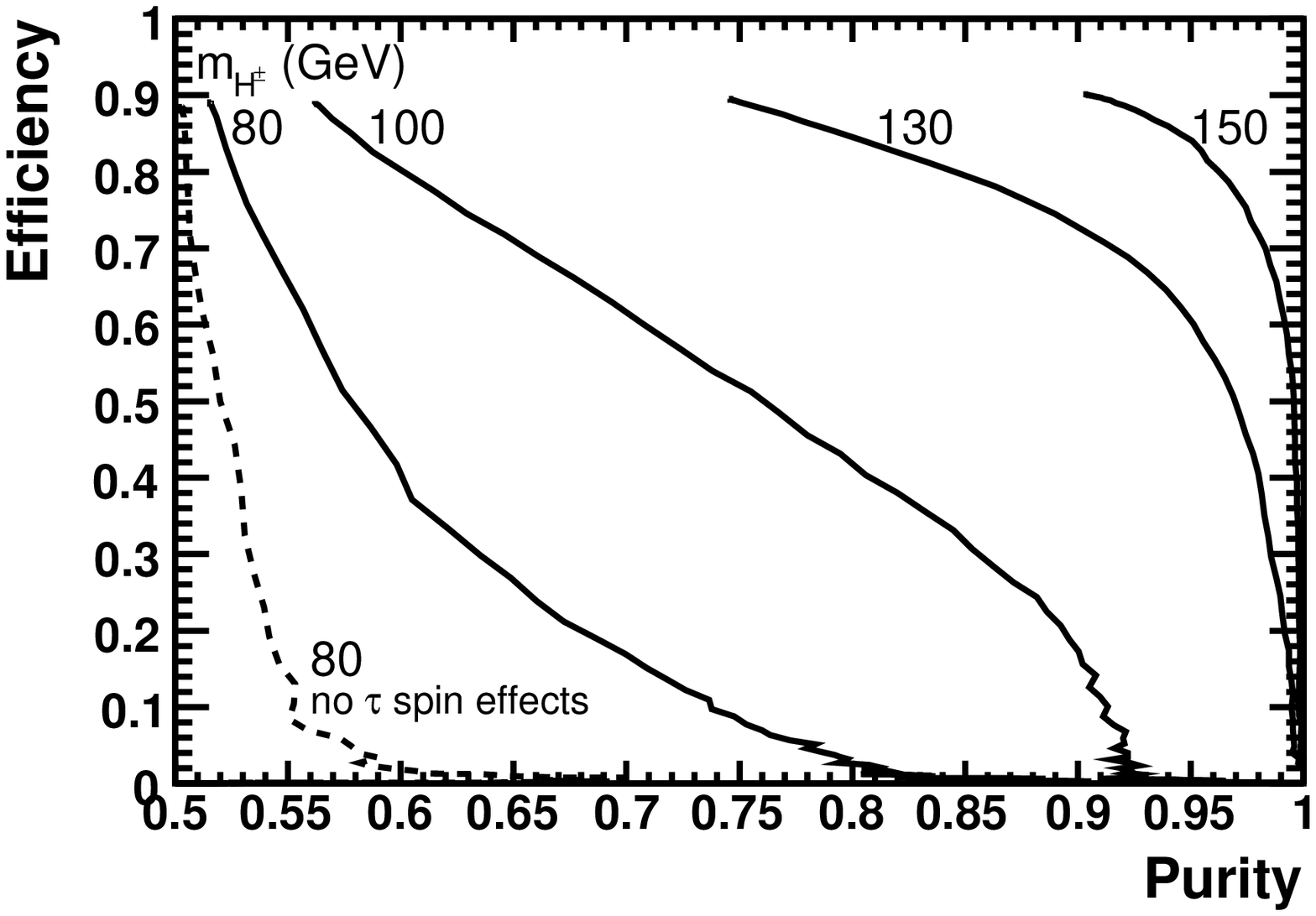, width=0.5\textwidth}
\caption{
Upper row, middle row and lower left figure:
distributions of the IDA output variable 
in the second IDA step
for 90\% efficiency in the first IDA
step (corresponding to a cut at 0 in Fig.~\ref{fig:ida1})
for the $tbH^\pm$ signal (solid, red)
and the $t\bar{t}$ background (dashed, blue).
Lower right figure: efficiency as a function of the purity
when not taking the spin effects in the $\tau$ decay into account for
$m_{H^\pm}=80$~GeV (dashed) and with 
spin effects in the $\tau$ decay for
$m_{H^\pm}=80,100,130,150$~GeV (solid, from left to right).
Results are for the Tevatron.
}
\label{fig:ida}
\end{figure}

%Two IDA iterations have been performed.
%The first iteration with a fixsed 90\% efficiency. 
%
The results of the IDA study are shown in Figs.~\ref{fig:ida1}
and \ref{fig:ida} for the event samples with spin effect in the
$\tau$ decays for $m_{H^\pm}=80, 100, 130, 150$~GeV and for the
reference samples without the spin effect for $m_{H^\pm}=80$~GeV
in order to illustrate the spin effect.
In all plots of the IDA output variable the number of 
background events has been normalized to the number of signal events.
Two IDA steps have been performed.
Figure~\ref{fig:ida1} shows the IDA output variable after the first
step, where 90\% of the signal is retained when a cut at zero is applied. 
The signal and background events after this cut are then passed to the 
second IDA step. 
Figure~\ref{fig:ida} shows the
IDA output variable distributions after the second step.
A cut on these distributions leads to the
efficiency and purity (defined as ratio of the number of signal events
divided by the sum of 
signal and background events) combinations as shown in the
lower right plot in Fig.~\ref{fig:ida}.
These combinations define the working point (number of expected
background events for a given 
signal efficiency) and the latter can be optimized to maximize the
discovery potential. 
The difference between the dashed (no spin effects in $\tau$ decay)
and solid (with spin effects in $\tau$ decay) lines for $m_{H^\pm}=80$~GeV
in the lower right plot in Fig.~\ref{fig:ida} stresses again the importance
of the spin effects to separate signal and background.

\begin{figure}[htp]
\begin{center}
\epsfig{file=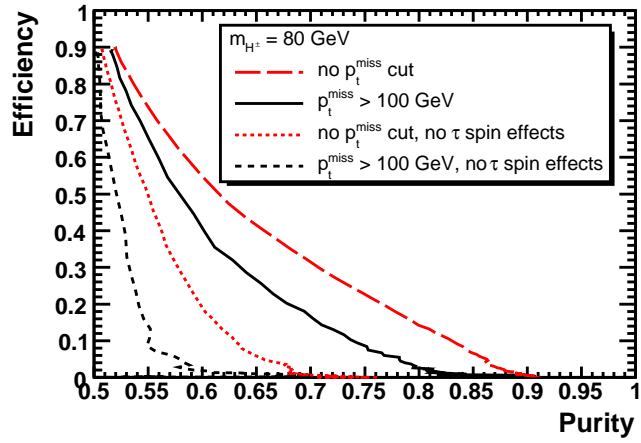, width=0.6\textwidth}
\end{center}
\caption{
Efficiency as a function of purity for $m_{H^\pm}=80$~GeV and
$\sqrt{s}=1.96$ TeV.
The black lines are the results after applying the hard cut
\mbox{$p_t^{\rm miss} > 100$~GeV}
when not taking the spin effects in the $\tau$ decay into account
(dashed) and with spin effects in the $\tau$ decay (solid),
as also shown in Fig.\ref{fig:ida}.
The red lines are the results without applying the hard cut
on \mbox{$p_t^{\rm miss}$}
when not taking the spin effects in the $\tau$ decay into account
(dotted) and with spin effects in the $\tau$ decay (long dashed).
}
\label{fig:ida_hard_cut}
\end{figure}

In order to illustrate the effect of the hard cut on the missing
transverse momentum ($p_t^{\rm miss}>100$~GeV), which is imposed to
suppress the QCD background, 
the final efficiency-purity plot of the IDA analysis is shown 
in Fig.~\ref{fig:ida_hard_cut} for $m_{H^\pm}=80$~GeV for two
reference samples (red, long dashed: with spin effects in the $\tau$
decay; red, dotted: without spin effects) without imposing the hard
cut. The black lines (dashed and solid) are for the samples with the
hard cut as also shown in the lower right plot in Fig.~\ref{fig:ida}.
As expected the achievable purity for a given efficiency decreases
with the hard cut,
therefore the spin effects become even more important to
separate signal and background.
In principle, by choosing the signal reduction rates in the previous IDA iterations, 
the signal and background rates in the final distributions can be varied appropriately.
However, we have checked that a different number of IDA iterations and/or different efficiencies 
for the first IDA iteration have only a minor effect on the final result.

\clearpage

\section{LHC ENERGY}

The simulation procedure and the emulation of the detector response are the same as 
those outlined in Sect.~2.1 for the Tevatron, as well as, for the preselection and IDA method, 
as described in Sects. 2.3 and 2.4, respectively. Hence, only the expected LHC rates are 
discussed, followed by the description of changes in the distributions of the variables and 
the final IDA results.

Unlike the case of the Tevatron, where only charged Higgs masses smaller
than the top quark mass can be explored, and 2HDM/MSSM signatures
practically rely on $\tau\nu_\tau$ pairs only, at the LHC the phenomenology 
is more varied. Here, the  search strategies depend strongly 
on the charged Higgs boson mass.
If $m_{H^\pm} < m_{t} - m_{b}$ (later referred to as a light Higgs boson), 
the charged Higgs boson can be produced in top \mbox{(anti-)}\-quark decay. The main source of 
top (anti-)quarks at the LHC is again $t \bar{t}$ pair production ($\sigma_{t\bar{t}}=850$ pb at 
NLO)~\cite{beneke00}. 
For the whole ($\tan\beta, m_{H^\pm}$) parameter space there is a competition between the $ bW^\pm$ 
and $ bH^\pm$ channels in top decay keeping the sum 
$\mathrm{BR}(t \to b W^+) + \mathrm{BR}(t \to b H^+)$
at almost unity.
The top quark decay to $ bW^\pm$ is however the dominant mode for most of the parameter space. 
Thus, the best way to search for a (light) charged Higgs boson is by requiring that the top
quark produced in the $tbH^\pm$ process decays to a $W^\pm$. 
While in the case of $H^\pm$ decays $\tau$'s will be tagged via their hadronic decay producing low-multiplicity narrow jets 
in the detector, there are two different $W^\pm$ decays that can be explored. The leptonic signature
$ b \bar{b} H^\pm W^\mp \to b \bar{b} \tau \nu l \nu $ provides a clean selection 
of the signal via the identification of the lepton $l=e,\mu$.
In this case the charged Higgs transverse mass cannot be reconstructed because
of the presence of two neutrinos with different origin. In this channel charged Higgs 
discovery will be determined 
by the observation of an excess of such events over SM expectations through a simple counting experiment. In the case of hadronic decays 
$ b \bar{b}H^\pm W^\mp  \to  b \bar{b}\tau \nu jj$ the transverse mass can instead be 
reconstructed since all neutrinos are arising from the charged Higgs boson decay. 
This allows for an efficient separation of the signal and the main 
$t\bar{t} \to b \bar{b}W^\pm W^\mp  \to  b \bar{b}\tau \nu jj$ background
(assuming $m_{H^\pm}\OOrd m_{W^\pm}$). 
The absence of a lepton ($e$ or $\mu$) provides a less 
clean environment but the use of the transverse mass makes it possible to reach the same mass discovery region as 
in the previous case and also to extract the charged Higgs boson mass. Both these channels show that after an 
integrated luminosity of 30 fb$^{-1}$ the discovery could be possible up to a mass of 150 GeV 
for all tan$\beta$ values in both ATLAS and CMS~\cite{Mohn:2007fd,biscarat,abdullin}. 

If the charged Higgs is heavier than the top quark, the dominant decay
channels are 
$H^\pm \to \tau \nu$ and $H^\pm \to tb$ depending on $\tan\beta$. They
have both been studied by  
ATLAS and CMS~\cite{assamagan,kinnunen,salmi,lowette}.
The charged Higgs bosons are produced in the $pp \to tbH^\pm$ channel. For the 
$H^\pm \to tb$ decay, a charged Higgs boson can be discovered up to 
high masses ($m_{H^\pm} \sim 400$~GeV) in the case of very large $\tan\beta$ values and this reach
cannot be much improved because of the large multi-jet environment. For the 
$H^\pm \to \tau \nu$ decay mode this reach is larger due to a cleaner signal despite a 
lower BR. In this case the 5$\sigma$ reach ranges from $\tan\beta=20$ for 
$m_{H^\pm}=200$ GeV to $\tan\beta=30$ for $m_{H^\pm}=400$ GeV.

For the LHC, signal and background events have been simulated in the
same way as for the
Tevatron as described before, however, without implying any rescaling factor to
match a measured $t\bar{t}$ cross section.
Table~\ref{tab:Hpm:crosssecLHC} lists 
the resulting cross sections before 
($\sigma^{\rm th}$) and after ($\sigma$) applying the basic cuts 
$p_t^{\rm jets} > 20$~GeV and the hard cut $p_t^{\rm miss} > 100$~GeV.
The LHC rates allow for the discovery to be less challenging than at 
the Tevatron in the region $m_{H^\pm} \sim m_{W^\pm}$, 
yet the separation of  signal events from
background remains crucial for the measurement of the charged Higgs mass.

\begin{table}[htbp]
\vspace*{-0.3cm}
\caption{\label{tab:Hpm:crosssecLHC}
LHC cross sections of background $q\bar q, gg \to  t\bar{t}$
and signal $q\bar q, gg \to  tbH^\pm$
for $\tan\beta = 30$ and $m_{H^\pm} = 80, 100, 130$ and $150$~GeV
into the final state
$2 b + 2 j + \tau_\mathrm{jet} + p_t^{\rm miss}$
before ($\sigma^{\rm th}$) and after ($\sigma$)
the basic cuts ($p_t > 20$~GeV for all jets)
and the hard cut ($p_t^{\rm miss} > 100$ GeV).
}
\centering
\begin{tabular}{c|c|c|c|c|c}
 & $q\bar q, gg \to  t\bar{t}$ &
  \multicolumn{4}{c}{$q\bar q, gg \to  tbH^\pm$} \\
$m_{H^\pm}$ (GeV) & 80 & 80 & 100 & 130 & 150 \\
  \hline
$\sigma^{\rm th}$ (pb) & 45.5 &  72.6    &  52.0    &  24.5 & 9.8  \\
$\sigma$ (pb) for $p_t^\mathrm{jets} > 20$ GeV 
   & 17.3 &  33.9   &  25.7    &  12.2  &  3.8 \\
$\sigma$ (pb) for $(p_t^\mathrm{jets},p_t^{\rm miss}) > (20,100)$ GeV
   & 4.6 &   6.0    &   4.8    &  2.9   &  1.2
\end{tabular}
\end{table}

\begin{figure}[htbp]
\epsfig{file=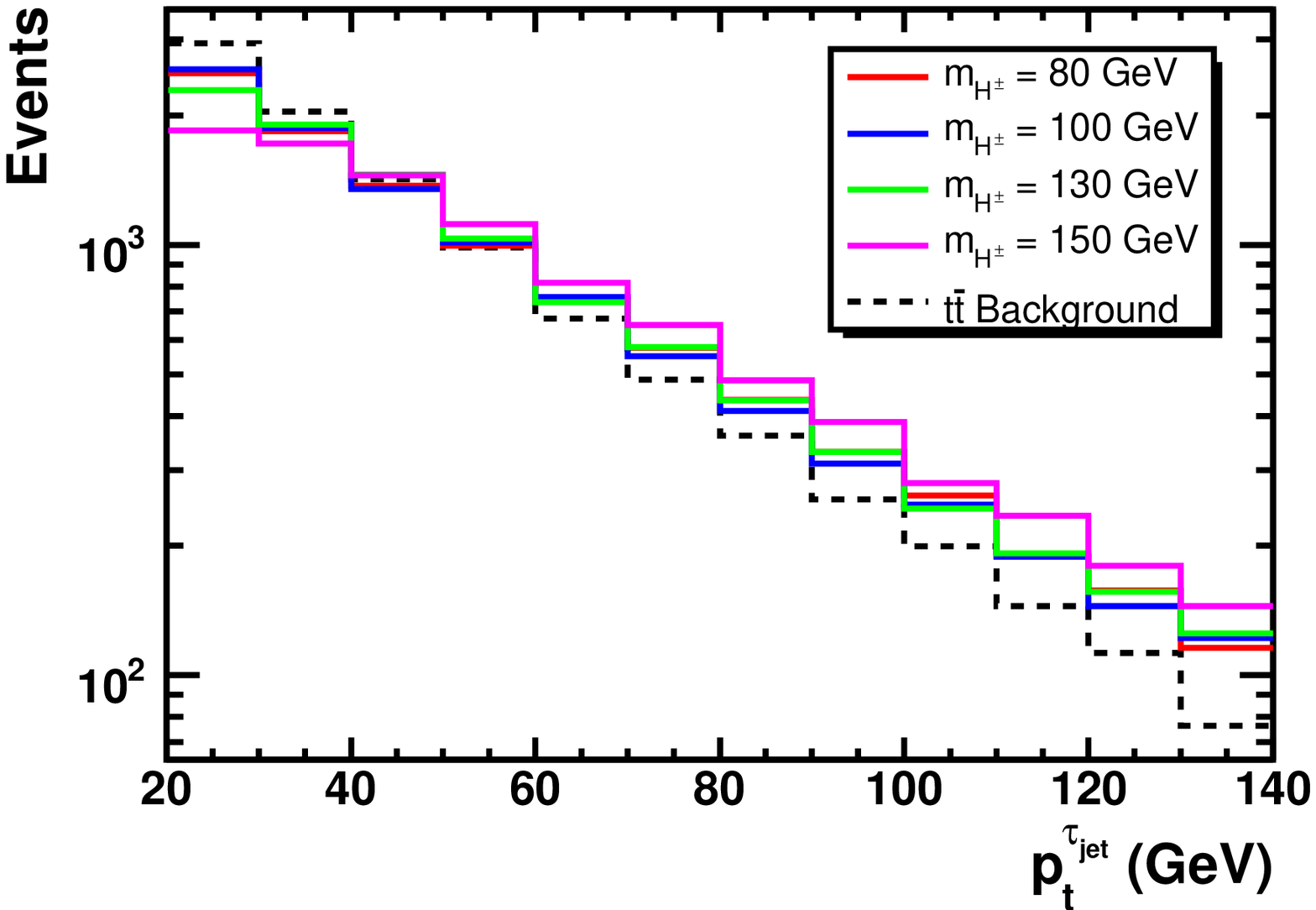, width=0.5\textwidth} \hfill
\epsfig{file=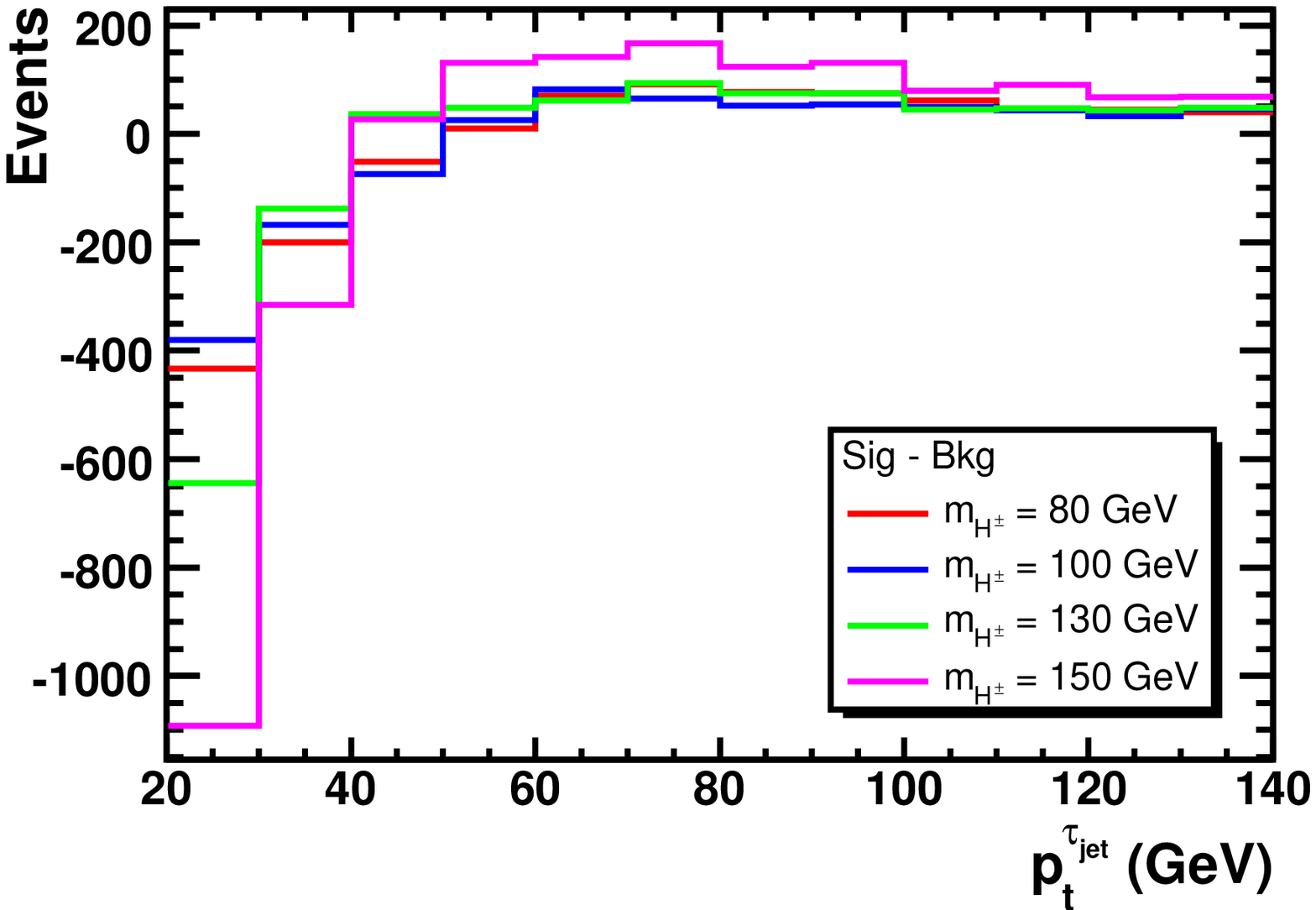, width=0.5\textwidth}
\epsfig{file=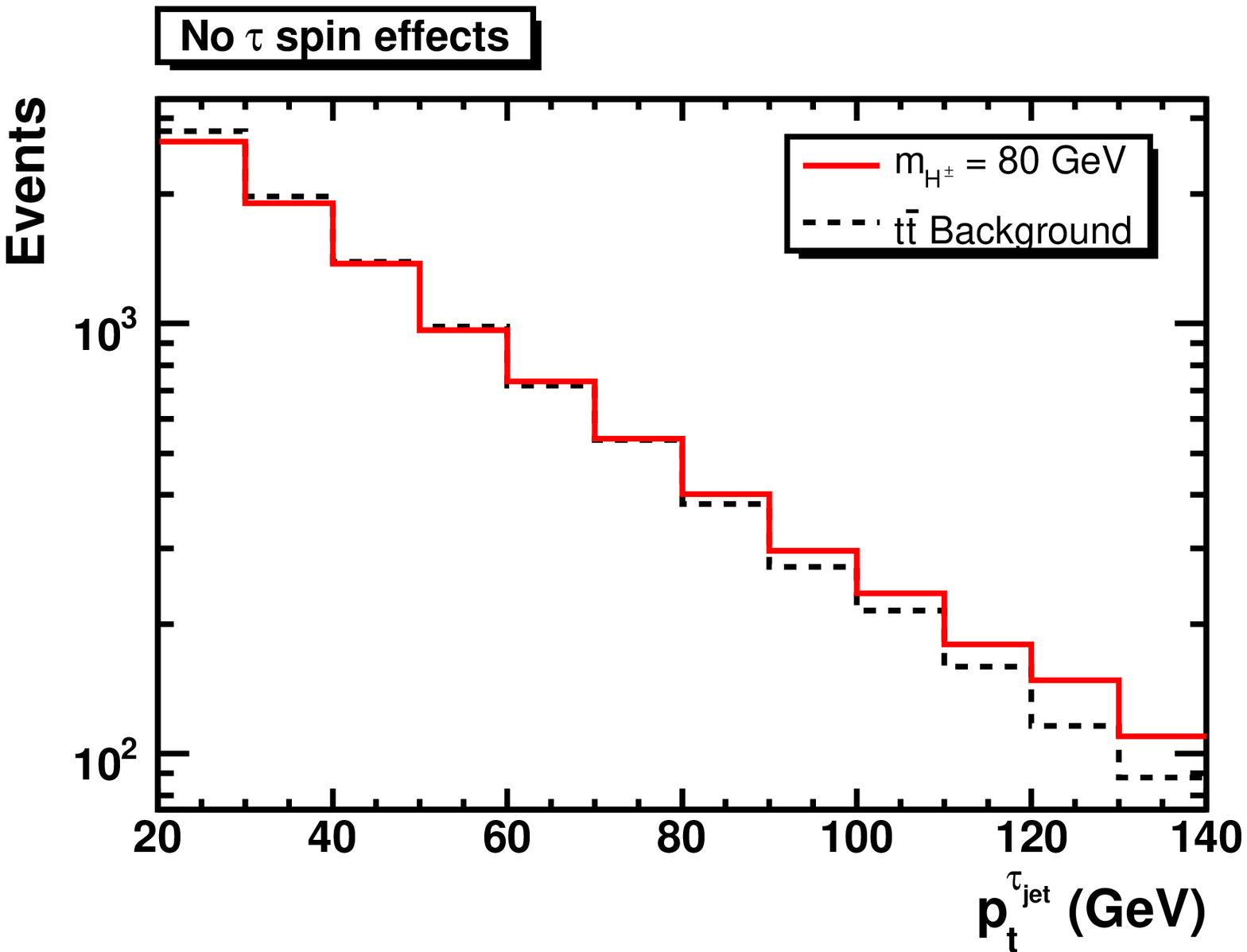, width=0.5\textwidth} \hfill
\epsfig{file=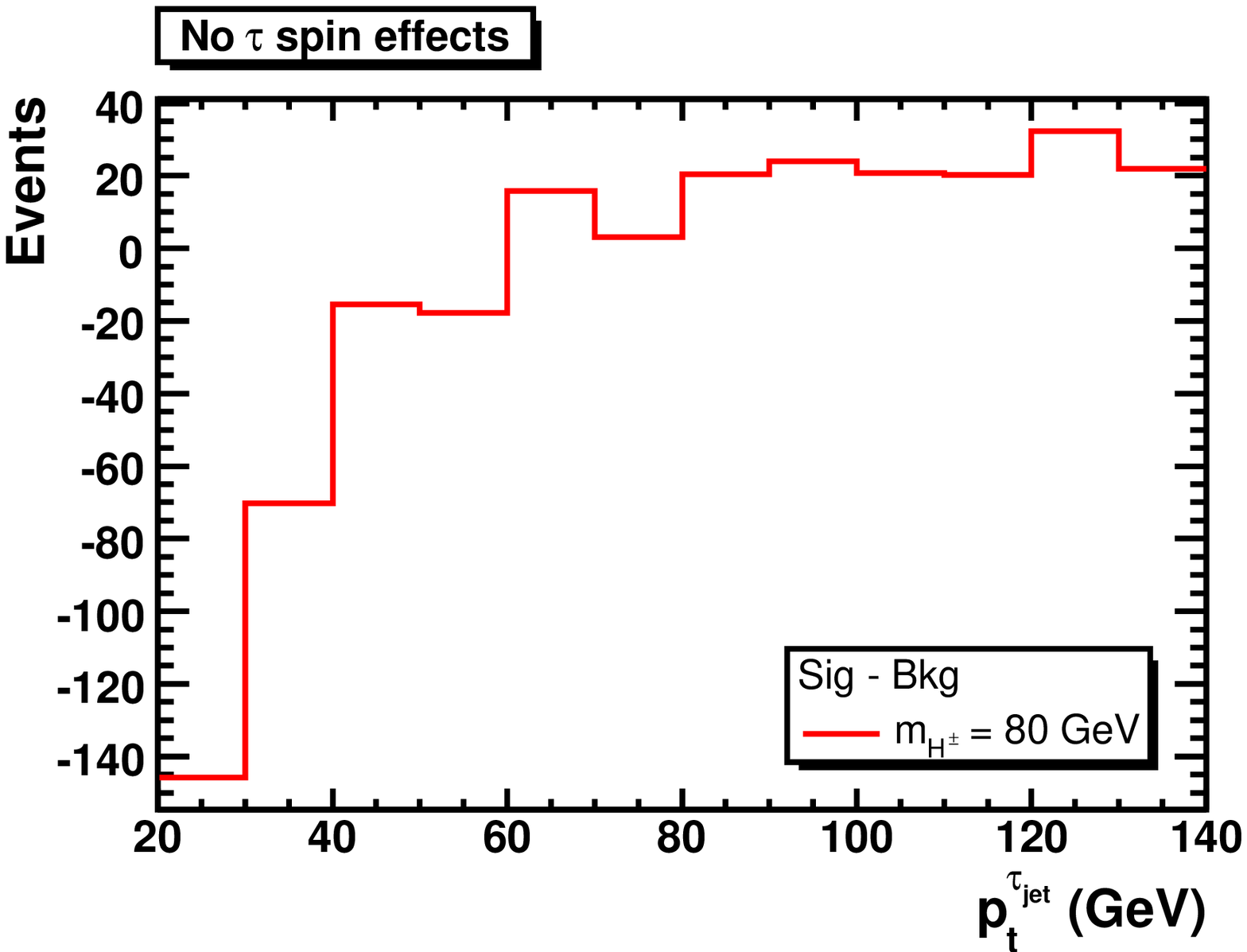, width=0.5\textwidth}
\caption{ 
$p_t$ distributions 
of the $\tau~{\mathrm{jet}}$ for the $tbH^\pm$ signal
and the $t\bar{t}$ background for $\sqrt{s}=14$~TeV (left)
and the respective differences between signal and background (right).
The lower plots show distributions without spin effects in the $\tau$
decays.
}
\label{fig:lhc_pttau}
\end{figure}

\begin{figure}[htbp]
\epsfig{file=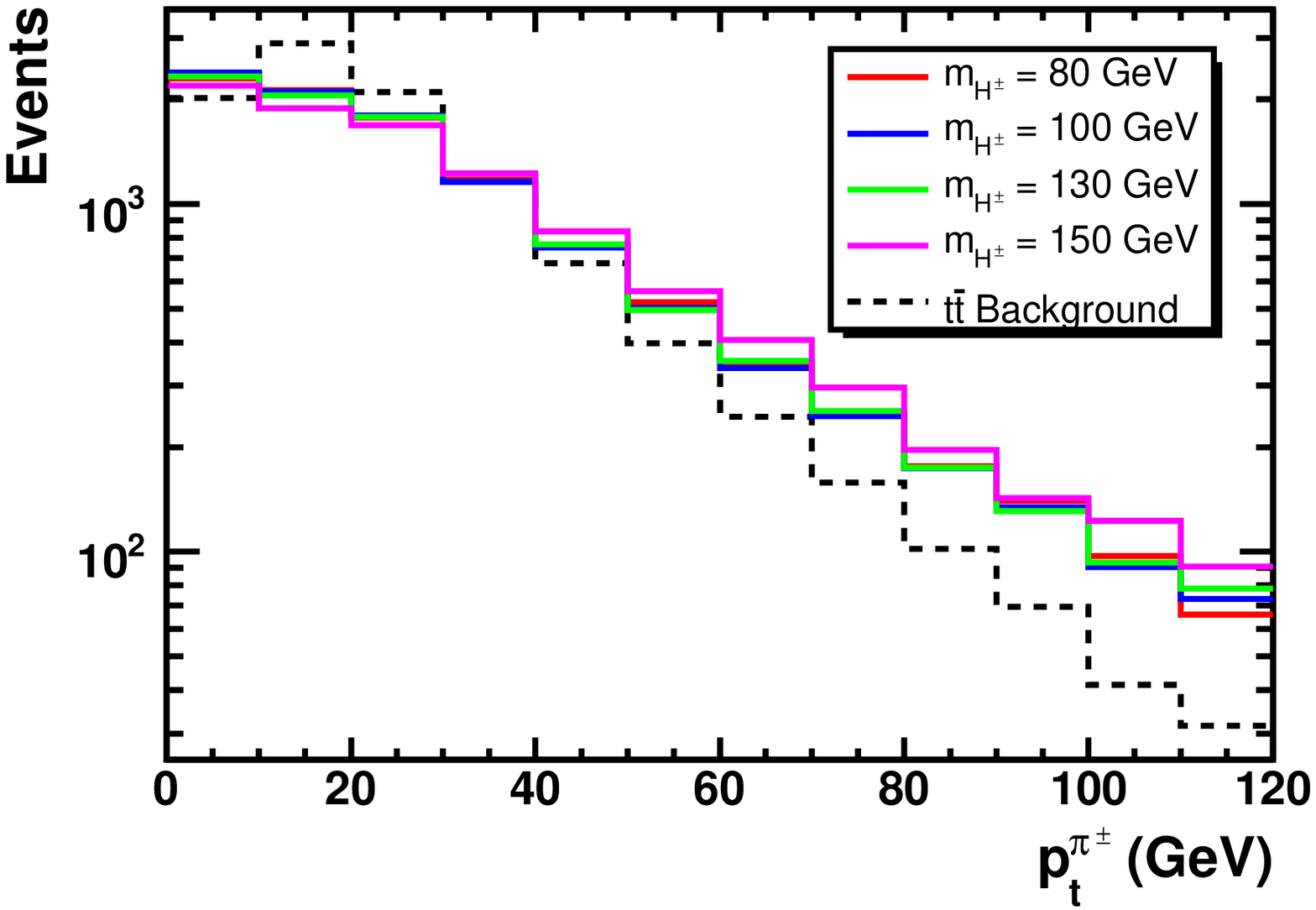, width=0.5\textwidth}  \hfill
\epsfig{file=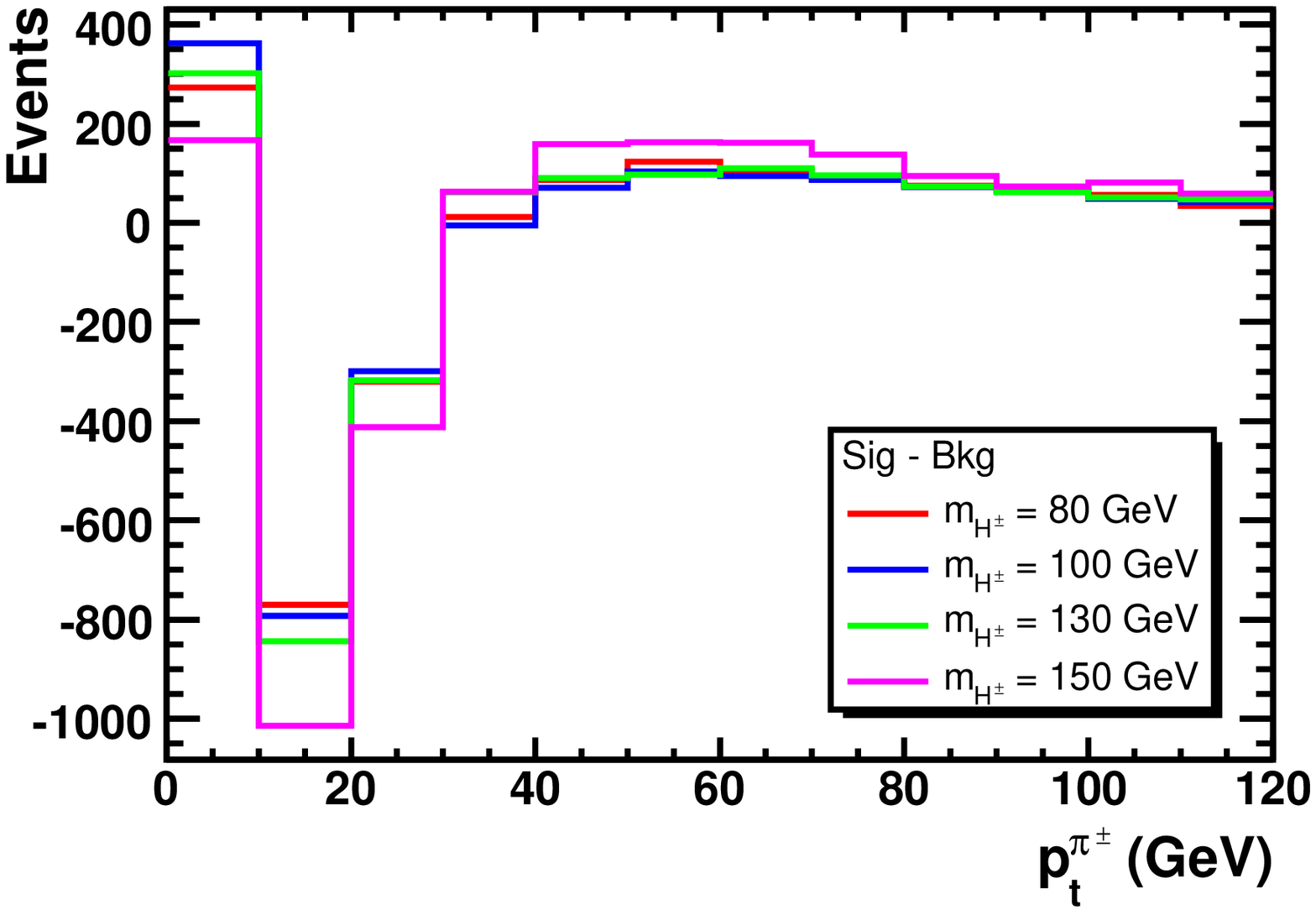, width=0.5\textwidth}
\epsfig{file=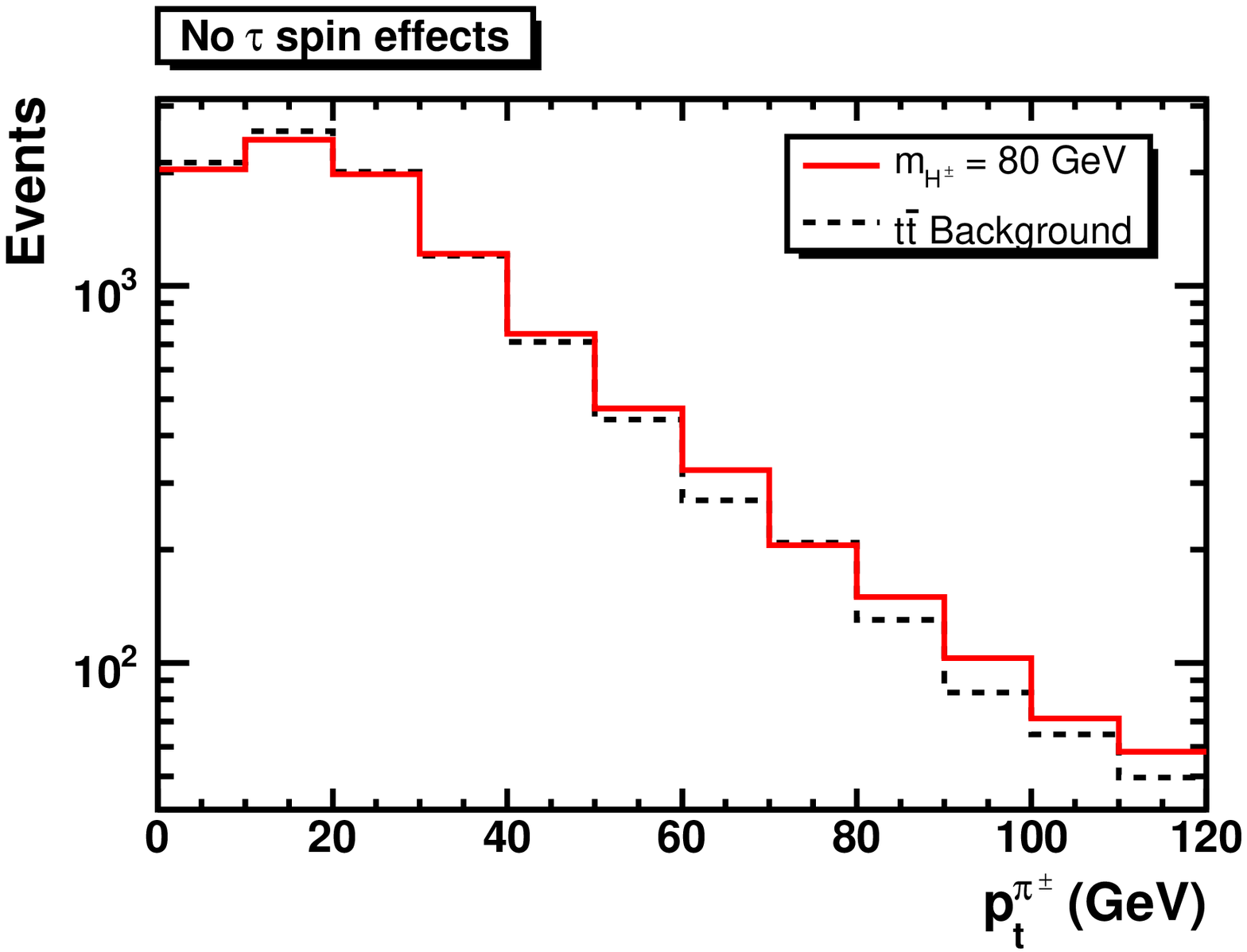, width=0.5\textwidth}  \hfill
\epsfig{file=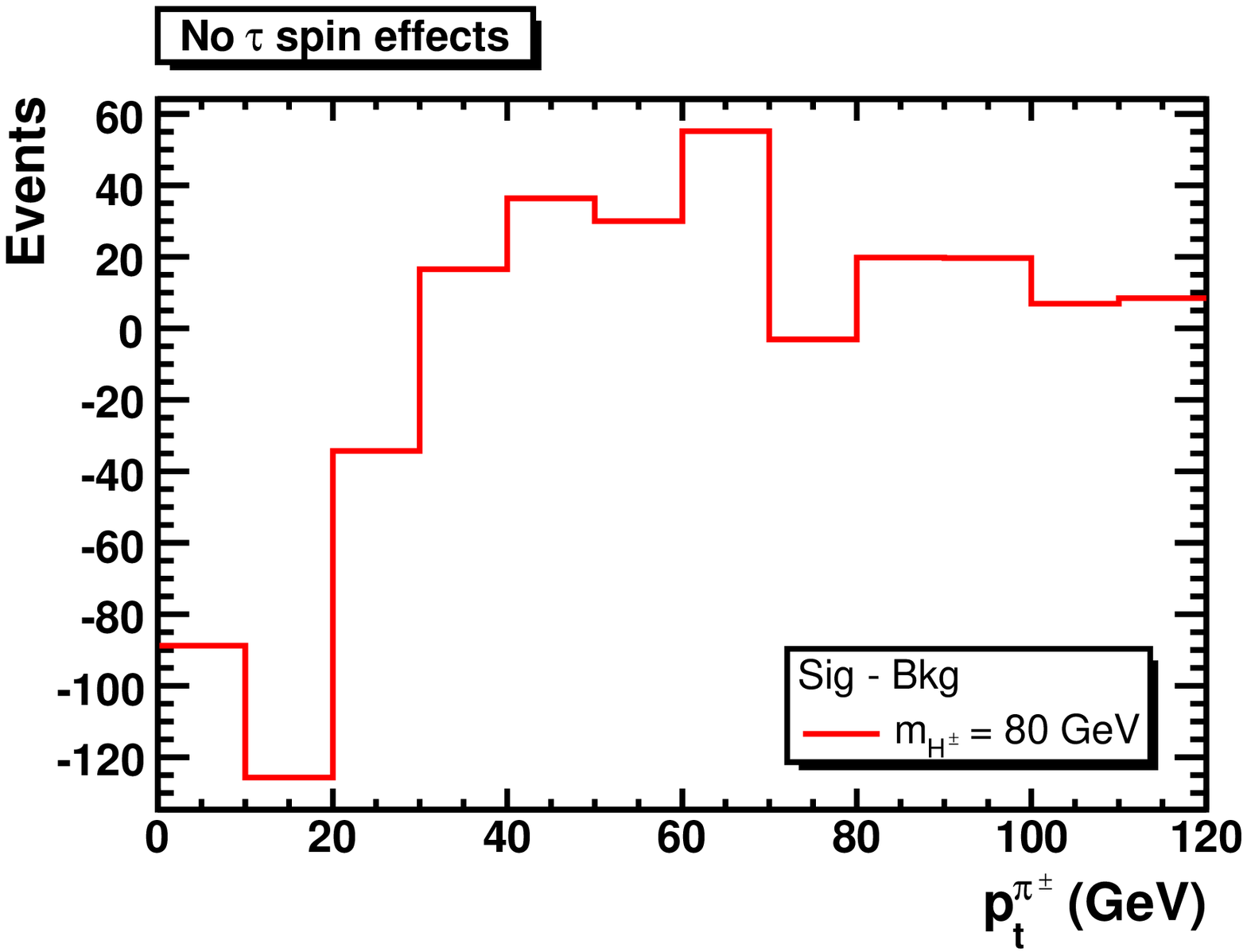, width=0.5\textwidth}
\caption{
$p_t$  distributions of the leading $\pi^\pm$ from the $\tau$ decay 
for the $tbH^\pm$ signal and the 
$t\bar{t}$ background for $\sqrt{s}=14$~TeV (left)
and the respective differences between signal and background (right).
The lower plots show distributions without spin effects in the $\tau$
decays.
}
\label{fig:lhc_ptpi}
\end{figure}

\begin{figure}[htbp]
\epsfig{file=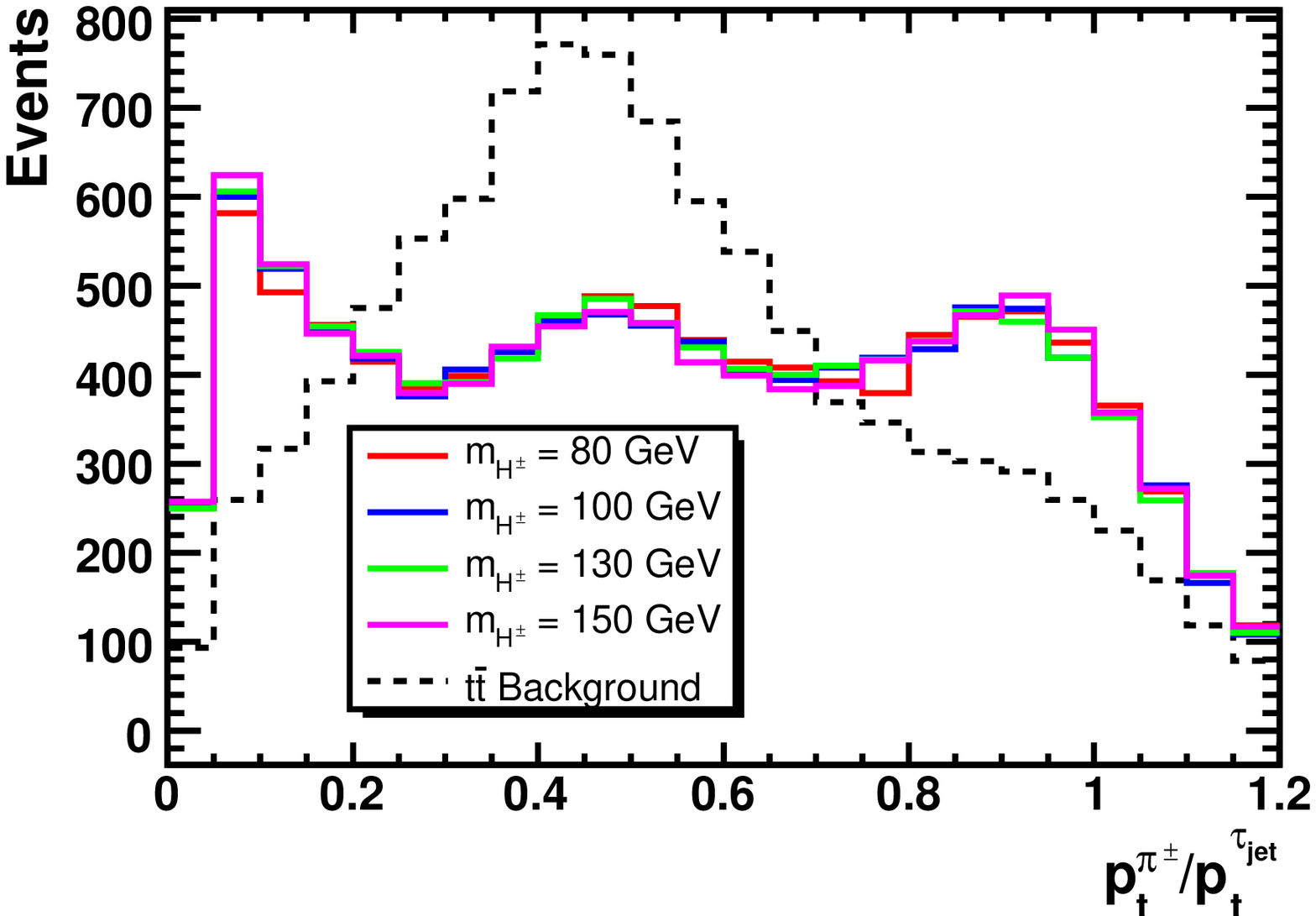, width=0.5\textwidth} \hfill
\epsfig{file=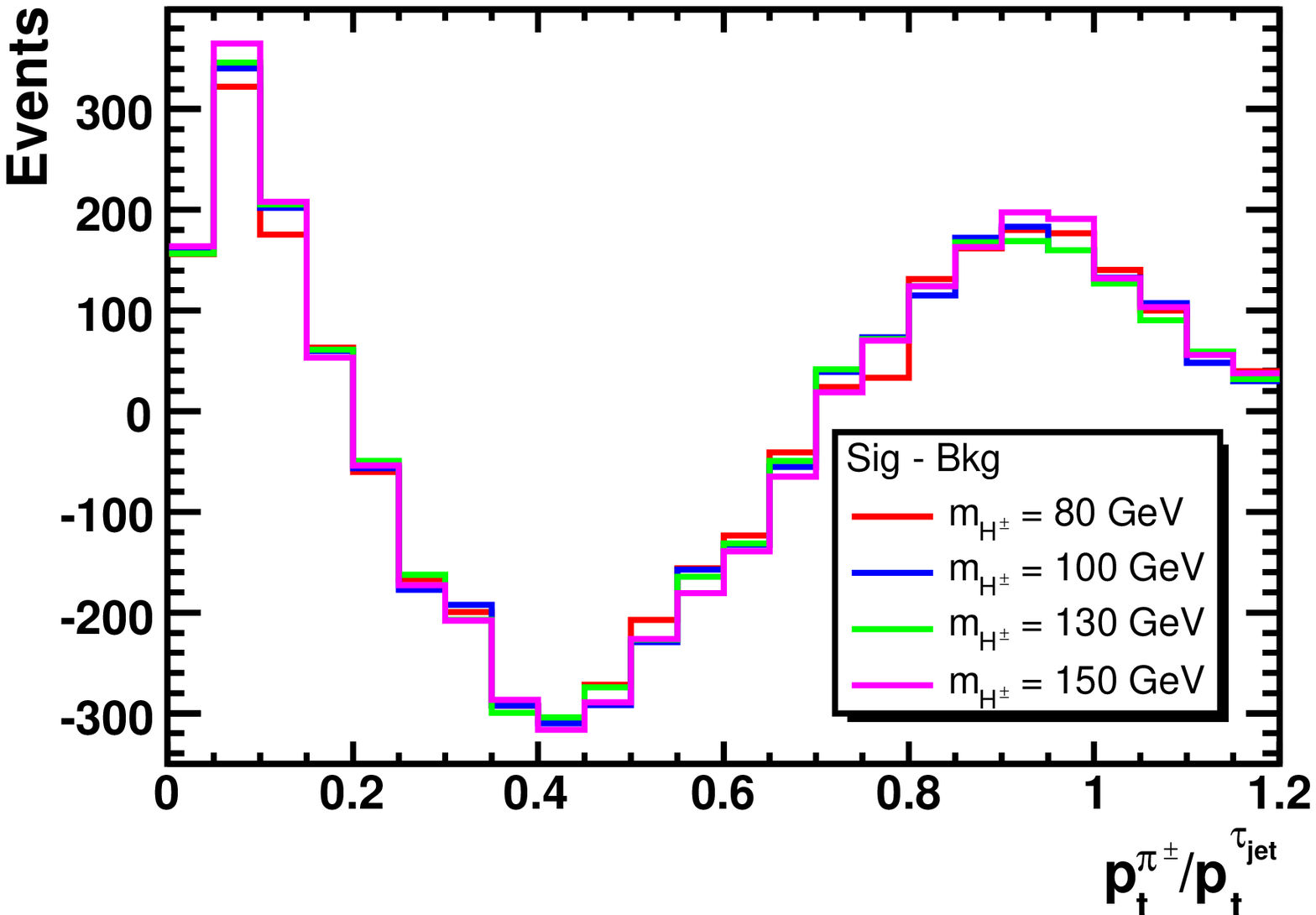, width=0.5\textwidth}
\epsfig{file=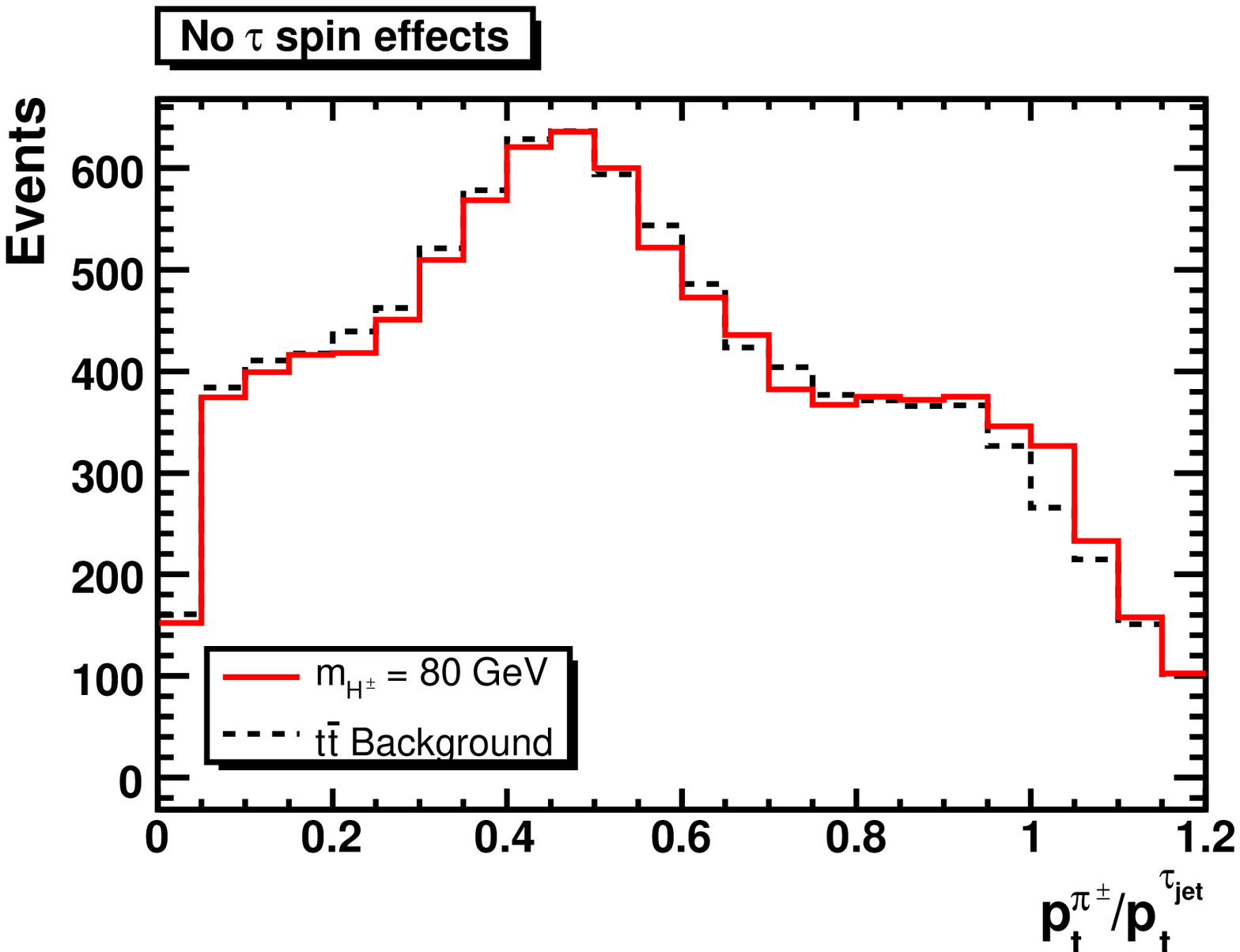, width=0.5\textwidth} \hfill
\epsfig{file=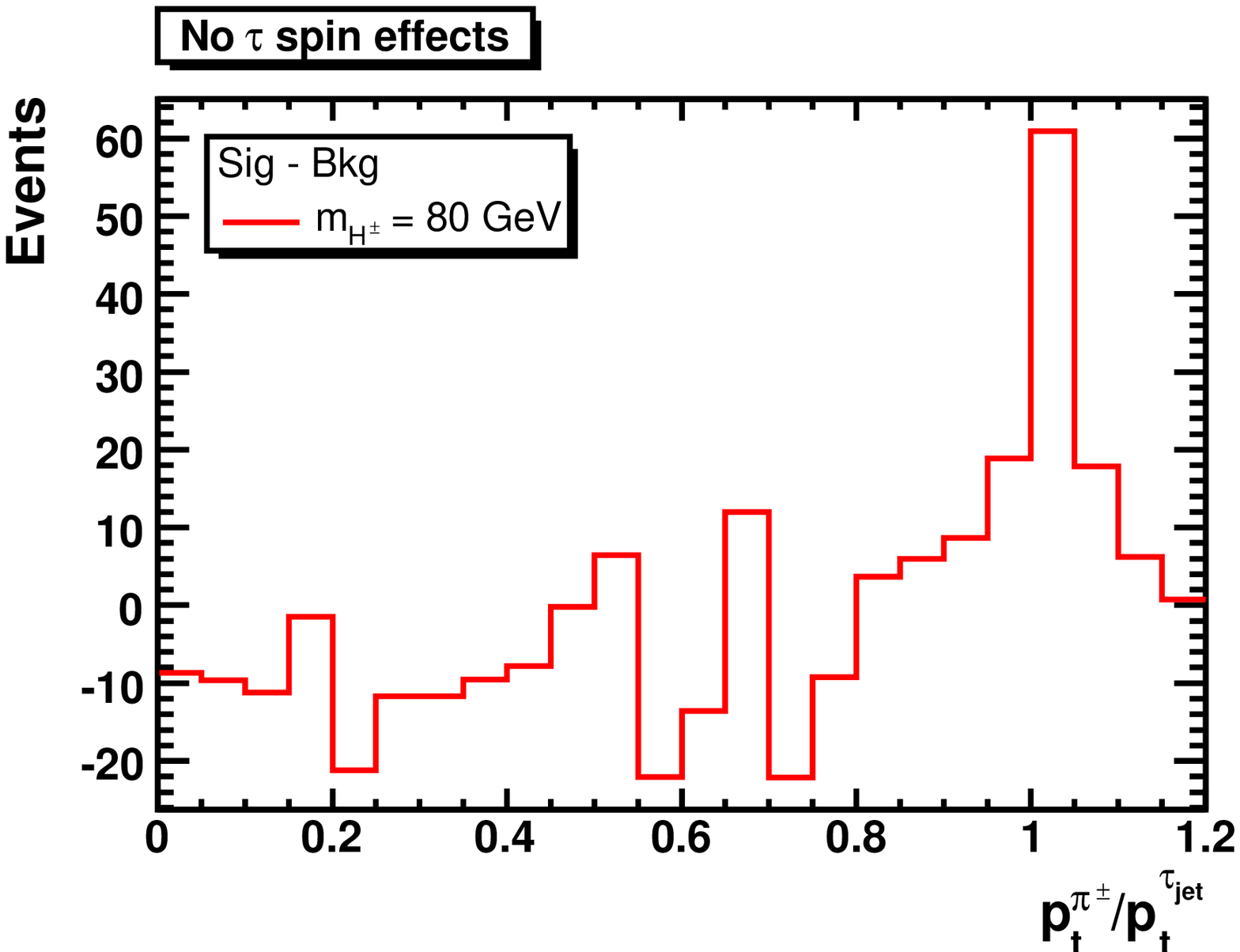, width=0.5\textwidth}
\caption{Distributions of the  
ratio $p_t^{\pi^\pm}/p_t^{\tau_\mathrm{jet}}$
for the $tbH^\pm$ signal
and the $t\bar{t}$ background for $\sqrt{s}=14$~TeV (left)
and the respective differences between signal and background (right).
The lower plots show distributions without spin effects in the $\tau$
decays.
}
\label{fig:lhc_r1}
\end{figure}

\begin{figure}[htbp]
\epsfig{file=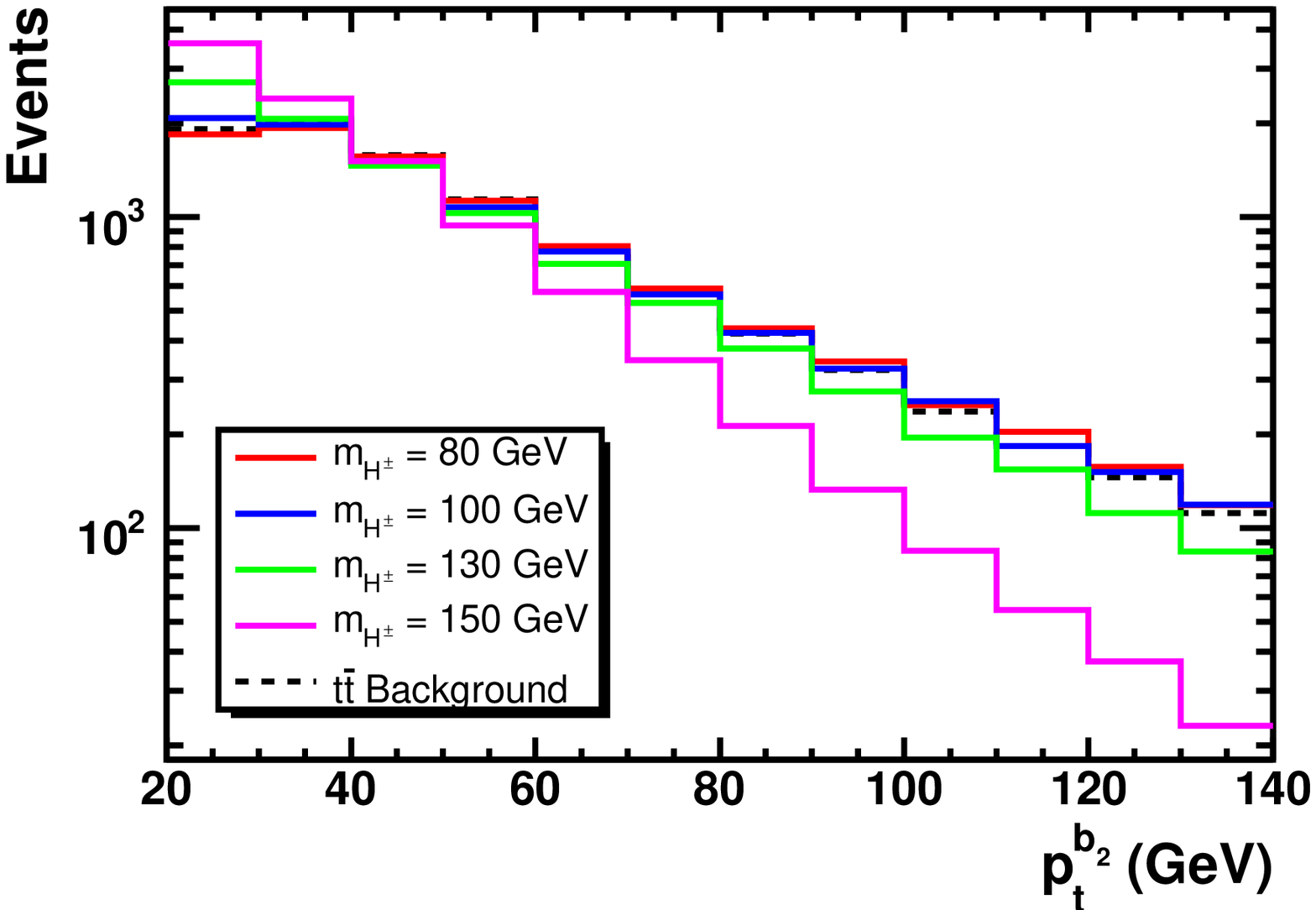, width=0.5\textwidth}  \hfill
\epsfig{file=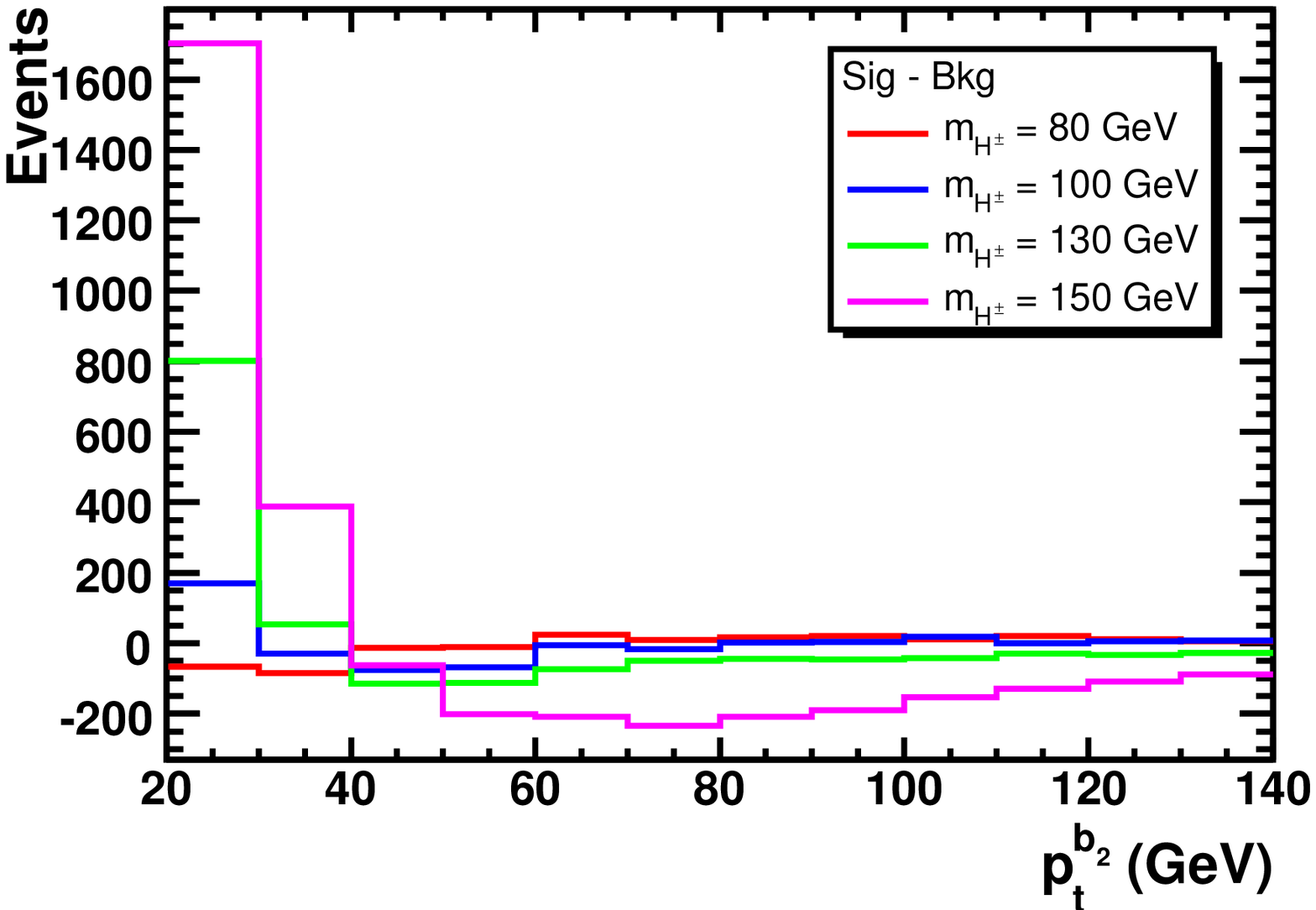, width=0.5\textwidth}
\caption{
$p_t$ distributions of the second (least energetic) $b$ quark jet
for the $tbH^\pm$ signal
and the $t\bar{t}$ background for $\sqrt{s}=14$~TeV (left)
and the respective differences between signal and background (right).
}
\label{fig:lhc_ptb2}
\end{figure}

\begin{figure}[htbp]
\epsfig{file=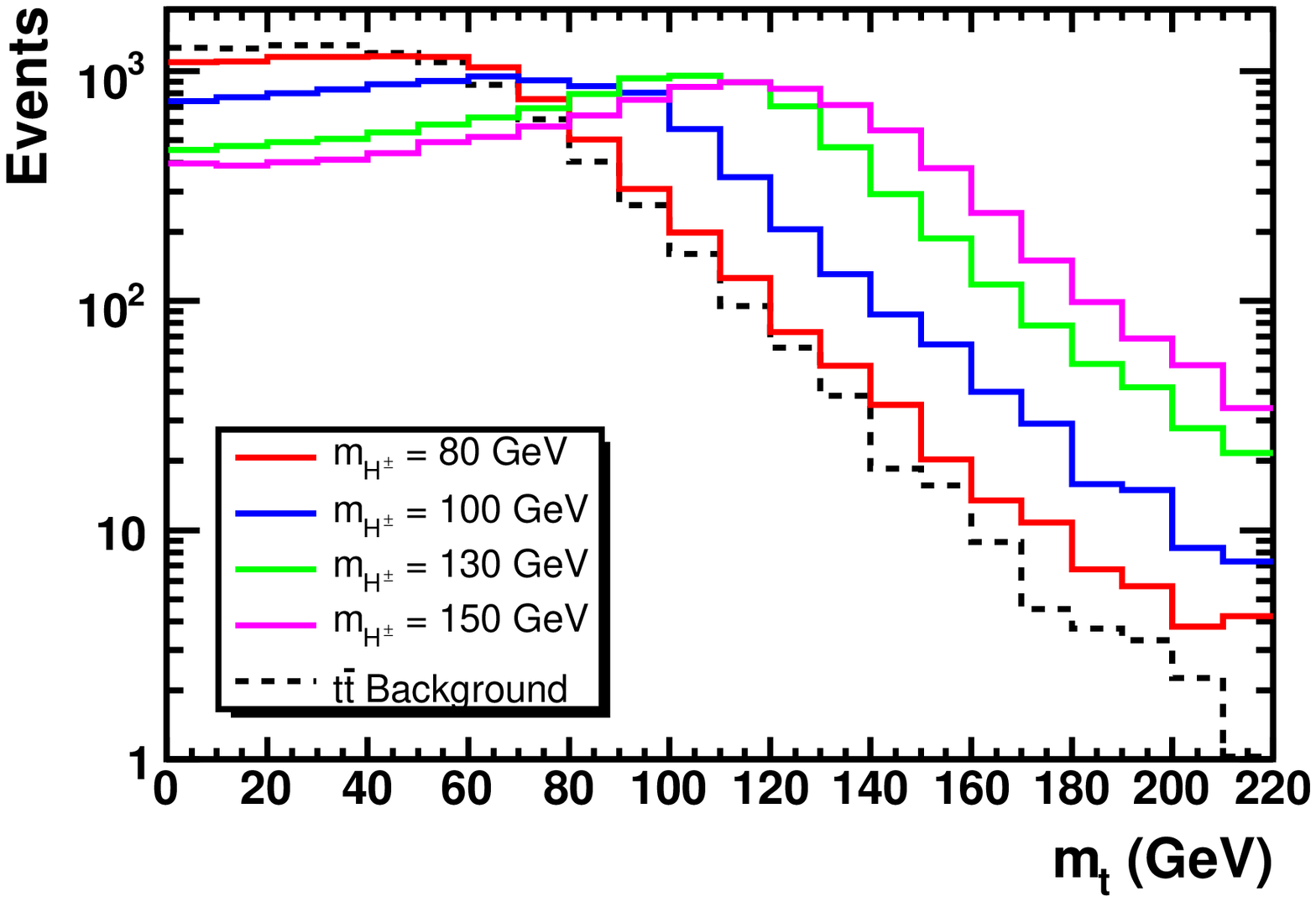, width=0.5\textwidth}  \hfill
\epsfig{file=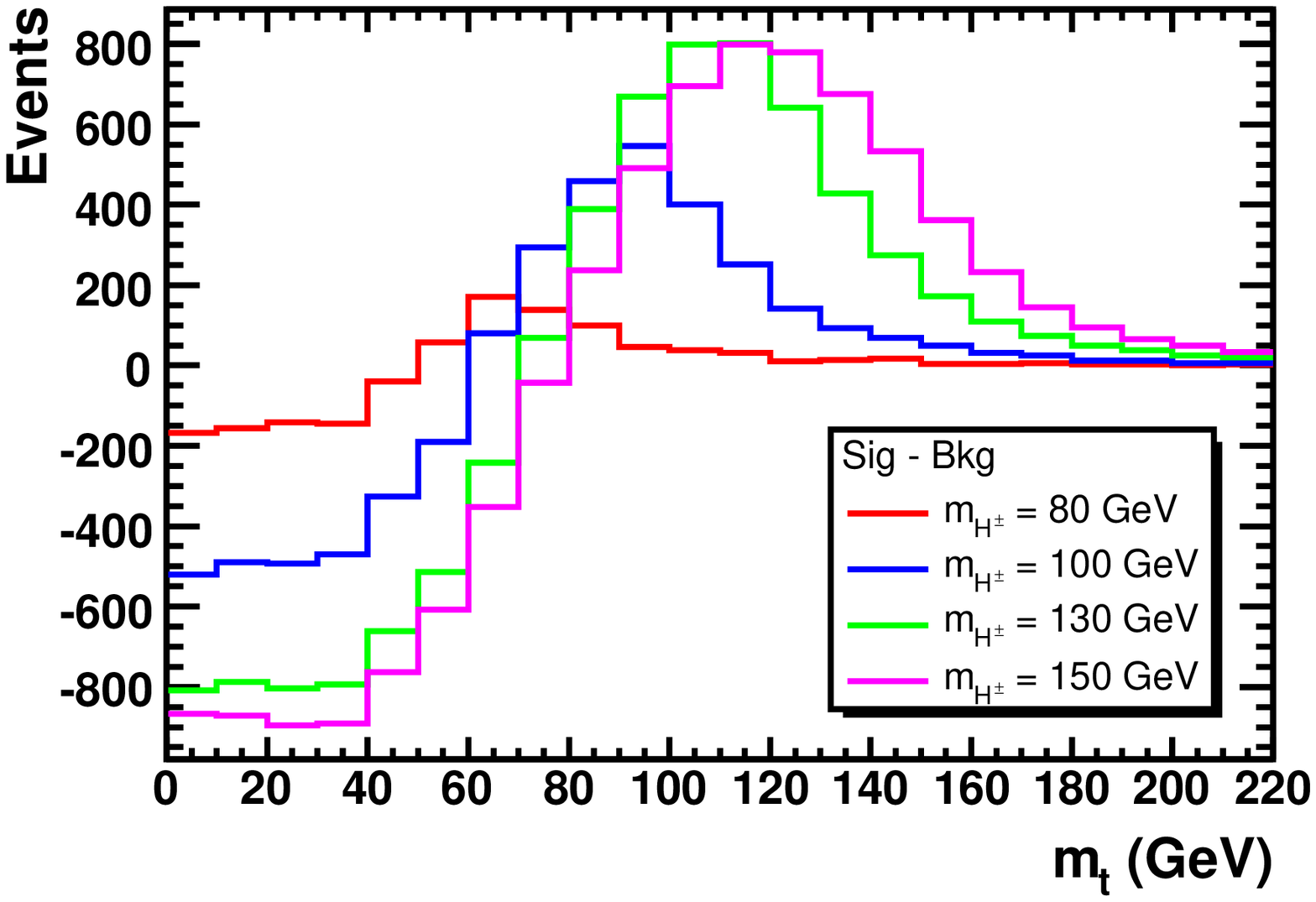, width=0.5\textwidth}
\caption{
Transverse mass
$m_t = \sqrt{2 p_t^{\tau_\mathrm{jet}} p_t^{\rm miss}
  [1-\cos(\Delta\phi)]}$
distributions of the
$\tau_{\rm jet} + p_t^{\rm miss}$ system
($\Delta\phi$ is the azimuthal angle
between $p_t^{\tau_\mathrm{jet}}$ and $p_t^{\rm miss}$)
for the $tbH^\pm$ signal
and the $t\bar{t}$ background for $\sqrt{s}=14$~TeV (left)
and the respective differences between signal and background (right).
}
\label{fig:lhc_mtransverse}
\end{figure}

\begin{figure}[htbp]
\epsfig{file=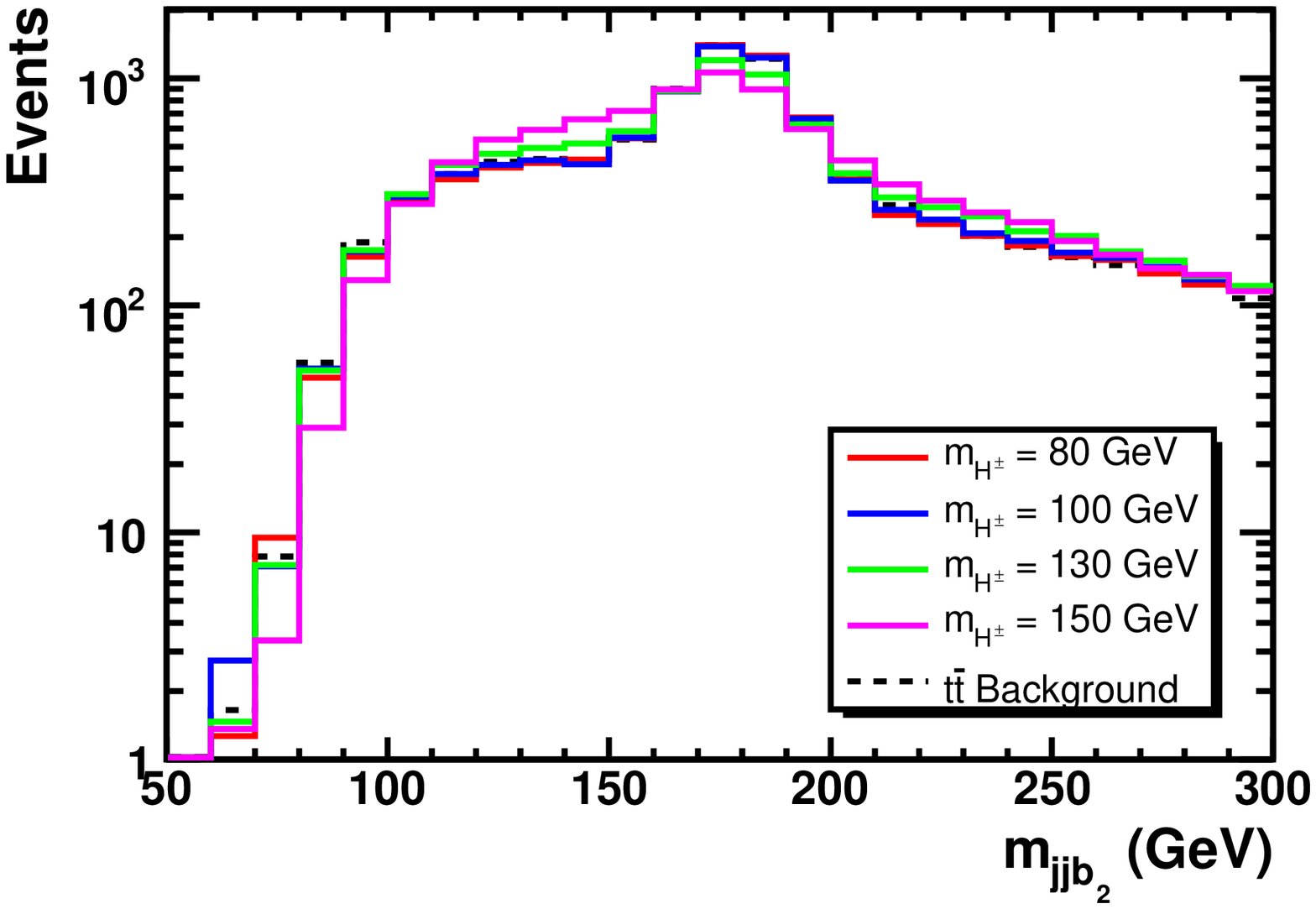, width=0.5\textwidth}  \hfill
\epsfig{file=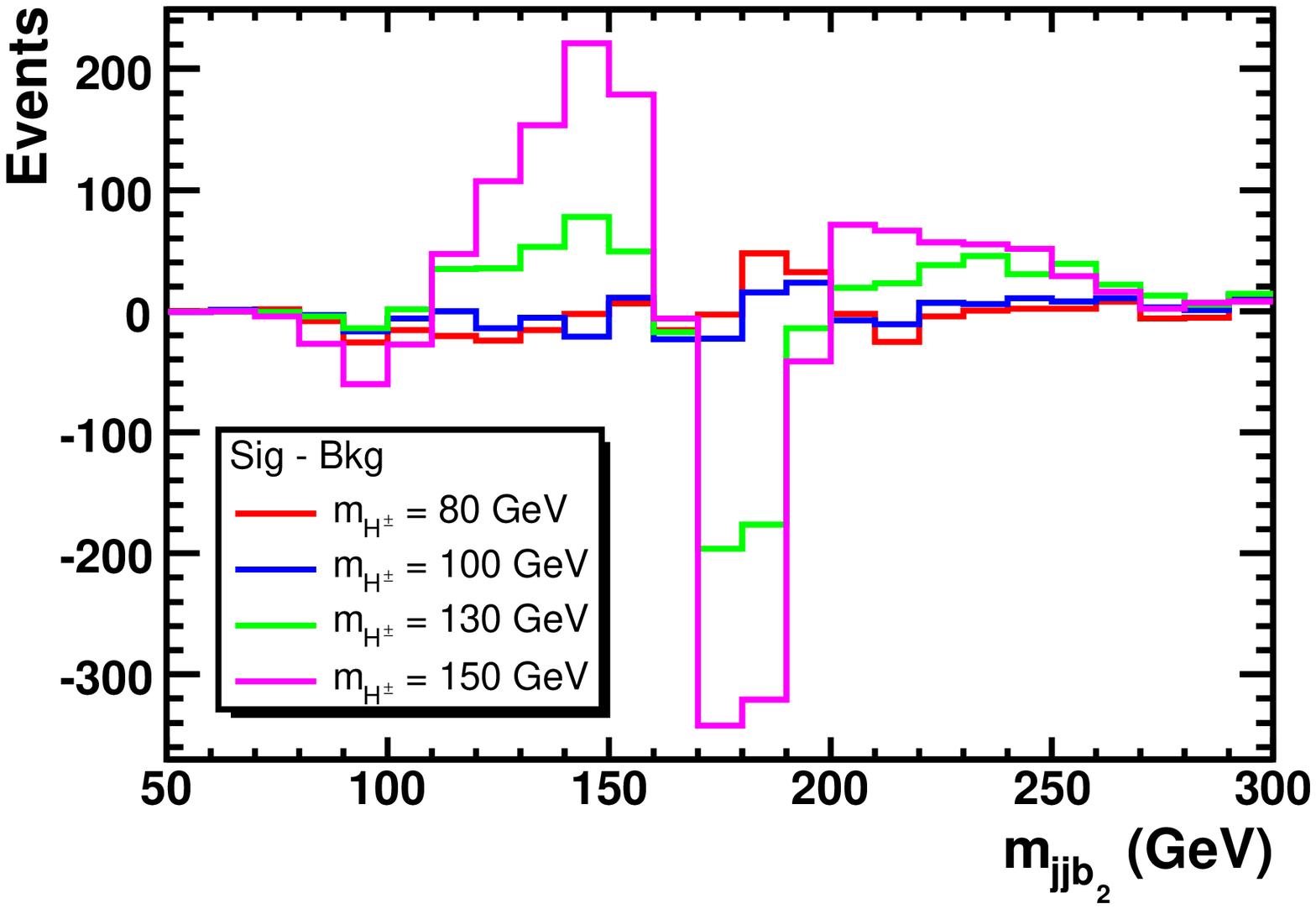, width=0.5\textwidth}
\caption{
Invariant mass distributions of the two light quark jets and the 
second (least energetic) $b$ quark jet
for the $tbH^\pm$ signal
and the $t\bar{t}$ background for $\sqrt{s}=14$~TeV (left)
and the respective differences between signal and background (right).
}
\label{fig:lhc_mjjb2}
\end{figure}

\begin{figure}[htbp]
\epsfig{file=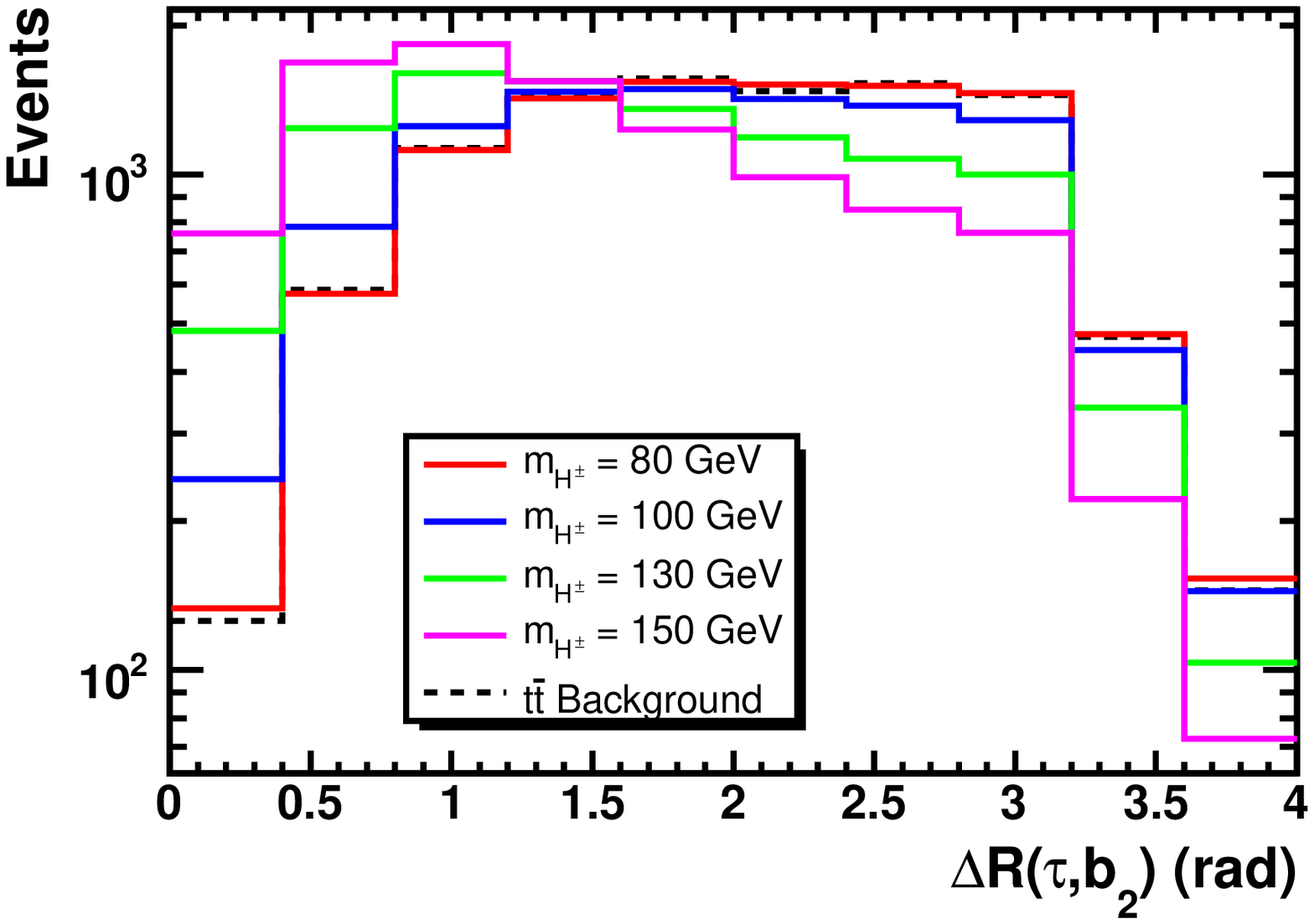, width=0.5\textwidth}  \hfill
\epsfig{file=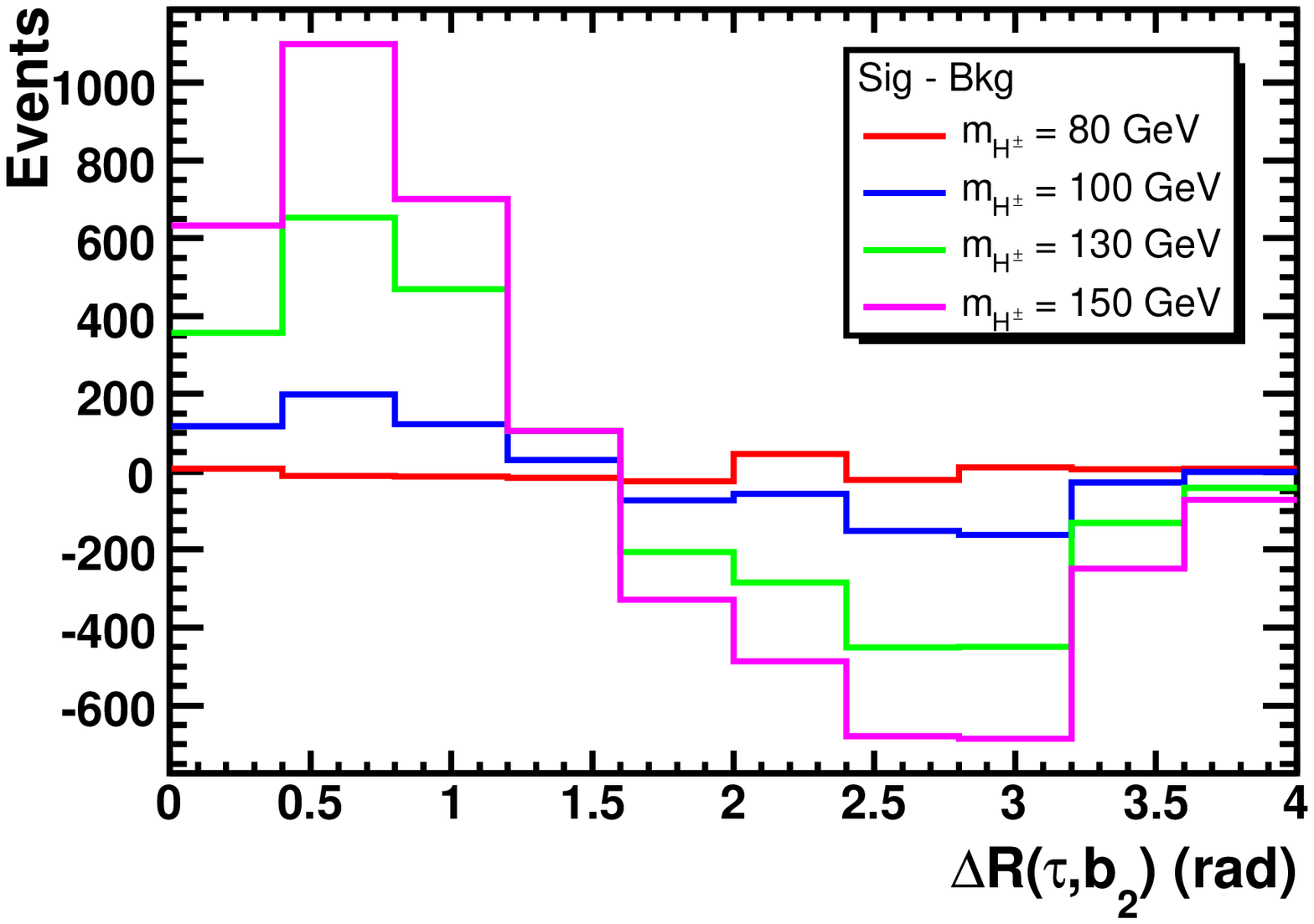, width=0.5\textwidth}
\caption{
Spatial distance
$\Delta R(\tau,b_2) = \sqrt{(\Delta\phi)^2 + (\Delta\eta)^2}$
distributions
(where $\Delta\phi$ is the azimuthal angle in rad between
the $\tau$ and $b$ jet)
for the $tbH^\pm$ signal
and the $t\bar{t}$ background for $\sqrt{s}=14$~TeV (left)
and the respective differences between signal and background (right).
}
\label{fig:lhc_distance-tau-b}
\end{figure}

\begin{figure}[htbp]
\epsfig{file=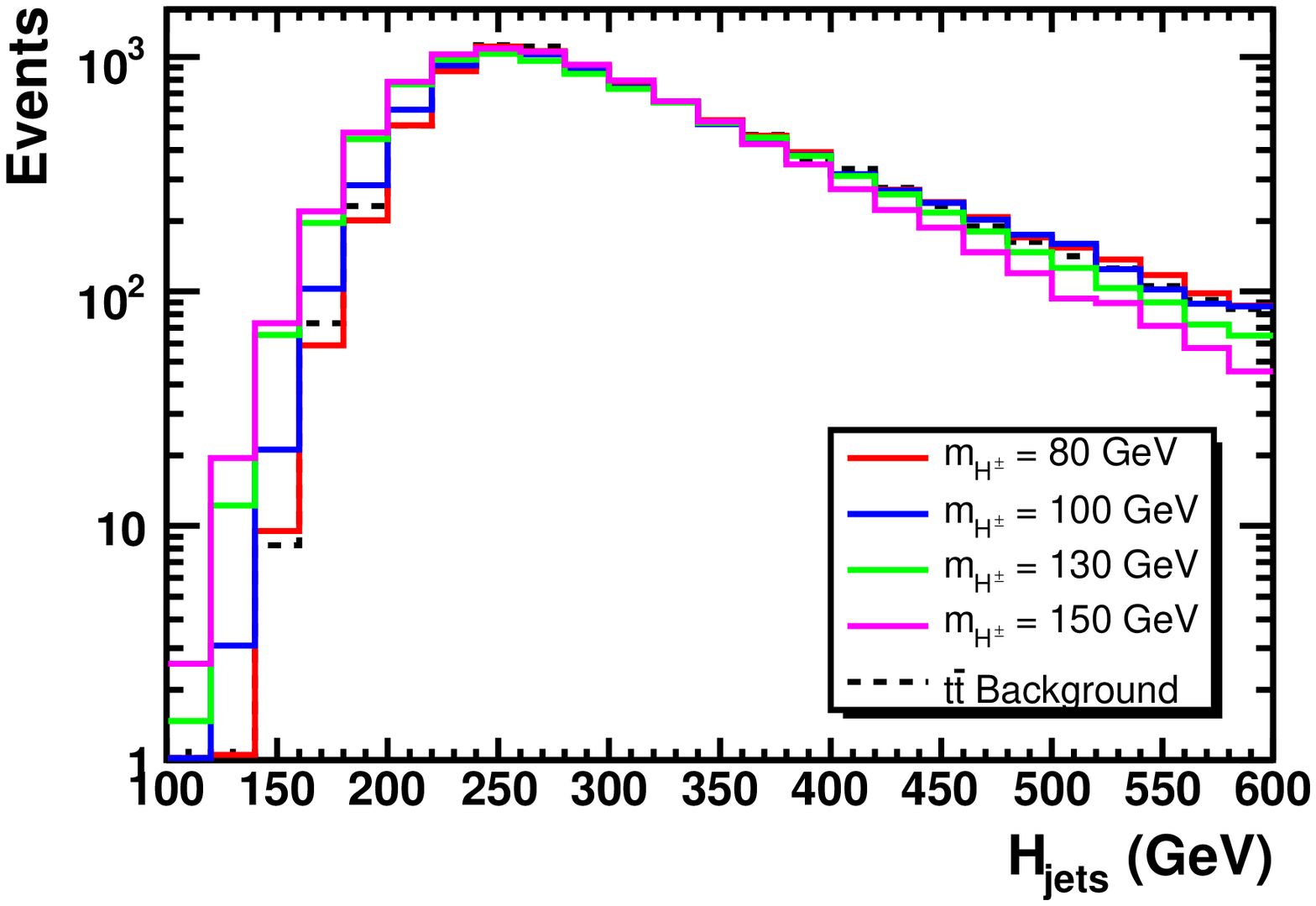, width=0.5\textwidth}  \hfill
\epsfig{file=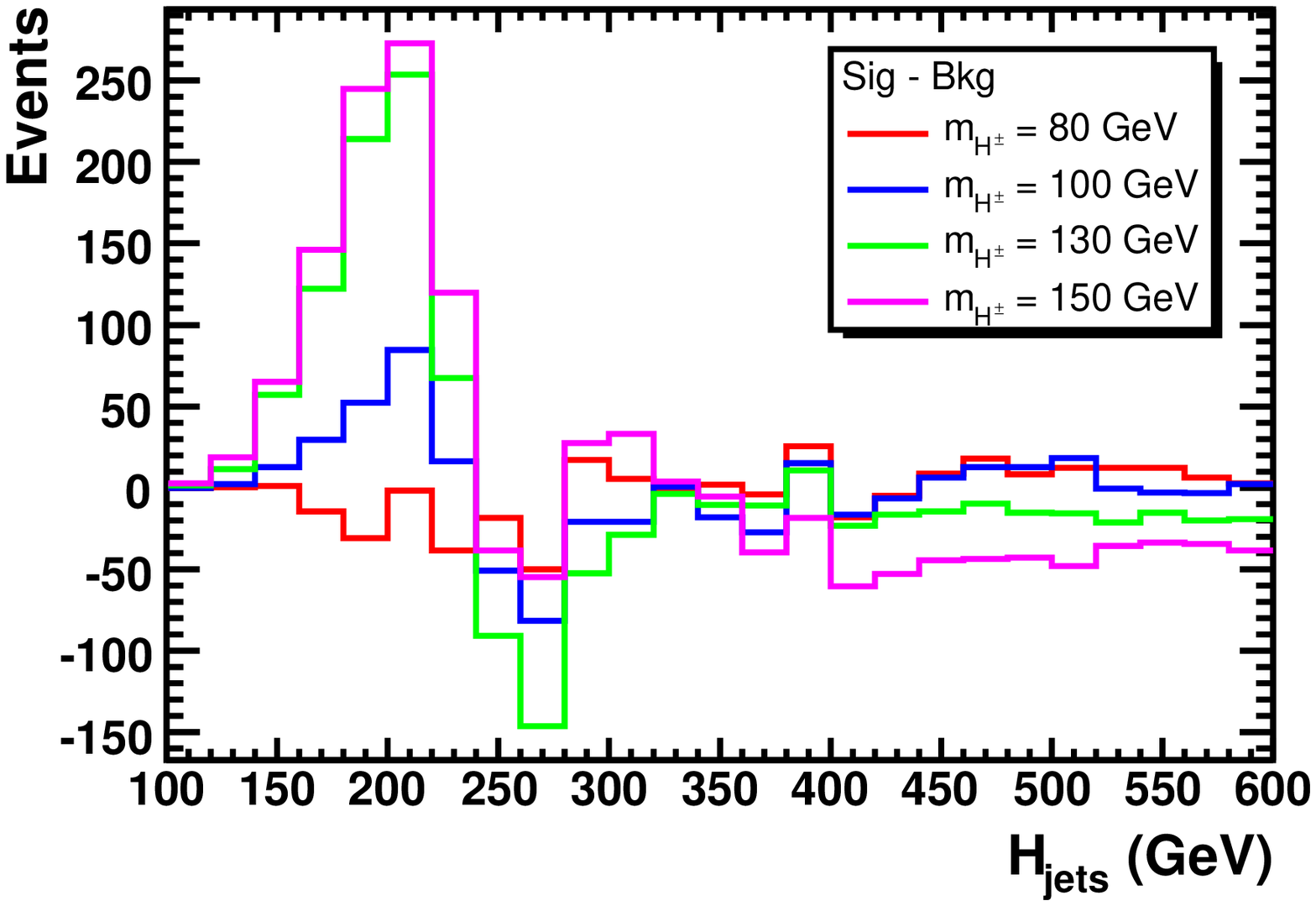, width=0.5\textwidth}
\caption{
Distributions of the
total transverse momentum of all quark jets,
$H_{\rm jets} = p_t^{j_1} + p_t^{j_2} + p_t^{b_1} + p_t^{b_2}$,
for the $tbH^\pm$ signal
and the $t\bar{t}$ background for $\sqrt{s}=14$~TeV (left)
and the respective differences between signal and background (right). 
}
\label{fig:lhc_hjet}
\end{figure}

The kinematic distributions are shown in Figs.~\ref{fig:lhc_pttau} to
\ref{fig:lhc_hjet} 
for $\sqrt{s}=14$~TeV.
The choice of variables is identical to the one for the Tevatron and
allows for a one-to-one comparison, 
the differences being  due to a change in CM energy (and, to a
somewhat lesser extent, due to the leading partonic mode of the 
production process\footnote{As the latter
is dominated by $q\bar q$ annihilation at the Tevatron and $gg$ fusion
at the LHC.}). 
The main differences with respect to
Figs.~\ref{fig:pttau}--\ref{fig:hjet} are that 
the various transverse momenta and
invariant masses have longer high energy tails. 
In particular, it should be noted
that the effect of the spin differences between $W^\pm$ and $H^\pm$
events can be explored very effectively also at LHC energies,
e.g. the ratio $p_t^{\pi^\pm}/p_t^{\tau_\mathrm{jet}}$ is shown
in Fig.~\ref{fig:lhc_r1} which is very sensitive to the spin effects.
These observations lead to the conclusion
that the same method using spin differences can be used to separate
signal from background at both the Tevatron and the LHC.

\begin{figure}[htbp]
\epsfig{file=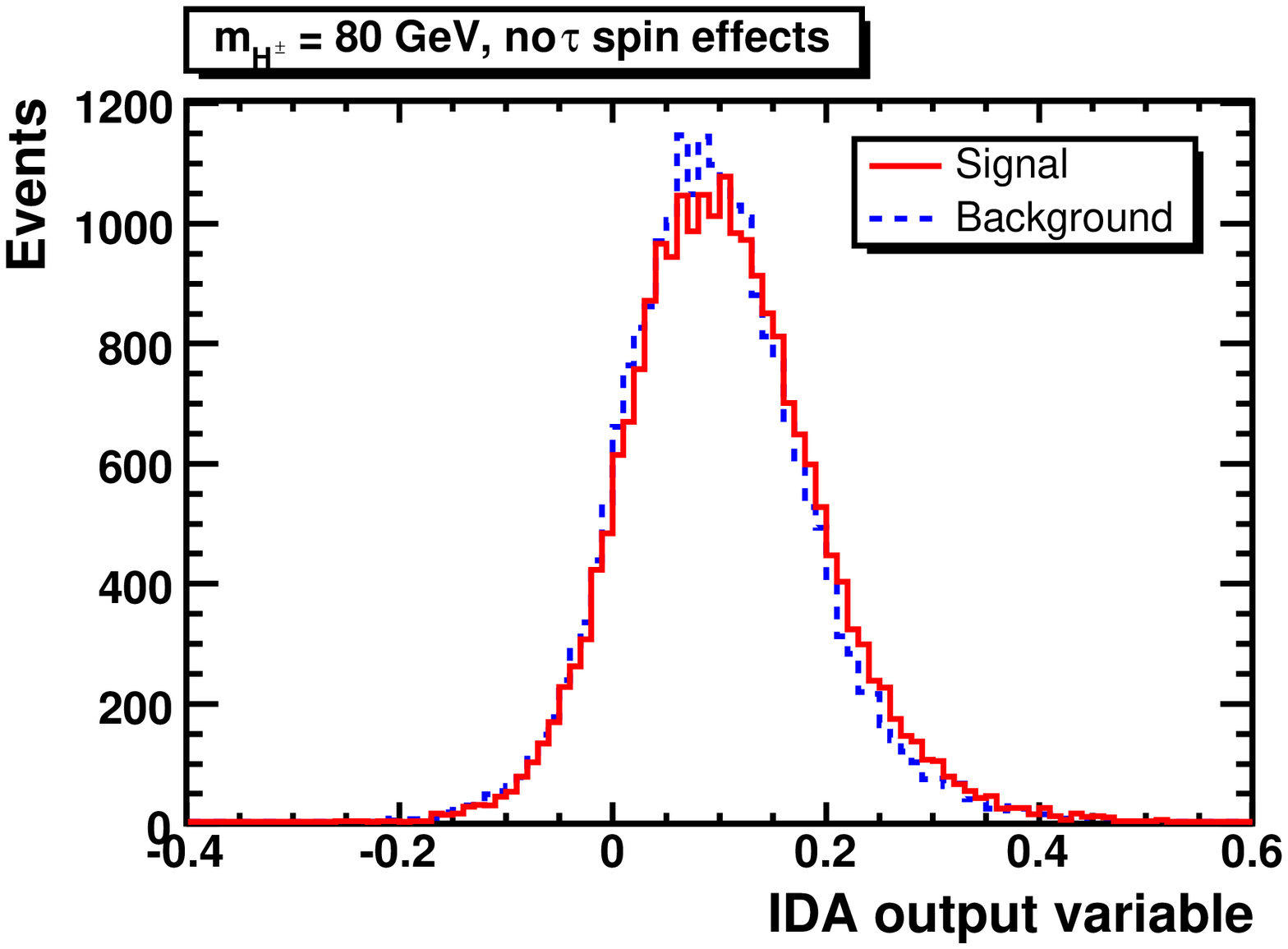, width=0.5\textwidth}  \hfill
\epsfig{file=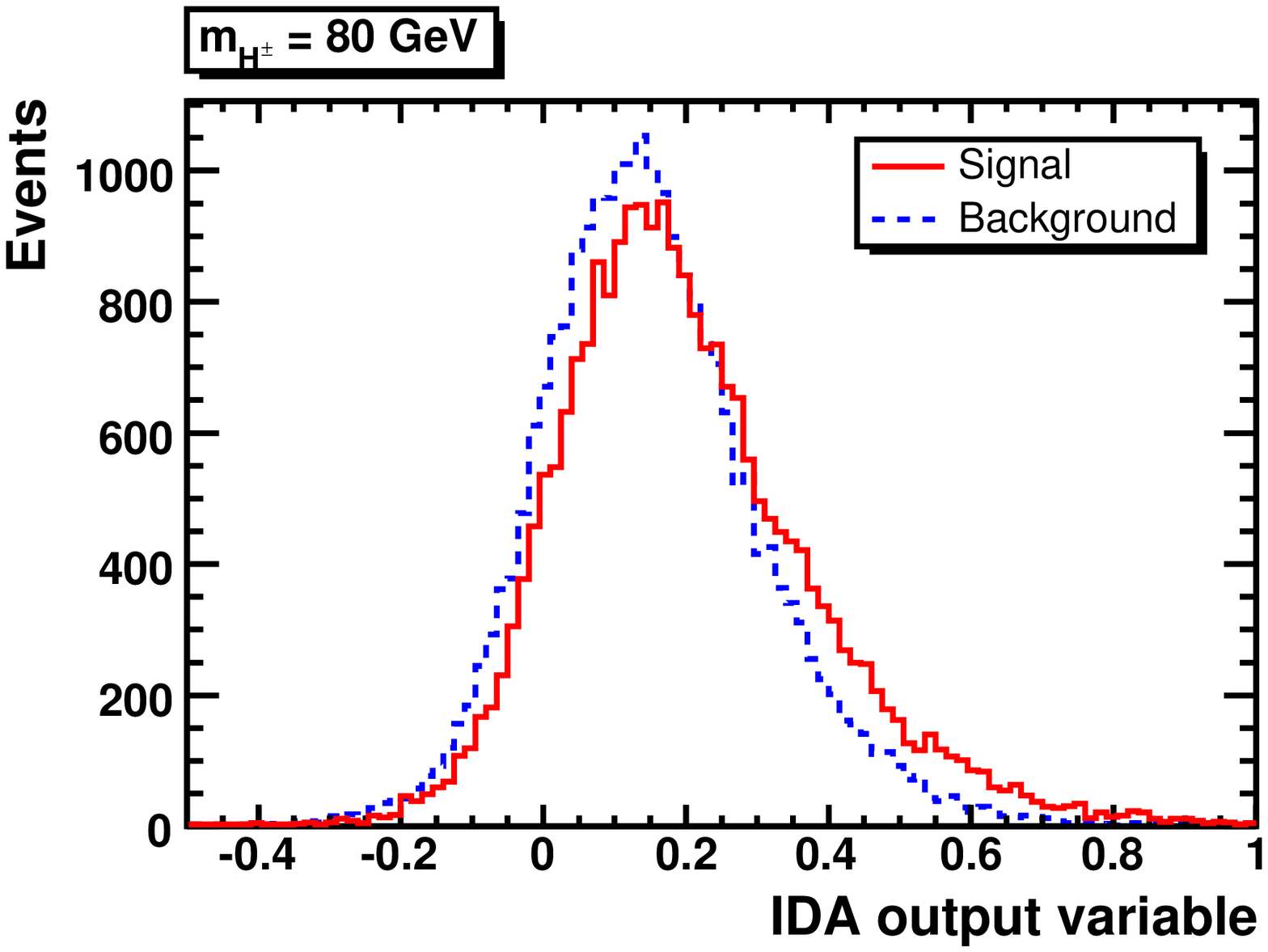, width=0.5\textwidth}
\epsfig{file=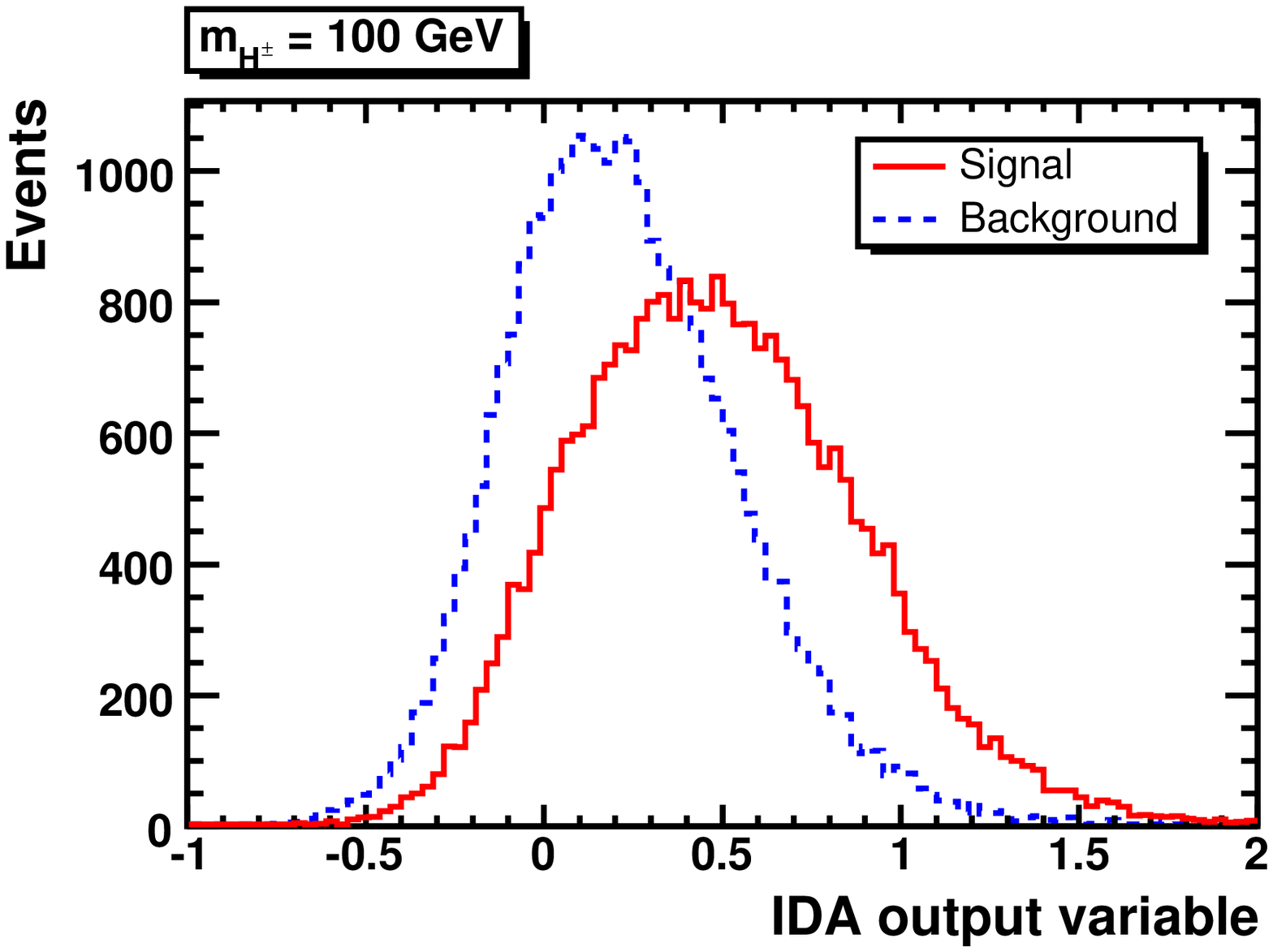, width=0.5\textwidth}  \hfill
\epsfig{file=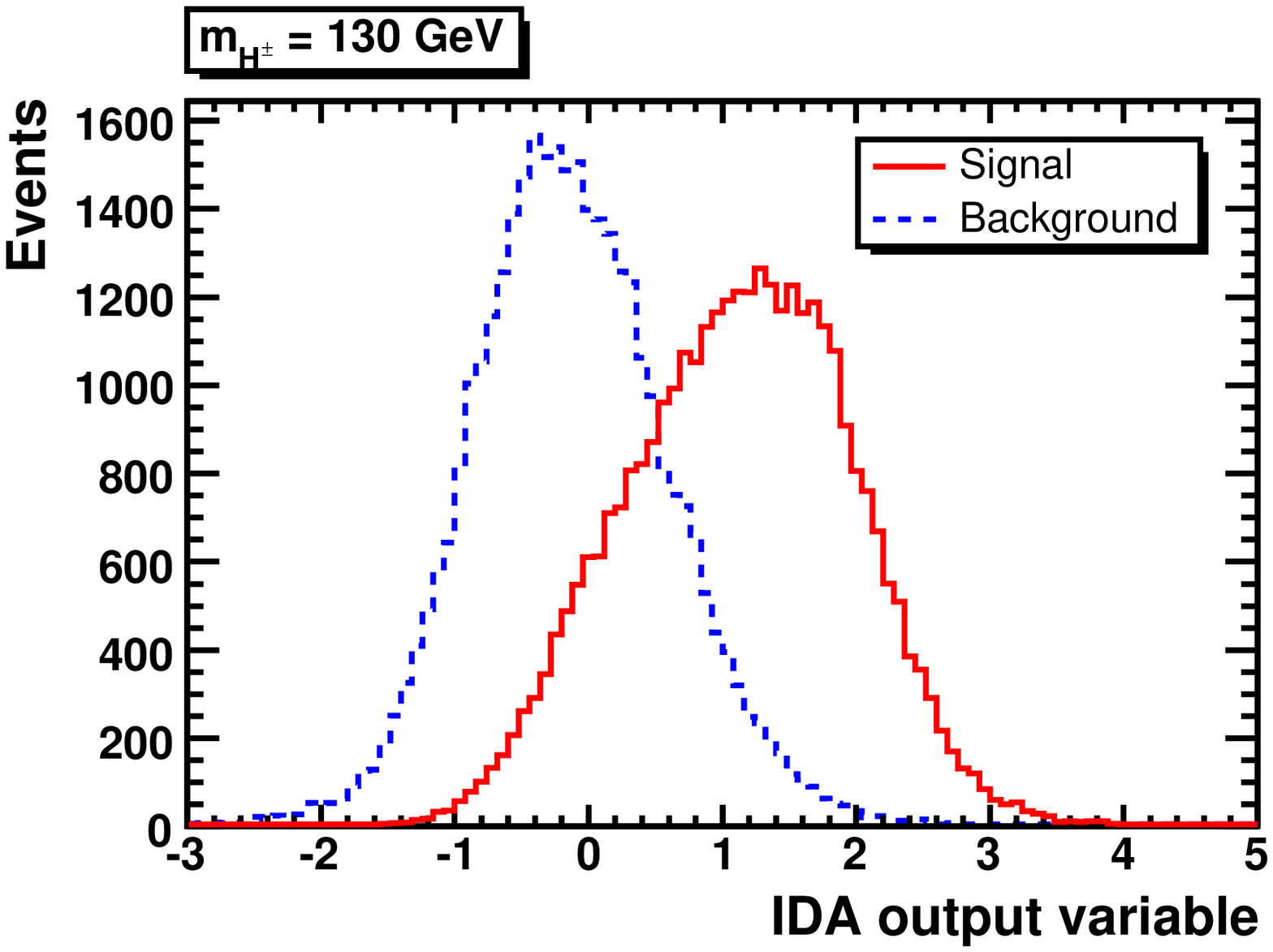, width=0.5\textwidth}
\epsfig{file=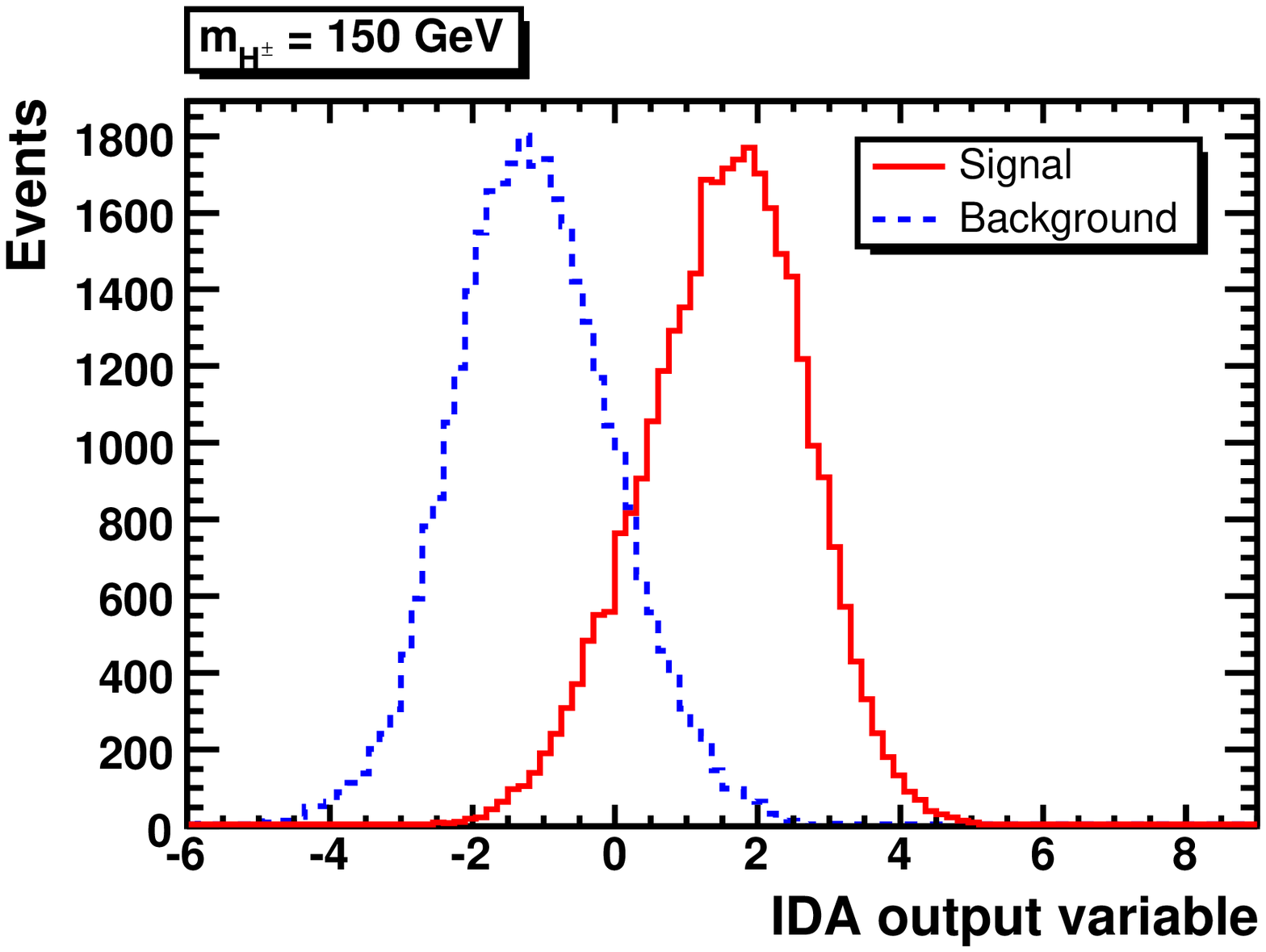, width=0.5\textwidth}  \hfill
\caption{
Distributions of the IDA output variable in the first IDA
step for the $tbH^\pm$ signal (solid, red)
and the $t\bar{t}$ background (dashed, blue) for $\sqrt{s}=14$~TeV.
}
\label{fig:lhc_ida1}
\end{figure}

\begin{figure}[htbp]
\epsfig{file=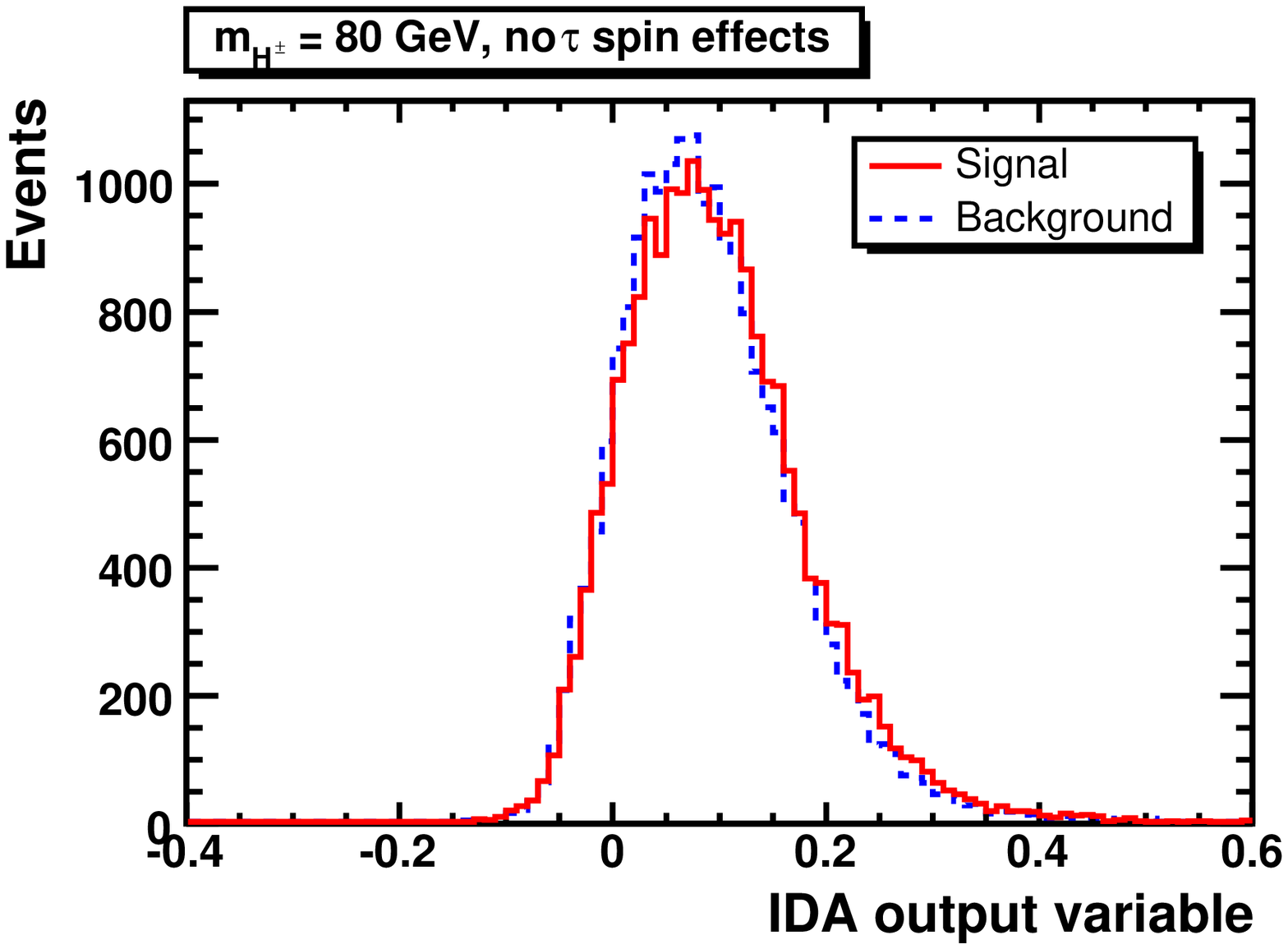, width=0.5\textwidth}  \hfill
\epsfig{file=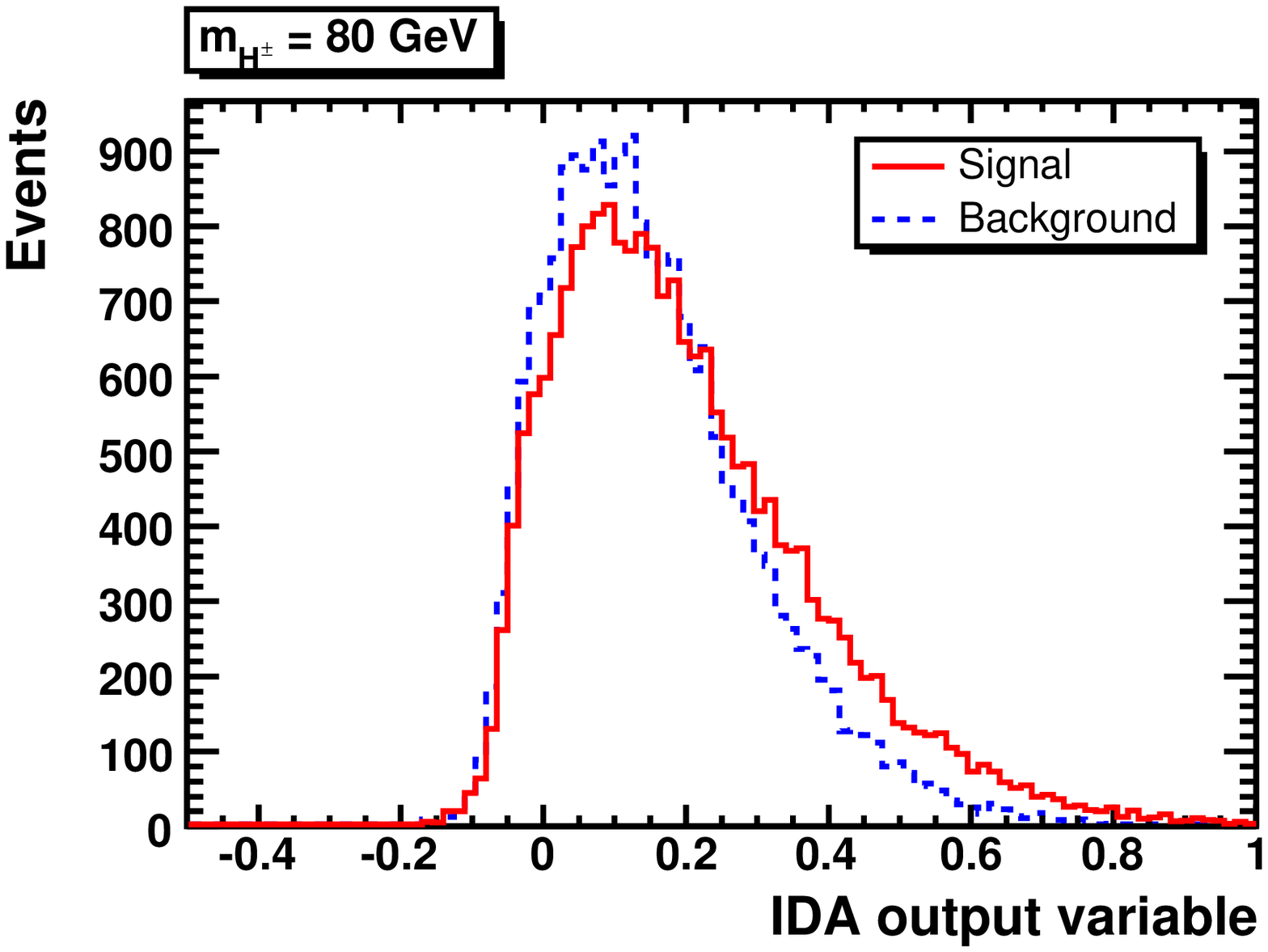, width=0.5\textwidth}
\epsfig{file=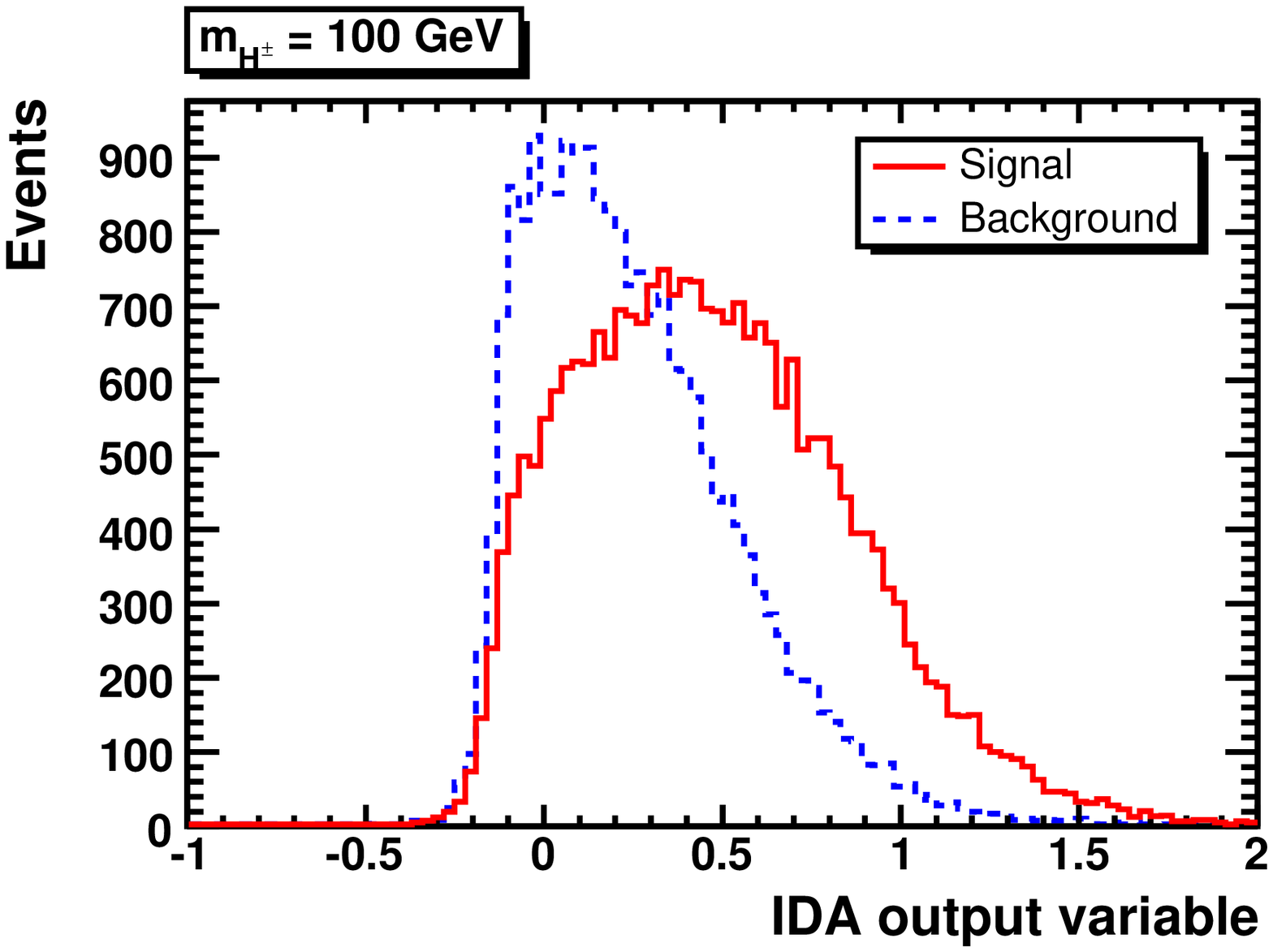, width=0.5\textwidth}  \hfill
\epsfig{file=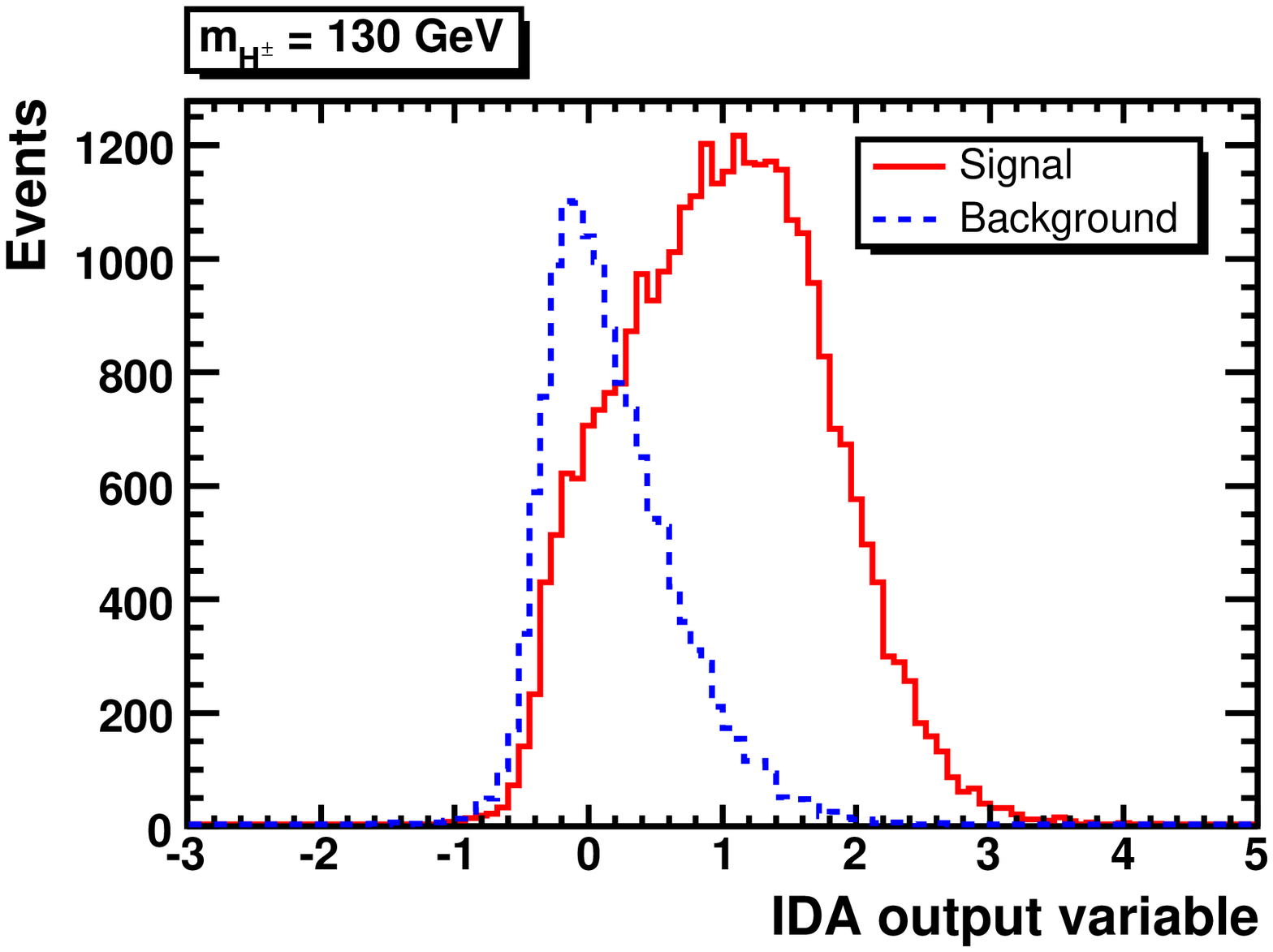, width=0.5\textwidth}
\epsfig{file=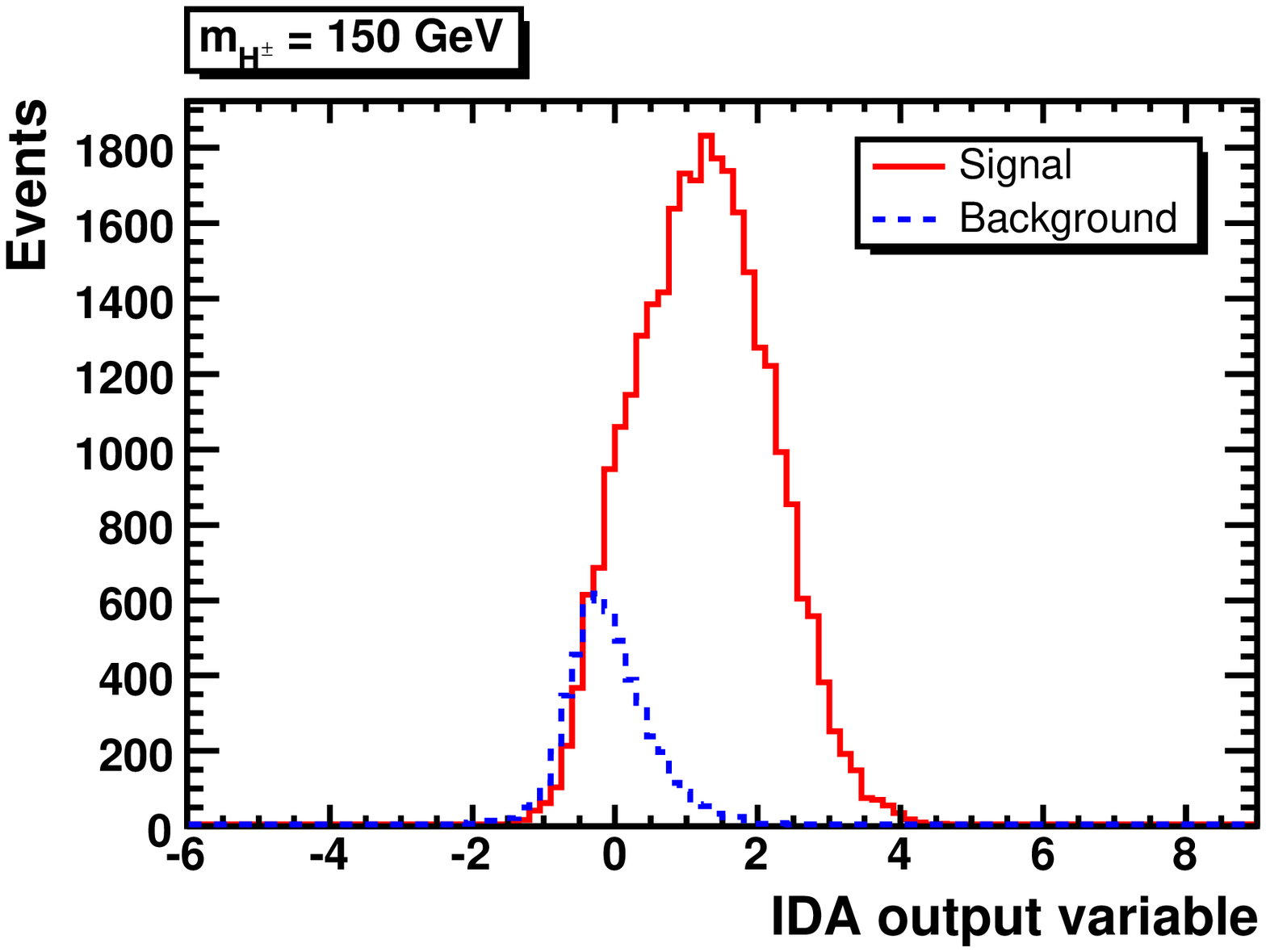, width=0.5\textwidth}  \hfill
\epsfig{file=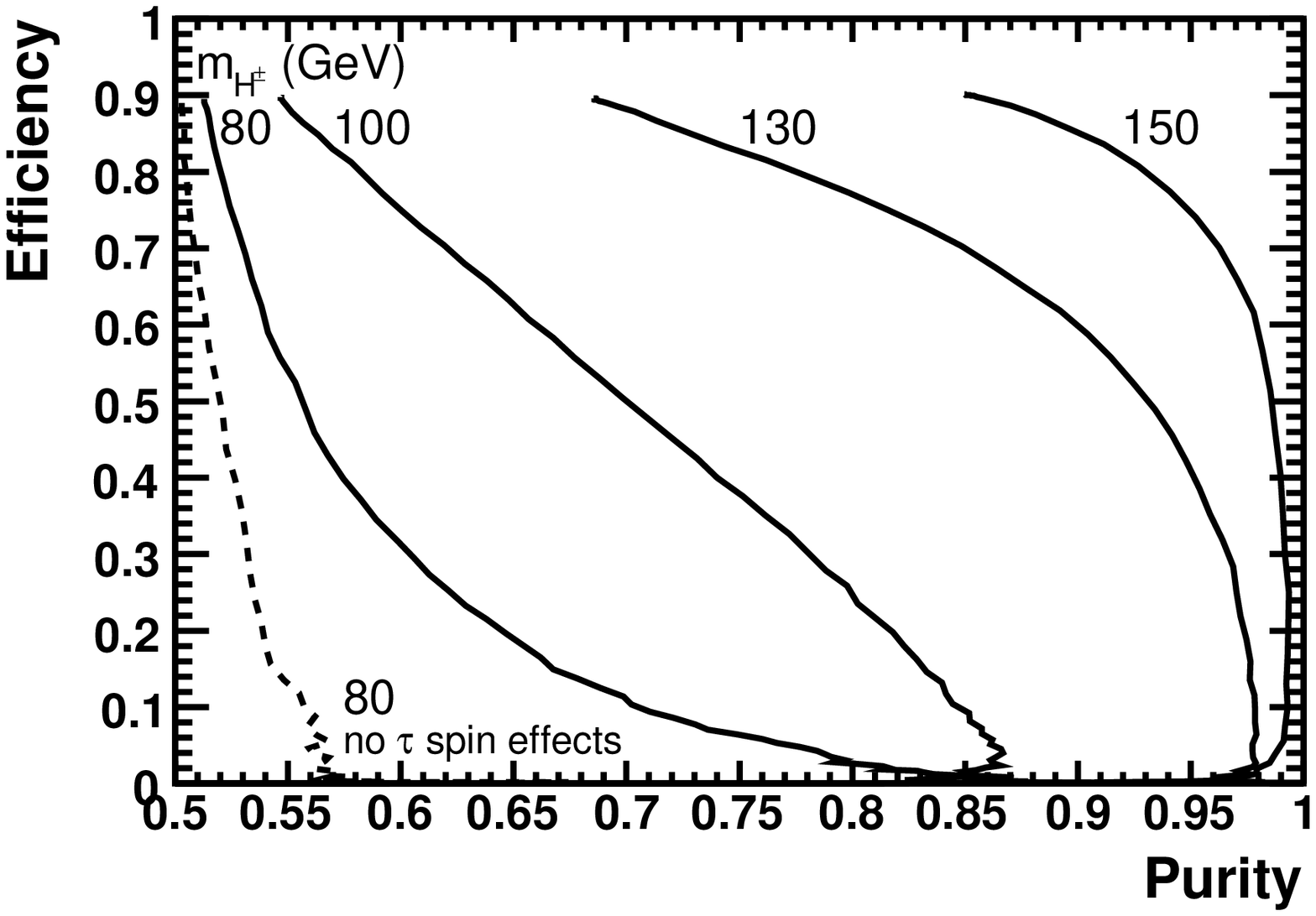, width=0.5\textwidth}
\caption{
Upper row, middle row and lower left figure:
distributions of the IDA output variable 
in the second IDA step
for 90\% efficiency in the first IDA
step (corresponding to a cut at 0 in Fig.~\ref{fig:lhc_ida1})
for the $tbH^\pm$ signal (solid, red)
and the $t\bar{t}$ background (dashed, blue).
Lower right figure: efficiency as a function of the purity
when not taking the spin effects in the $\tau$ decay into account for
$m_{H^\pm}=80$~GeV (dashed) and with spin effects in the $\tau$ decay for
$m_{H^\pm}=80,100,130,150$~GeV (solid, from left to right).
Results are for the LHC.
}
\label{fig:lhc_ida}
\end{figure}

The distributions of the IDA output variables for the study at
$\sqrt{s}=14$~TeV for two steps with 90\% efficiency in the first step
are shown in Figs.~\ref{fig:lhc_ida1} and \ref{fig:lhc_ida}.
These distributions are qualitatively similar to those for the  Tevatron 
(Figs.~\ref{fig:ida1} and \ref{fig:ida}) and the final achievable purity for a 
given efficiency is shown in Fig.~\ref{fig:lhc_ida}. As for the Tevatron energy 
a good separation of signal and background events can be achieved with the spin 
sensitive variables and the IDA method even in case $m_{H^\pm} \sim m_{W^\pm}$.
For heavier $H^\pm$ masses the separation of signal and background events increases due to 
the kinematic differences of the event topology.

\clearpage
\section{CONCLUSIONS}
The discovery of charged Higgs bosons 
would be a clear sign of physics beyond the SM.
In this case study we have investigated charged Higgs boson topologies 
produced at the current Tevatron and LHC energies and compared
them against the irreducible SM background due to top-antitop production
and decay. 
While sizable differences between signal and background
are expected whenever $m_{H^\pm}\ne m_{W^\pm}$, 
near the current mass limit of about $m_{H^\pm}\approx 80$ GeV the
kinematic spectra are very similar between SM decays and
those involving charged Higgs bosons. In this case,
spin information will significantly
distinguish between signal and irreducible SM background. In fact,
we have considered hadronic $\tau\nu_\tau$ decays of charged Higgs bosons, 
wherein the $\tau$ polarization induced by a decaying (pseudo)scalar object
is significantly different from those emerging in the vector ($W^\pm$) decays
onsetting in the top-antitop case. 
For a realistic analysis which is not specific for a particular detector, 
a dedicated Monte Carlo event generation and a simplified multipurpose 
detector response approximation have been applied.
The identification of a hadronic tau-lepton will be an experimental challenge 
in an environment with typically four jets being present. 
We have demonstrated how an IDA method can be an applied to separate signal and background
when the differences between the signal and background distributions are small. 
Our results show that the IDA method 
will be equally effective at both the Tevatron and LHC. 
While only the dominant irreducible $t\bar{t}$ background has been dealt 
with in detail, we have also specifically addressed the QCD background.
A suitably hard missing transverse momentum cut has been applied to
reject such jet activity 
and we have demonstrated that 
although the discriminative power is reduced by such a cut, the reduction is
small compared to the gain from including the $\tau$ polarization effects.
Using the differences in $\tau$
polarization between the signal and the dominant SM irreducible  $t\bar{t}$
background is crucial for disentangling the former
from the latter.

\section*{ACKNOWLEDGMENTS}

We would like to thank the organizers of the 2005 and 2007 editions of the
Les Houches workshops ``Physics at TeV Colliders' (where part of this work
was carried out) 
for their kind invitation and Johan Alwall for fruitful discussions.
SH and AS thank the Department of Nuclear and Particle Physics, Uppsala
University, for hospitality and financial support during the Charged2006 
workshop where part of this work was carried out.
SM thanks The Royal Society (London, UK) for partial financial support in the form 
of a `Conference Grant' to attend all the aforementioned workshops.

\end{document}